\documentclass[twocolumn,superscriptaddress,nofootinbib,notitlepage,longbibliography,aps]{revtex4-1}
\pdfoutput=1

\usepackage{graphicx}
\usepackage{amsmath}
\usepackage{amssymb}
\usepackage{amsfonts}
\usepackage[dvipsnames]{xcolor}
\usepackage[linktoc=none]{hyperref}
\usepackage[utf8]{inputenc}
\usepackage[normalem]{ulem}
\usepackage{comment}

\hypersetup{colorlinks=true,linkcolor=blue,citecolor=teal,filecolor=magenta, urlcolor=cyan}

\newcommand{\INFN}{INFN - Sezione di Napoli, Complesso Universitario Monte S. Angelo, I-80126 Napoli, Italy}
\newcommand{\INFNSA}{INFN - Gruppo Collegato di Salerno, Via Giovanni Paolo II, 132 - 84084 Fisciano (SA), Italy.}
\newcommand{\UNINA}{Dipartimento di Fisica "Ettore Pancini", Università degli studi di Napoli ``Federico II'', Complesso Universitario Monte S. Angelo, I-80126 Napoli, Italy}
\newcommand{\UNISA}{Dipartimento di Fisica ``E.R. Caianiello'', Università degli Studi di Salerno, Via Giovanni Paolo II, 132 - 84084 Fisciano (SA), Italy}
\newcommand{\SSM}{Scuola Superiore Meridionale, Università degli studi di Napoli ``Federico II'', Largo San Marcellino 10, 80138 Napoli, Italy}
\newcommand{\NBIA}{Niels Bohr International Academy, Niels Bohr Institute, University of Copenhagen, Copenhagen, Denmark}

\newcommand{\gag}[1]{g_{a\gamma}}

\def\XXint#1#2#3{{\setbox0=\hbox{$#1{#2#3}{\int}$ }
\vcenter{\hbox{$#2#3$ }}\kern-.6\wd0}}

\begin{document}

\title{Constraining axion-like particles with the diffuse gamma-ray flux \\ measured by the Large High Altitude Air Shower Observatory}

\author{Leonardo Mastrototaro}
\email{lmastrototaro@sa.infn.it}
\affiliation{\UNISA}
\affiliation{\INFNSA}

\author{Pierluca Carenza}
\email{pierluca.carenza@fysik.su.se}
\affiliation{The Oskar Klein Centre, Department of Physics, Stockholm University, Stockholm 106 91, Sweden}

\author{Marco Chianese}
\email{chianese@na.infn.it}
\affiliation{\UNINA}
\affiliation{\INFN}

\author{Damiano F.G. Fiorillo}
\email{damiano.fiorillo@nbi.ku.dk}
\affiliation{\UNINA}
\affiliation{\INFN}
\affiliation{\NBIA}

\author{Gennaro Miele}
\email{miele@na.infn.it}
\affiliation{\UNINA}
\affiliation{\INFN}
\affiliation{\SSM}

\author{Alessandro Mirizzi}
\email{alessandro.mirizzi@ba.infn.it}
\affiliation{Dipartimento Interateneo di Fisica ``Michelangelo Merlin'', Via Amendola 173, 70126 Bari, Italy.}
\affiliation{Istituto Nazionale di Fisica Nucleare - Sezione di Bari,
Via Orabona 4, 70126 Bari, Italy.}

\author{Daniele Montanino}
\email{daniele.montanino@le.infn.it }
\affiliation{Dipartimento di Matematica e Fisica ``Ennio De Giorgi'', Universit\`a del Salento, Via Arnesano, 73100 Lecce, Italy}
\affiliation{Istituto Nazionale di Fisica Nucleare - Sezione di Lecce,
Via Arnesano, 73100 Lecce, Italy.} 

\date{\today}
\begin{abstract}
The detection of very high-energy neutrinos by IceCube experiment supports the existence of a comparable gamma-ray counterpart from the same cosmic accelerators. {Under the likely assumption that the sources of these particles are of extragalactic origin,  the emitted photon flux would be significantly absorbed during its propagation over cosmic distances. However, in the presence of photon mixing with  ultra-light axion-like-particles (ALPs), this expectation would be strongly modified.} Notably,  photon-ALP conversions in the  host galaxy would produce an ALP flux which propagates unimpeded in the extragalactic space. Then, the back-conversion of ALPs in the Galactic magnetic field leads to a diffuse high-energy photon flux. In this context, the recent detection of the diffuse high-energy photon flux by the Large High Altitude Air Shower Observatory (LHAASO) allows us to exclude at the $95\%$ CL an ALP-photon coupling $g_{a\gamma}\gtrsim 3.9-7.8 \times 10^{-11}~\mathrm{GeV^{-1}}$  for $m_{a}\lesssim 4\times10^{-7}~\mathrm{eV}$, depending on the assumptions on the  magnetic fields and on  the original gamma-ray spectrum. This new bound is   complementary with other ALP constraints from very-high-energy gamma-ray experiments and sensitivities of future experiments.
\end{abstract}

\maketitle

\section{Introduction}

Ultra-light Axion-Like-Particles (ALPs) often appear in various Standard Model extensions, including effective models derived from string theory~\cite{Svrcek:2006yi,Arvanitaki:2009fg,Cicoli:2012sz}, or in the context of ``relaxion'' models~\cite{Graham:2015cka}. 
Similarly to the QCD  axion, ALPs can also be thought as pseudo-Nambu-Goldstone bosons of broken, approximate symmetries, (see e.g. Sec.~6.7 of Ref.~\cite{DiLuzio:2020wdo} for a recent review). 
A minimal model of ALPs accounts for their electromagnetic coupling through the Lagrangian term~\cite{Raffelt:1987im}
\begin{equation}
{\cal L}_{a\gamma}=-\frac{1}{4} \,g_{a\gamma}
F_{\mu\nu}\tilde{F}^{\mu\nu}a=g_{a\gamma} \, {\bf E}\cdot{\bf B}\,a~,
\label{eq:lagrangian}
\end{equation}
where $g_{a\gamma}$ is the ALP-photon coupling constant, $F_{\mu \nu}$ is the electromagnetic field tensor and $\tilde{F}_{\mu\nu} = \frac{1}{2}\epsilon_{\mu\nu\rho\sigma}F^{\rho\sigma}$ is its dual, and the ALP, $a$, is assumed to have a mass $m_a$. The coupling in Eq.~\eqref{eq:lagrangian} induces a mixing between ALPs and photons in a background electromagnetic field, leading the two states to oscillate into one another~\cite{Anselm:1987vj,Raffelt:1987im}. This conversion phenomenon is the basis for the majority of the experimental and observational ALP searches (see, e.g., Ref.~\cite{Graham:2015ouw,Irastorza:2018dyq,Irastorza:2021tdu,Sikivie:2020zpn} for recent reviews). 
Conversions of very-high-energy (VHE) cosmic photons into ultra-light ALPs (with $m_a \lesssim 10^{-7}$~eV)  in cosmic magnetic fields of Galactic or extragalactic origin have been proposed as an intriguing possibility to perform ALP searches with  gamma-ray telescopes (see, e.g., Refs.~\cite{Mirizzi:2007hr,DeAngelis:2007wiw,DeAngelis:2007dqd,Hooper:2007bq,Simet:2007sa,DeAngelis:2011id} for seminal papers). 
VHE photon-ALP conversions would  imprint peculiar modulations in astrophysical spectra from faraway sources, such as blazars, active galactic nuclei, pulsars and galaxy clusters, see, e.g., Refs.~\cite{Mirizzi:2009aj,Horns:2012kw,Meyer:2013pny,HESS:2013udx, Meyer:2014epa,Meyer:2014gta, Fermi-LAT:2016nkz, Montanino:2017ara, Zhang:2018wpc, Xia:2018xbt,Majumdar:2018sbv, Galanti:2018upl,Liang:2018mqm, Bu:2019qqg, Cheng:2020bhr,Carenza:2021alz,Marsh:2021ajy,Reynes:2021bpe,Matthews:2022gqi,Jacobsen:2022swa} for an incomplete list of studies that pointed out intriguing hints and bounds on ALP parameter space (see also Ref.~\cite{Galanti:2022ijh} for a recent review).

The production of VHE gamma-rays in astrophysical environments is strictly connected with the emission of neutrinos. VHE astrophysical neutrinos ($E\gtrsim 10~{\rm TeV}$) detected in IceCube~\cite{IceCube:2013cdw,IceCube:2013low,IceCube:2014stg,IceCube:2015qii,IceCube:2016umi,Ahlers:2018fkn,IceCube:2020wum} are produced in connection with high-energy gamma-rays via the $pp$ and $p\gamma$ interactions, and they have comparable energies.
The energy and angular distribution of the neutrino flux detected at IceCube points to the extragalactic origin of these neutrinos and gamma-rays.
Photons with energies between a few TeV and a few PeV have a short mean free path (a few Mpc to about 10 kpc) compared to the extragalactic distances where the emitters are located. As such, these photons are not expected to reach the Earth. In presence of ALPs, however, this may become possible. Indeed, photons might convert into ALPs in the magnetic field of the source, travel unabsorbed until our Galaxy, and then convert back in the Galactic magnetic field.
This setup would allow one to realize a sort of cosmic ``light-shining-through-the-Universe'' experiment, as proposed in different papers, see e.g.~\cite{Simet:2007sa,Horns:2012kw}.

Ref.~\cite{Vogel:2017fmc} pointed out the physics potential of current and upcoming gamma-ray detectors to constrain the photon-ALP mixing, through a measurement of the diffuse gamma-ray flux generated by extragalactic sources of 100 TeV-PeV photons. 
In this regard, the recent Ref.~\cite{Eckner:2022rwf} used the diffuse gamma-ray signal measured by Tibet AS$\gamma$ and HAWC to search for ALPs, constraining $g_{a\gamma} \lesssim 2.1\times 10^{-11}$~GeV$^{-1}$ for $m_a < 10^{-7}$~eV.
Following this interesting result, we take advantage of the recent preliminary measurement of the diffuse gamma-ray flux by the Large High Altitude Air Shower Observatory (LHAASO)~\cite{LHAASO:2019qtb,Zhao:2021dqj} to present a new bound on ALPs, complementary to the one of Ref.~\cite{Eckner:2022rwf}.
  
The plan of our work is as follows.
In Sec.~\ref{sec:flux} we discuss the expected photon spectra from the extragalactic sources exploiting the connection with the measured neutrino flux at IceCube.
Then in Sec.~\ref{sec:ALPconv} we revise the photon-ALP conversion mechanism, recalling the equations of motion for the photon-ALP ensemble and modelling the photon-ALP conversions in the magnetic field of the host galaxy and the back-conversions in the Milky-Way.
We arrive in this way at characterizing the diffuse gamma-ray flux produced by the ALP conversions.
In Sec.~\ref{Sec:Analysis and results} we show how to obtain a bound on the ALP-photon coupling, requiring that 
the diffuse photon flux produced by the ALP-photon oscillations in the Galactic magnetic field does not exceed the flux observed in LHAASO.
In Sec.~\ref{sec:comparison} we discuss the complementarity of our bound with other ones from very-high-energy photon observations.
Finally, in Sec.~\ref{sec:conclusions}  we summarize our results and we conclude.
In Appendix 
we comment on the dependence of our bound on the host galaxy magnetic field, on the assumption on the star formation rate, and on the impact of a photon Galactic background on our results.

\section{Gamma-ray sources and initial fluxes}
\label{sec:flux}

The ALP flux entering our Galaxy originates from the gamma-ray production in the extragalactic sources of IceCube neutrinos. The extragalactic origin of IceCube neutrinos is well motivated by their energies of the order of 1~PeV, which require acceleration of cosmic-rays up to tens of PeV, a feature that is much more naturally realized in extragalactic rather than Galactic sources.  Since the sources are outside of the Galaxy, they can naturally be taken as isotropically distributed. Therefore, even though the regions probed by IceCube and LHAASO are different, we can naturally extend our inferred source properties from the IceCube data to the region probed by LHAASO.

We are assuming photohadronic $p\gamma$ (or hadronic $pp$) neutrino production in a compact region inside a host galaxy. This neutrino production is accompanied by a corresponding gamma-ray production which escapes the compact region. There are two crucial assumptions here: first of all, we are assuming the compact region to be  transparent for gamma-rays with energy of the order of $100$~TeV. This is a tricky assumption, especially in view of the fact that the sources of IceCube neutrinos in the $10-100$~TeV are likely to be gamma-ray opaque~\cite{Murase:2015xka,Capanema:2020rjj,Capanema:2020oet} because of the tension between the IceCube data and the diffuse gamma-ray background Fermi-LAT measurements. In fact, using a broken power-law spectrum close to our parameterization in Eq.~(\ref{neutrino spectrum}) with a break energy $E_b$, Refs.~\cite{Capanema:2020rjj,Capanema:2020oet} show that with $E_b\lesssim 60$~TeV explaining the IceCube data would imply a cascaded gamma-ray flux that would exceed the Fermi-LAT measurements of the extragalactic gamma-ray background.
Furthermore, as noted in Ref.~\cite{Murase:2015xka}, it is quite natural for $p\gamma$ sources of neutrinos between $25$~TeV and $3$~PeV to be opaque to gamma-rays between $1$~GeV and $100$~GeV, since the target photons for $p\gamma$ interactions also act as a target for photon attenuation. However, this does not mean that gamma-rays above $100$~TeV should be absorbed. In this higher energy range, which is of interest here, the neutrino production can be explained by gamma-ray transparent sources without exceeding the measured extragalactic gamma-ray background. Thus, current data do not suggest sources opaque to gamma-rays with these energies, and we consider it likely that gamma-rays above $100$~TeV manage to escape the compact region into the host galaxy.

Our second assumption is indeed that the neutrino sources are embedded in a galactic environment. This assumption is likely verified in most candidates proposed for explaining the IceCube data, including star-forming and starburst galaxies~\cite{Loeb:2006tw,Thompson:2006np,Tamborra:2014xia,Chang:2014hua,Senno:2015tra,Chakraborty:2015sta,Peretti:2018tmo,Peretti:2019vsj,Ambrosone:2020evo,Ambrosone:2021aaw}, active galactic nuclei~\cite{Stecker:1995th,Atoyan:2001ey,Murase:2014foa,Padovani:2016wwn,Alvarez-Muniz:2004xlu,Peer:2009vnw,Palladino:2018lov}, and gamma-ray bursts~\cite{Paczynski:1994uv,Waxman:1997ti,Murase:2005hy,Baerwald:2010fk,Bustamante:2016wpu}.

Concerning the neutrino flux, we follow the model in Ref.~\cite{Vogel:2017fmc}. In order to exploit the neutrino-gamma connection, we consider at first photohadronic sources. In each source the neutrino spectrum is given by~\cite{Vogel:2017fmc}
\begin{equation}
    Q_\nu(E)=\frac{dN_\nu}{dE dt} \propto
\left[1+\left(\frac{E}{E_b}\right)^{2\alpha}\right]^{-1/2}.
    \label{neutrino spectrum}
\end{equation}
We consider three models from three different IceCube data analyses: the 9.5-year through-going (TG) muon neutrinos data sample~\cite{IceCube:2021uhz}, the 6-year cascades data sample~\cite{Aartsen:2020aqd}, and the 7.5-year high-energy starting events (HESE) one~\cite{Abbasi:2020jmh}. The three models mainly differ for the spectral index $\alpha$, namely
\begin{equation}
\label{cases} \alpha =  \left\{
\begin{array}{ll}
2.37 & \text{TG \,$\nu_\mu$} \,\ , \\
2.48 & \text{cascades} \,\ , \\
2.92 & \text{HESE} \,\ . \\
\end{array}\right.
\end{equation}
For all the models, the break energy is taken as $E_b=60$ TeV to avoid exceeding Fermi-LAT data~\cite{Capanema:2020rjj,Capanema:2020oet}. We want to point out that the different choice of $E_b$ respect to Ref.~\cite{Eckner:2022rwf,Vogel:2017fmc}, where $E_b=25~\mathrm{TeV}$, has no impact on the final result of this work because it does not strongly affect the detected flux at the energies of interest, close to $300~{\rm TeV}$.
Indeed, the normalization of the neutrino spectrum in Eq.~\eqref{neutrino spectrum} is fixed from the measurement of the diffuse neutrino flux $d \phi_\nu/d E$ at 100~TeV by inverting the following equation~\cite{Eckner:2022rwf}:
\begin{eqnarray}
    \frac{d\phi_\nu}{dE}&=&\int_0^{\infty} \Big[(1+z)Q_{\nu}(E(1+z)) \Big] 
      n_s(z) \left|c\frac{dt}{dz}\right|\, dz \, .
    \label{eq:neutrino_flux}
\end{eqnarray}
where the first term in large brackets is the emission neutrino spectrum $Q_\nu$ for neutrinos emitted at redshift $z$ with the prefactor of $(1+z)$ accounting for the compression of the energy scale, and the second term $n_s(z)$ is the comoving source density  that we assume proportional to the star formation rate (SFR), that we describe it by the functional fit of Ref.~\cite{Yuksel:2008cu} so that
\begin{equation}
	n_{s}(z) = n_{s}(0) \biggl[ (1+z)^{\alpha \eta}+\biggl(\frac{1+z}{B}\biggr)^{\beta\eta}+\biggl(\frac{1+z}{D}\biggr)^{\gamma\eta}\biggr]^{1/\eta}\;,
	\label{Eq:RSF}
	\end{equation}
where $n_{s}(0)$ is the normalization (in units of $ 10^{-6} \,\ \textrm{Mpc}^{-3}$), $B$ and $D$ encode the redshift breaks, the transitions are smoothed by the choice $\eta \simeq -10$, and $\alpha$, $\beta$, and $\gamma$ are the logarithmic slopes of the low, intermediate, and high redshift regimes, respectively. The constants $B$ and $D$ are defined as
\begin{eqnarray}
	B &=&(1+z_1)^{1-\alpha/\beta} \,\ , \nonumber \\
	D &=& (1+z_1)^{(\beta-\alpha)/\gamma}(1+z_2)^{1-\beta/\gamma} \,\ ,
\end{eqnarray}
where $z_1$ and $z_2$ are the redshift breaks. All the parameters of the model are collected in Tab.~\ref{tab:RSFparameters} based on~\cite{Horiuchi:2008jz}. In Tab.~\ref{tab:RSFparameters} we have also reported the source  neutrino spectrum that we consider for each SFR, defining Upper, Fiducial, and Lower benchmark initial neutrino fluxes.  
\begin{table}[!t]
	\caption{Model parameters for the the source density $n_s$, Eq.~(\ref{Eq:RSF}), values taken from~\cite{Horiuchi:2008jz}. The last column reports the neutrino data we adopt to compute the source neutrino  spectrum $Q_\nu$ in the three cases.}
	\begin{center}
	\begin{tabular}{lccccccc}
		\hline
		Analytic fits & \,\ $n_{s}(0)$ \,\ & \,\ $\alpha$ \,\ & \,\ $\beta$  \,\  & \,\ $ \gamma$  \,\  &  \,\ $ z_1$  \,\  &   \,\ $z_2$ \,\ & \,\ $Q_\nu$\,\ \\
		\hline
		\hline
		Upper & 0.0213 & 3.6 & -0.1 & -2.5 & 1 & 4 & HESE \\
		Fiducial & 0.0178 & 3.4 & -0.3 & -3.5 & 1 & 4 & Cascades \\
		Lower & 0.0142 & 3.2 & -0.5 & -4.5 & 1 & 4 & TG \,$\nu_\mu$\\
		\hline
	\end{tabular}
	\label{tab:RSFparameters}
	\end{center}
\end{table}

\begin{figure}
    \centering
    \includegraphics[scale=0.36]{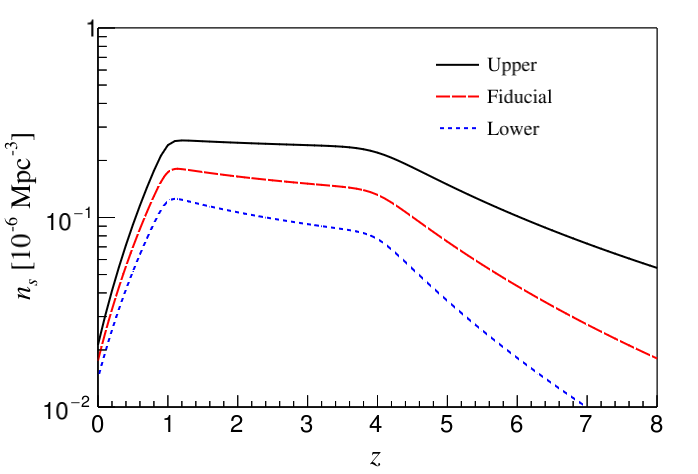}
    \caption{
    Source density $n_s$ in function of the redshift $z$. We show $n_s$ for the three sets of parameters in Tab.~\ref{tab:RSFparameters}:  Upper (black continuous curve), Fiducial (red dashed curve) and Lower (blue dotted curve). 
    }
    \label{fig:sfr}
\end{figure}

In Fig.~\ref{fig:sfr} we show the
source  density of Eq.~(\ref{Eq:RSF}) as a function of the redshift $z$ for the three different set of parameters of Table~\ref{tab:RSFparameters}: Upper (black continuous curve), Fiducial (red dashed curve) and Lower (blue dotted curve). It is possible to observe how the maximum  contribution comes from $1\lesssim z \lesssim 4$, for all the cases.

\begin{figure}
    \centering
    \includegraphics[scale=0.37]{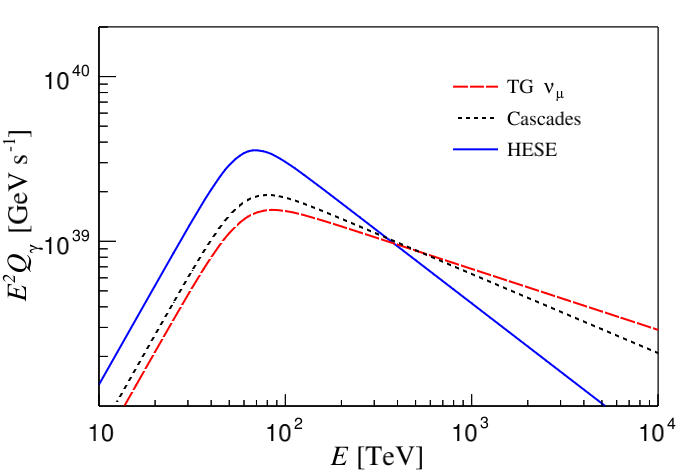}
    \caption{Photon spectrum $Q_{\gamma}$ from a single source as a function of the energy, according to three different neutrino data-sets: TG muon neutrinos (red dashed line), cascades (black dotted line) and HESE (blue continuous line).}
    \label{fig:Q_gamma}
\end{figure}

Having ascertained the number of neutrinos emitted from each source, we obtain the corresponding gamma-ray spectrum by the multi-messenger relation for $p\gamma$ interactions~\cite{Halzen:2018iak}:
\begin{equation}
   Q_\gamma(E_\gamma)= \frac{2}{3}Q_\nu\left(\frac{E_\gamma}{2}\right) \, .
   \label{eq:source}
\end{equation}
While this relation specifically applies to neutrinos produced by photohadronic interactions, it is still valid as order of magnitude for $pp$ sources. In Fig.~\ref{fig:Q_gamma} we show $Q_{\gamma}$ as a function of the energy $E$, obtained considering our fiducial SFR (see Sec.~III C) and the three data-sets, HESE (black dotted line), cascades (blue continuous line), and TG $\nu_\mu$ (red dashed line).

\section{Photon-ALP conversions}
\label{sec:ALPconv}

\subsection{Equations of motion}
The initial gamma-ray flux, in presence of ALPs, is strongly modified compared to the standard case. The Lagrangian describing the interaction between ALPs and photons is shown in Eq.~\eqref{eq:lagrangian} and it allows for ALP-photon conversions in the  magnetic field of the host galaxy and the back-conversions in the large-scale Galactic magnetic field.

Assuming a monochromatic photon/ALP beam of energy $E$ propagating along the $x_3$ direction in a cold ionized and magnetized medium, 
for very relativistic ALPs and  photons,
the evolution can be described in terms of the 
Liouville equation~\cite{Raffelt:1987im,Kartavtsev:2016doq}
\begin{equation}
    i\frac{d}{dx_3}\mathbf{\rho}=[\mathcal{H}_0,\rho]-\frac{i}{2}\{ \mathcal{H}_{\rm abs},\rho \} \,\ ,
    \label{Eq:rho}
\end{equation}
for the polarization density matrix
\begin{equation}
\rho (x_3) = \left(\begin{array}{c}A_1(x_3)  \\A_{2} (x_3) \\
a (x_3) \end{array}\right) \otimes \left(\begin{array}{c}A_1
(x_3)\  A_{2} (x_3)\ a (x_3)\end{array}\right)^{*}
\end{equation}
where $A_1(x_3)$ and $A_2 (x_3)$ are the  photon linear polarization amplitudes along the $x_1$ and $x_2$ axis, respectively, and $a (x_3)$ denotes the ALP amplitude. In Eq.~\eqref{Eq:rho} the first commutator at right-hand-side contains the ALP-photon mixing Hamiltonian  
$\mathcal{H}_0$ and the second anticommutator contains the photon absorption Hamiltonian $\mathcal{H}_{\rm abs}$.

The mixing Hamiltonian  ${\cal H}_0$  simplifies if we restrict our attention to the case in which ${\bf B}$ is homogeneous. We denote by ${\bf B}_T$ the transverse magnetic field, namely its component in the plane normal to the beam direction and we choose the $y$-axis along ${\bf B}_T$ so that $B_x$ vanishes. The linear photon polarization state parallel to the transverse field direction ${\bf B}_T$ is then denoted by $A_{\parallel}$ and the orthogonal one by $A_{\perp}$. Correspondingly, the mixing matrix can be written as~\cite{Mirizzi:2005ng,Mirizzi:2006zy}
\begin{equation}
{\cal H}_0 =   \left(\begin{array}{ccc}
\Delta_{ \perp}  & 0 & 0 \\
0 &  \Delta_{ \parallel}  & \Delta_{a \gamma}  \\
0 & \Delta_{a \gamma} & \Delta_a 
\end{array}\right)~,
\label{eq:massgen}
\end{equation}
whose elements are~\cite{Raffelt:1987im}: $\Delta_\perp \equiv \Delta_{\rm pl} + \Delta_{\perp}^{\rm CM} + \Delta_{\rm CMB},$ $ \Delta_\parallel \equiv \Delta_{\rm pl} + \Delta_{\parallel}^{\rm CM} + \Delta_{\rm CMB}$, $\Delta_{a\gamma} \equiv {g_{a\gamma} B_T}/{2} $ and $\Delta_a \equiv - {m_a^2}/{2E}$, where $m_a$ is the ALP mass. The term $\Delta_{\rm pl} \equiv -{\omega^2_{\rm pl}}/{2E}$ takes into account  plasma effects, in terms of the plasma frequency $\omega_{\rm pl}$ expressed as a function of the electron density in the medium $n_e$ as $\omega_{\rm pl} \simeq 3.69 \times 10^{- 11} \sqrt{n_e /{\rm cm}^{- 3}} \, {\rm eV}$. The terms $\Delta_{\parallel,\perp}^{\rm CM}$ represent the Cotton-Mouton  effect, accounting for the birefringence
of fluids in the presence of a transverse magnetic field.  A
vacuum Cotton-Mouton effect is expected from QED one-loop
corrections to the photon polarization in the presence of an
external magnetic field $\Delta_\mathrm{QED} \propto
|\Delta_{\perp}^{\rm CM}- \Delta_{\parallel}^{\rm CM}|
\propto B^2_T$, precisely $\Delta_\parallel=\frac{7}{2}\Delta_\mathrm{QED}$ and $\Delta_\perp=2\Delta_\mathrm{QED}$~\cite{Raffelt:1987im}. 
Finally, the term $\Delta_{\rm CMB}
\propto \rho_{\rm CMB}$ represents the 
background photon contribution to the photon
polarization \cite{Dobrynina:2014qba}.  An
off-diagonal $\Delta_{R}$ would induce the Faraday rotation,
which is however totally irrelevant at VHE, and so it has
been dropped. 
For relevant parameters at redshift $z=0$ we use
\begin{eqnarray}  
\Delta_{a\gamma}&\simeq &   1.5\times10^{-2} 
\left(\frac{g_{a\gamma}}{10^{-11}\textrm{GeV}^{-1}} \right)
\left(\frac{B_T}{10^{-6}\,\rm G}\right) {\rm kpc}^{-1}
\nonumber\,,\\  
\Delta_a &\simeq &
 -0.8 \times 10^{-4} \left(\frac{m_a}{10^{-8} 
        {\rm eV}}\right)^2 \left(\frac{E}{10^2 \,\ {\rm TeV}} \right)^{-1} 
        {\rm kpc}^{-1}
\nonumber\,,\\  
\Delta_{\rm pl}&\simeq & 
  -1.1\times10^{-12}\left(\frac{E}{10^2 \,\ {\rm TeV}}\right)^{-1}
         \left(\frac{n_e}{10^{-3} \,{\rm cm}^{-3}}\right) 
         {\rm kpc}^{-1}
\nonumber\,,\\
\Delta_{\rm QED}&\simeq & 
6.1\times10^{-4}\left(\frac{E}{10^2 \,\ {\rm TeV}}\right)
\left(\frac{B_T}{10^{-6}\,\rm G}\right)^2 {\rm kpc}^{-1} \nonumber\,, \\
\Delta_{\rm CMB}&\simeq & 
8.0\times 10^{-3}\left(\frac{E}{10^2 \,\ {\rm TeV}}\right)
{\rm kpc}^{-1} \,,
\label{eq:Delta0}\end{eqnarray}
where the expression for $\Delta_{\rm CMB}$ is a valid approximation for $E<100~\mathrm{TeV}$.
We will provide later the complete 
expression used in our work.

VHE photons undergo  pair production absorptions by background low energy photons $\gamma_{\rm VHE}+\gamma_{\rm EBL}\to e^+ + e^-$. We emphasize that we are  assuming the compact source inside the host galaxy to be gamma-ray transparent, and are only accounting for absorption in the larger host environment and in the Milky-Way. The absorptive part of the Hamiltonian in Eq.~\eqref{Eq:rho} can be
written in the form
\begin{equation}
{\cal H}_{\rm abs} =   \left(\begin{array}{ccc}
\Gamma  & 0 & 0 \\
0 &  \Gamma  & 0  \\
0 & 0 & 0 
\end{array}\right)~,
\label{eq:abs}
\end{equation}
where $\Gamma$ is the VHE  photon absorption rate, which as
a function of the incident photon  energy $E$
is given by~\cite{Gould:1967zzb} (see also~\cite{Mirizzi:2009aj})
\begin{equation}
\Gamma(E)=\int_{m_e^2/E}^{\infty} d\epsilon\, \frac{dn^{\rm bkg}_\gamma}{d\epsilon}
\int_{-1}^{1-\frac{2m_e^2}{E\epsilon}} d\xi\frac{1-\xi}{2}\sigma_{\gamma\gamma}(\beta) \,\ ,
\, \label{eq:gamma}\end{equation}
where the limits of integration in both integrals are determined by the kinematical threshold
of the process and
\begin{equation}
\nonumber
\sigma_{\gamma\gamma}(\beta) = \sigma_0
(1-\beta^2)\left[2\beta(\beta^2-2)+(3-\beta^4)\log\frac{1+\beta}{1-\beta}\right]
\, ,
\end{equation}
with $\sigma_0=1.25\times 10^{-25}$ cm$^2$, is the cross section for the pair production process~\cite{Heitler:1936jqw}
as a function of the electron velocity in the center of mass frame
$\beta=[1-2m_e^2/E\epsilon(1-\xi)]^{1/2}$. 
Here $\epsilon$ is the background photon energy, and 
$\xi$ is the cosine of the angle between the incident 
and the background photon. The photon background spectrum $dn^{\rm bkg}_\gamma/{d\epsilon}$
takes into account the $\gamma$ absorption caused by cosmic microwave background (CMB), and in the Milky Way we also account for the presence of starlight (SL) and infrared (IR) backgrounds. The SL+IR background is extracted from the \texttt{GALPROP} code~\cite{Porter:2017vaa}.

\begin{figure}
    \centering
    \includegraphics[scale=0.36]{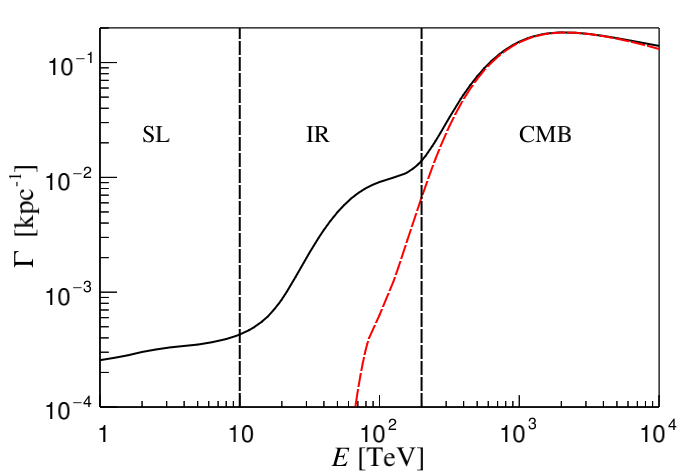}
    \includegraphics[scale=0.36]{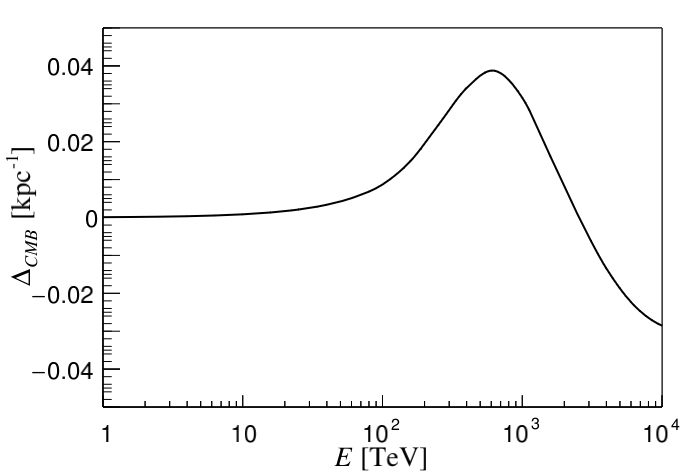}
    \caption{Upper panel: VHE photon absorption rate as a function of the photon energy $E$ for the host galaxy (red dashed curves) and the Milky Way near the Sun, using the  $dn^{\rm bkg}_\gamma/d\epsilon$ in Ref.~\cite{Dobrynina:2014qba} (black continuous curves). Lower panel: $\Delta_{\rm CMB}$ factor as a function of the photon energy $E$ obtained with Eq.~(\ref{eq:DeltaCMB}).}
    \label{fig:Gammafit}
\end{figure}

In the upper panel of Fig.~\ref{fig:Gammafit} we show the VHE photon absorption rate, obtained from Eq.~(\ref{eq:gamma}), as a function of the photon energy $E$, for the host galaxy (red dashed lines) and the Milky Way near the Sun (black continuous line).
The changing of the slope of the Milky Way $\Gamma$  reflects the three components of $dn^{\rm bkg}_\gamma/d\epsilon$ (CMB, SL and IR). Due to these components, the absorption rate is monotonically increasing for almost all the energy range considered. Only  at very-high-energies ($E_{\gamma}>3\times 10^3~\mathrm{TeV}$), $\Gamma$  decreases as expected, reflecting the decreasing behaviour of $\sigma_{\gamma\gamma}$ for $\beta\rightarrow 1$. The absorption term $\Gamma$ dominates the mixing term $\Delta_{a\gamma}$ for $E> 10^3$~TeV
for $B=5~\mathrm{\mu G}$ and $g_{a\gamma}<2.0\times 10^{-11}~\mathrm{GeV^{-1}}$.

Once obtained $\Gamma$, we can calculate the $\Delta_{\rm CMB}$ parameter as discussed in Ref.~\cite{Dobrynina:2014qba}:
\begin{equation} 
\Delta_{\rm{CMB}}= \frac{E}{\pi} \times \, \textrm{p.v.}\int_0^{\infty}dE'\frac{\Gamma(E')}{E'^2-E^2} \,\ ,
\label{eq:DeltaCMB}
\end{equation}
where p.v. indicates the Cauchy principal value integral.
This expression is valid in all the range of energy of our interest, i.e. $1~\mathrm{TeV}<E<10^4~\mathrm{TeV}$, and we use it in the conversion probability evaluation. In the lower panel of Fig.~\ref{fig:Gammafit} we show the $\Delta_{\rm CMB}$ parameter as a function of the photon energy $E$. One realizes that 
at $E>10^2~\rm TeV$ there are strong deviations with respect to the naive linear dependence of 
 Eq.~(\ref{eq:Delta0}), reflecting the behaviour of the absorption factor $\Gamma$.

\subsection{Photon-ALP conversions in the host galaxy}
\label{sec:Photon-ALP conversions in the source}

In principle the cosmic accelerators producing neutrinos and photons may host strong magnetic fields $B\sim\mathcal{O}(1-10)~{\rm \mu G}$ (see, e.g.,~\cite{moss1996turbulence,Tavecchio:2012um,Galanti:2018upl}).
However, due to the severe uncertainties in the characterization of these fields, we conservatively neglect this possibility. Besides being conservative, this is also pretty realistic: within the compact region, one needs rather strong magnetic fields to achieve an efficient conversion. However, large magnetic fields actually inhibits, rather than enhance, the photon-ALP conversion in the compact region, because of the large photon refraction in the magnetic field due to $\Delta_{\rm QED}$. Therefore, the conversion in the compact region is likely subdominant compared to the conversion in the host galaxy.
Therefore, we consider at first the conversions of the gamma-ray flux in the 
host galaxies where the sources are embedded. Concerning the strength of the magnetic fields, one 
expects values and morphology similar to our 
Milky Way (see Refs.~\cite{Fletcher:2011fn,Beck:2013bxa,Pakmor_2017} for recent studies). 
Furthermore, a combination of regular and turbulent components might be present. 
However, due to the presence of large uncertainties in the description of these source fields, we assume
two simplified models. Namely \emph{(a)} a box of constant magnetic field, mimicking in this way a regular field, \emph{(b)} a cell model where in each domain the magnetic field can change strength and direction, like in Ref.~\cite{Vogel:2017fmc}. This latter model would represent the pure turbulent case.

\subsubsection{Box model}
\label{sec:box model}

We start considering the propagation 
of photons in a single magnetic
domain with a uniform ${\bf B}$-field with $B_x = 0$, the 
component $A_{\perp}$ decouples away, and the propagation
equations reduce to a 2-dimensional problem. Its solution
follows from the diagonalization of the Hamiltonian through
a similarity transformation performed with an orthogonal
matrix, parametrized by the (complex) rotation angle
$\Theta$ which takes the
value~\cite{Raffelt:1987im,Kartavtsev:2016doq}
\begin{equation}
\Theta = \frac{1}{2}\textrm{arctan}\left(\frac{2 \Delta_{a
\gamma}}{\Delta_{\parallel}-\Delta_{a} - \frac{i}{2}
\Gamma}\right) \,\ . \label{theta}
\end{equation}
When $
\Delta_{a \gamma} \gg \Delta_{\parallel}-\Delta_{a}$ the
photon-ALP mixing is close to maximal, $\Theta \to \pi/4$
(if the absorption is small as well). On the other hand,
from Fig.~\ref{fig:Gammafit}
 one sees that
 for $E < 10^3$~TeV
 $\Delta_{\rm CMB}$
grows with the photon energy. Therefore  at
sufficiently high energies $ \Delta_{a \gamma} \ll
\Delta_{\parallel}-\Delta_{a}$  and the photon-ALP mixing is
suppressed. 

One can introduce a generalized (including absorption) 
photon-ALP oscillations frequency
\begin{equation}
\Delta_{\rm osc} \equiv \left[(\Delta_{\parallel}
-\Delta_{a} -\frac{i}{2} \Gamma)^2 + 4 \Delta_{a \gamma}^2
\right]^{1/2}~. \label{eq:deltaosc}
\end{equation}

As noticed before, for $E \lesssim 10^{3}$~TeV  the absorption effects are subleading.  
In this situation  the
probability for a photon emitted in the state
$A_{\parallel}$ to oscillate into an ALP after traveling a
distance $L$ is given by~\cite{Raffelt:1987im}
\begin{eqnarray}
P_{\gamma \to a}^{s}  &=& 
{\rm sin}^2 2 \Theta \  {\rm
sin}^2 \left( \frac{\Delta_{\rm osc} \, L}{2} \right)  \,\
\nonumber \\ &=&   (\Delta_{a \gamma} L)^2
\frac{\sin^2(\Delta_{\rm osc} L/2)}{(\Delta_{\rm osc}
L/2)^2} \,\ , \label{conv}
\end{eqnarray}
where in  the oscillation wave number and mixing angle we set $\Gamma=0$.
It is also useful to define a critical energy, above which $P_{a\gamma} \simeq 0$. Similarly to Ref.~\cite{Bassan:2010ya}, the terms in $\Delta_{\rm osc}$ can be rearranged such that a critical energy is defined as
\begin{equation}
\begin{split}
E_{\rm c}&=\frac{2\Delta_{a\gamma}E}{\Delta_{\parallel}+\Delta_{\rm CMB}}=\\
&=\frac{2.14\times10^{3} \,\ {\rm TeV}}{\left(\frac{B}{{\rm \mu G}}\right)^{2}+5.71}\left(\frac{g_{a\gamma}}{10^{-11}{\rm GeV}^{-1}}\right)\left(\frac{B}{\rm\mu G}\right) \,.
\end{split}
\label{eq:ec}
\end{equation}
Therefore, one expects that the conversion probability would be already 
strongly suppressed before absorption effects become relevant.

\begin{figure}
    \centering
    \includegraphics[scale=0.36]{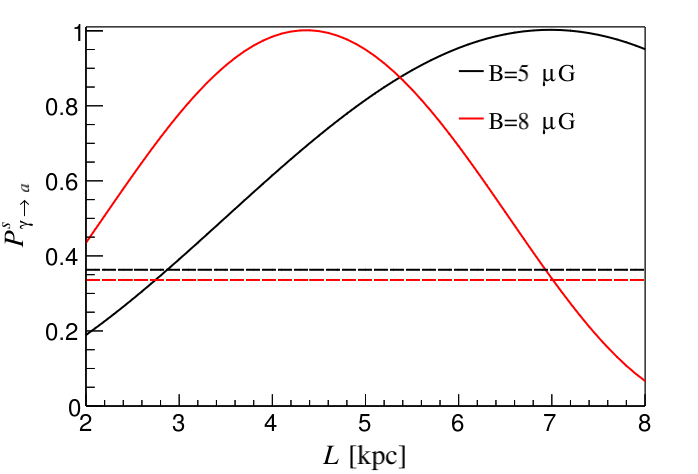}
    \caption{Photon-ALP conversion probability in the host-galaxy as a function of the source size $L$ for the case of a regular field, for $E=10~\mathrm{TeV}$, $g_{a \gamma}=3\times 10^{-11}~\mathrm{GeV^{-1}}$ and $z=0$. We consider two values of the magnetic field:  $B_T=8 \times 10^{-6} \,\ \mathrm{G}$ (red curves) and $B_T=5 \times 10^{-6} \,\ \mathrm{G}$ (black curves). The continuous curves represent the probability $P^{s}_{\gamma\rightarrow a} (L)$, while 
 the dashed ones represent the average $\langle P^{s}_{\gamma\rightarrow a}\rangle$.
 }
    \label{fig:prob_distance}
\end{figure}

For definiteness, we assume values of the magnetic field for the host galaxies in the ballpark of what suggested by observational constraints~\cite{Fletcher:2011fn,Beck:2013bxa,Pakmor_2017}. In particular we take ${ B}_T=(5\pm 3)  \,\ \mathrm{\mu G}$, constant on a box with $L=(5\pm 3)~\mathrm{kpc}$. In Fig.~\ref{fig:prob_distance} we show the value of the conversion probability $P_{\gamma\rightarrow a}^{s}$ as a function of the host-galaxy size for $E=10$ TeV, $g_{a \gamma}=3\times 10^{-11}~\mathrm{GeV^{-1}}$ and $B_T=8~ \mu\mathrm{G}$ (red continuous line), and $B_T=5~\mu\mathrm{G}$ (black continuous line).
This result must be averaged over the photon polarization, assuming unpolarized light. Because of the oscillating behavior with distance, with wavelength comparable to the propagation length, we choose to average over the path traversed by the photon as well:
 \begin{equation}
    \langle P^s_{\gamma\rightarrow a} \rangle= \frac{1}{2} \times \frac{\int_{L_{\rm min}}^{L_{\rm max}}\mathrm{d}L~ P^{s}_{\gamma\rightarrow a}}{\Delta L} \,\ ,
    \label{eq:aver}
 \end{equation}
 where $L_{\rm min}=2~\mathrm{kpc}$, $L_{\rm max}=8~\mathrm{kpc}$, $\Delta L=6~\mathrm{kpc}$,
 and the factor $1/2$ takes into account the average over the two photon polarization states.
 The averaged probabilities are also shown in 
 Fig.~\ref{fig:prob_distance} in dashed curves. 
It results that the case of $B_T=5~\mu\mathrm{G}$ presents a larger  $ \langle P^s_{\gamma\rightarrow a} \rangle$ than the case with  $B_T=8~\mu\mathrm{G}$ because of the choice of magnetic model, which gives an oscillation length comparable with the extension of the magnetic field.

\subsubsection{Cell model}
\label{sec:cell model}

As a second case, following Ref.~\cite{Vogel:2017fmc} 
we consider a different model for the source magnetic field to represent the turbulent case. The field is assumed to be divided into cells, whose length is fixed to $l=1$ kpc and where \textbf{B} has fixed strength and direction. In each cell, the ${\bf B}_T$ components are variable and follow a Gaussian distribution with zero mean and variance $2B_{T}^2/3$, such that $\langle |\mathbf{B}_{T}| \rangle=B_{T}$. We obtain the mean probability in each cell $\langle P_{\gamma\rightarrow a}\rangle$ by averaging over the Gaussian distribution of $B_{T}$. Then, to obtain the conversion probability for $n$ domains of length $l$, we follow the treatment in Ref.~\cite{Kartavtsev:2016doq} to which we address the reader for further details.
Finally, we obtain the total conversion probability in the source $\langle P^s_{\gamma\rightarrow a}\rangle$ averaging over the host-galaxy size.

\begin{figure}[!t]
    \centering
    \includegraphics[scale=0.36]{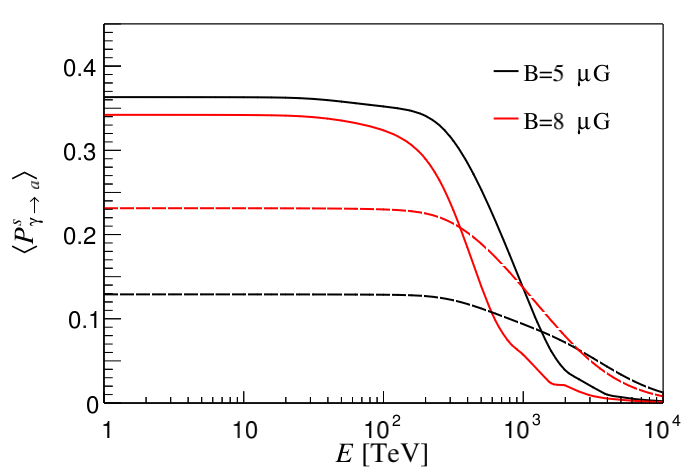}
    \caption{Average $\langle P^{s}_{\gamma\rightarrow a}\rangle$  conversion probabilities in the host galaxy at $z=0$ as a function of the energy $E$. We take $g_{a \gamma}=3\times 10^{-11}~\mathrm{GeV^{-1}}$ and consider $B_T=8 \times 10^{-6} \,\ \mathrm{G}$ (red curves) and $B_T=5 \times 10^{-6} \,\ \mathrm{G}$ (black curves). The continuous curves are for a single box regular $B$-field, while the dashed curves refer to the turbulent cell model.
 }
    \label{fig:prob_config}
\end{figure}

In Fig.~\ref{fig:prob_config} we show the average conversion probability $\langle P^s_{\gamma\rightarrow a}\rangle$ for the single box magnetic field (continuous curves) and for the  cell model (dashed curves). We take $g_{a\gamma}=3\times 10^{-11}~\mathrm{GeV^{-1}}$ and $z=0$. Concerning the magnetic field,  we take $B_T=5~\mu\mathrm{G}$ (black curves) and $B_T=8~\mu\mathrm{G}$ (red curves). The general trend of the probabilities is quite similar for the two models. However, the absolute value in the energy-independent region ($E< 10^2~\mathrm{TeV}$) is smaller in the cell model case due to the less efficient conversions because of the loss of coherence after each cell. Conversely, at higher energies one finds an opposite trend because of the reduced suppression proportional to the average magnetic field magnitude (see Eq.~\eqref{eq:ec}). On the other hand, when $E$ approaches $E_c \sim 10^3$ TeV [see Eq.~(\ref{eq:ec})] the conversion probability drops till it becomes negligible in both cases.\\

\subsection{Impact of the redshift}

Till now all the considerations we have done neglect the effect of redshift on the photon-ALP conversions. However, as evident from Fig.~\ref{fig:sfr} one expects the largest contributions from photon sources at $1 \lesssim z \lesssim 4$. Therefore, redshift effects should be taken into account as we describe in this Section.

The redshift effects leave unaltered the equations
of motion [Eq.~(\ref{Eq:rho})], after an appropriate rescaling
of the different parameters, as we now explain.
Since the number density of the electrons traces that of matter, and the average number density of electrons goes as the third power of the size of
the Universe, we obtain the relationship
\begin{equation}
n_e(z)= n_{e,0} (1+z)^3 \,\ .
\label{eq:nevolution}
\end{equation}
This is the redshift effect entering the plasma frequency
 $\omega_{\mathrm{pl}}$. Concerning the $B$-field in the host galaxy, in principle one should take into account a possible evolution as a function of the redshift~\cite{galaxies7020054}. However, there is no clear picture in the trend of this effect, see e.g. Ref.~\cite{10.1093/mnras/sty3270}. Therefore we prefer to assume no redshift dependence in the host magnetic field. Moreover, we have not assumed a scaling of the magnetic cells size in the source~\cite{Adams:1996xe}, while the energy of the beam scales as
\begin{equation}
E(z)=E_0(1+z) \,\, ,
\end{equation}
where with subscript $0$ we indicate the today values of the different quantities.
Considering these redshift relations, we find that the quantities in Eq.~(\ref{eq:Delta0}) evolve as
\begin{eqnarray}  
\Delta_{a\gamma}&=&  {\Delta^{0}_{a\gamma}}
\nonumber\,,\\  
\Delta_\textrm{a} &=& \frac{\Delta^0_\textrm{a}}{(1+z)}
\nonumber\,,\\  
\Delta_{\rm pl}&=& 
\Delta^0_{\rm pl} (1+z)^2
\nonumber\,,\\
\Delta_{\rm QED}&=& \Delta^0_{\rm QED} (1+z) \nonumber\,,
\label{eq:Deltaz}
\end{eqnarray}
where the supescript $0$ indicates the today value. 
Concerning the absorption factor $\Gamma$ in
Eq.~(\ref{eq:gamma}) one has to properly redshift the background photon energy $\epsilon$, the 
VHE photon $E$ and the photon background density 
${{\rm bkg}_\gamma}$. 
In particular, as discussed above for $E\gtrsim 10^3$~TeV the main contribution to the cosmic opacity is associated with CMB photon background. In this situation the 
redshifted expression of $\Gamma$ is analytical. Indeed, the CMB background spectra scales as 
\begin{equation}
\frac{dn^{\rm CMB}_\gamma}{d\epsilon}=\frac{\epsilon^2}{\pi^2}\frac{1}{e^{\epsilon/T_{\rm CMB}(1+z)}-1}\,\ .
\, 
\label{eq:CMBbg}
\end{equation}
Then, with the change of variable $\epsilon'=\epsilon/(1+z)$ in Eq.~(\ref{eq:gamma}) it can be shown that $\Gamma$ scales as
\begin{equation}
\begin{split}
\Gamma(E,z)=(1+z)^3\Gamma(E(1+z),z=0)\\=(1+z)^3\Gamma(E_0(1+z)^2,z=0)\,\ ,
\, 
\end{split}
\label{eq:gammaz1}
\end{equation}
where the factor $(1+z)^3$ accounts for the increasing for the background photon number density of the redshift and the scaling $E(1+z)$ in the energy is due to the increasing in energy of the background photon. The maximum of absorption thus shifts to lower energies when $z$ increases.
From the scaling of $\Gamma(E,z)$ and Eq.~\eqref{eq:DeltaCMB} one can obtain that the redshift of $\Delta_{\rm CMB}(E,z)$ follows the same law.

\begin{figure}
    \centering
    \includegraphics[scale=0.36]{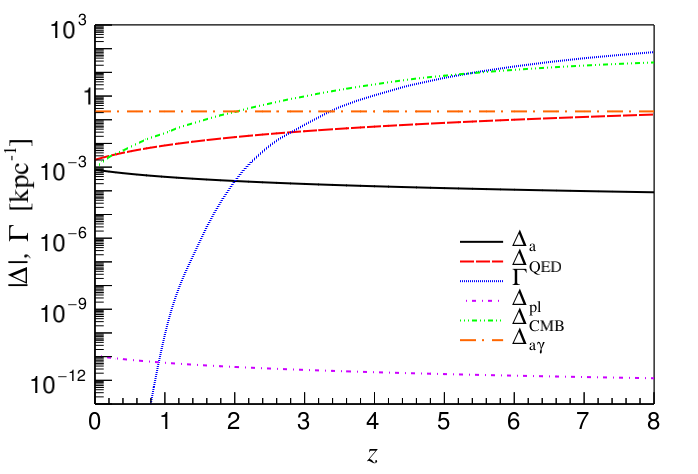}
    \caption{Redshift evolution of the factors $\Delta_{\rm a}$ (black continuous curve), $\Delta_{\rm QED}$ (red dashed curve), $\Gamma$ (blue dotted curve), $\Delta_{\rm pl}$ (magenta dot-dashed curve), $\Delta_{\mathrm{CMB}}$ (green dot-dashed curve) and $\Delta_{a\gamma}$ (orange long dot-dashed curve) evaluated at $E=10~\mathrm{TeV}$, $B=5~\mathrm{\mu G}$, $m_a\ll 10^{-7}~\mathrm{eV}$ and $g_{a\gamma}=3\times 10^{-11}~\mathrm{GeV^{-1}}$.
    }
    \label{fig:redshiftpar}
\end{figure}

In Fig.~\ref{fig:redshiftpar}, we show the evolution in $z$ of the relevant quantities, in Eq.~(\ref{eq:Delta0}), for $E=10~\mathrm{TeV}$, $B=5~\mathrm{\mu G}$, $m_a\ll 10^{-7}~\mathrm{eV}$ and $g_{a\gamma}=3\times 10^{-11}~\mathrm{GeV^{-1}}$. It is clear that the suppression of the conversion probability $\langle P_{\gamma\rightarrow a}\rangle$ starts at $z\sim 2-3$, when $\Gamma \gtrsim \Delta_{a \gamma},\Delta_{\rm QED}$.

\begin{figure}[!t]
    \centering
    \includegraphics[scale=0.36]{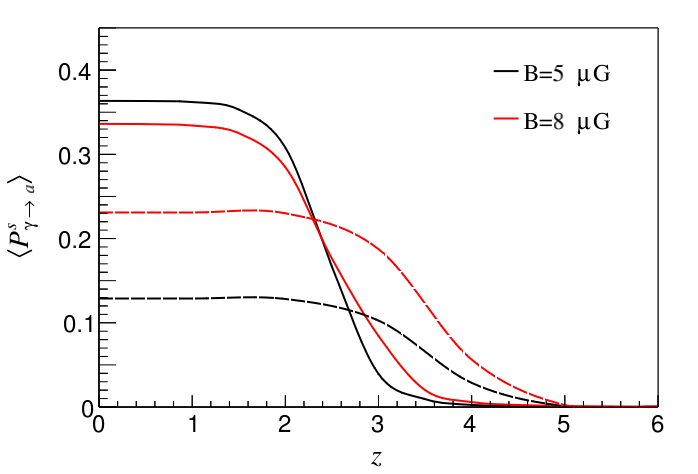}
    \caption{Average $\langle P^{s}_{\gamma\rightarrow a}\rangle$  conversion probabilities in the host galaxy at $E=10~\mathrm{TeV}$ as a function of the redshift $z$ of the source. We take $g_{a \gamma}=3\times 10^{-11}~\mathrm{GeV^{-1}}$ and consider $B_T=8 \times 10^{-6} \,\ \mathrm{G}$ (red curves) and $B_T=5 \times 10^{-6} \,\ \mathrm{G}$ (black curves). The continuous curve are for a single box regular $B$-field, while the dashed curves refer to the cell model.
 }
    \label{fig:prob_configz}
\end{figure}

In Fig.~\ref{fig:prob_configz} we show the averaged conversion probability $\langle P^s_{\gamma\rightarrow a}\rangle$
in function of the redshift $z$ of the host galaxy. We consider 
 the single box magnetic field (continuous curves) and for the  cell model (dashed curves).
We take $g_{a\gamma}=3\times 10^{-11}~\mathrm{GeV^{-1}}$ and $E=10~\mathrm{TeV}$. As in Fig.~\ref{fig:prob_config}, we take $B_T=5~\mu\mathrm{G}$ (black curves) and $B_T=8~\mu\mathrm{G}$ (red curves). The general trend is a reduction of the conversion probability at increasing  $z$ due to the increasing of the $\Gamma$ and $\Delta_{\rm CMB}$ factors over $z$, as it is possible to see in Fig.~\ref{fig:redshiftpar}. For $z>2$, $\langle P_{\gamma\rightarrow a}\rangle$ is suppressed as a consequence of $\Delta_{\rm CMB}>\Delta_{a\gamma}$.
Moreover $\langle P_{\gamma\rightarrow a}^s\rangle$ in the single box model drops to zero before that in the cell model, similarly to what happen for the energy evolution in Fig.~\ref{fig:prob_config}.

\subsection{Diffuse ALP flux}

As next step we determine the diffuse ALP flux produced by VHE gamma-ray conversions in the different host galaxies. In analogy with the diffuse neutrino flux of Eq.~\eqref{eq:neutrino_flux} we write~\cite{Eckner:2022rwf}
\begin{eqnarray}
    \frac{d\phi_a}{dE}&=&\int_0^{\infty} \Big[(1+z)Q_{\gamma}(E(1+z)) \Big] 
    \nonumber \\
    &\times& \Big\langle P^s_{a \gamma}(E(1+z)) \Big\rangle \,
      n_s(z)\, \left|c\frac{dt}{dz}\right|\,dz \,\ ,
    \label{eq:axion_flux}
\end{eqnarray}
where we added the conversion probability in the host galaxy $\langle P^s_{a\gamma} \rangle$ computed at the emission energy $E(1+z)$.

\subsection{ALP-photons back-conversions in the Milky-Way}
\begin{figure*}[ht!]
    \centering
    \includegraphics[scale=0.74]{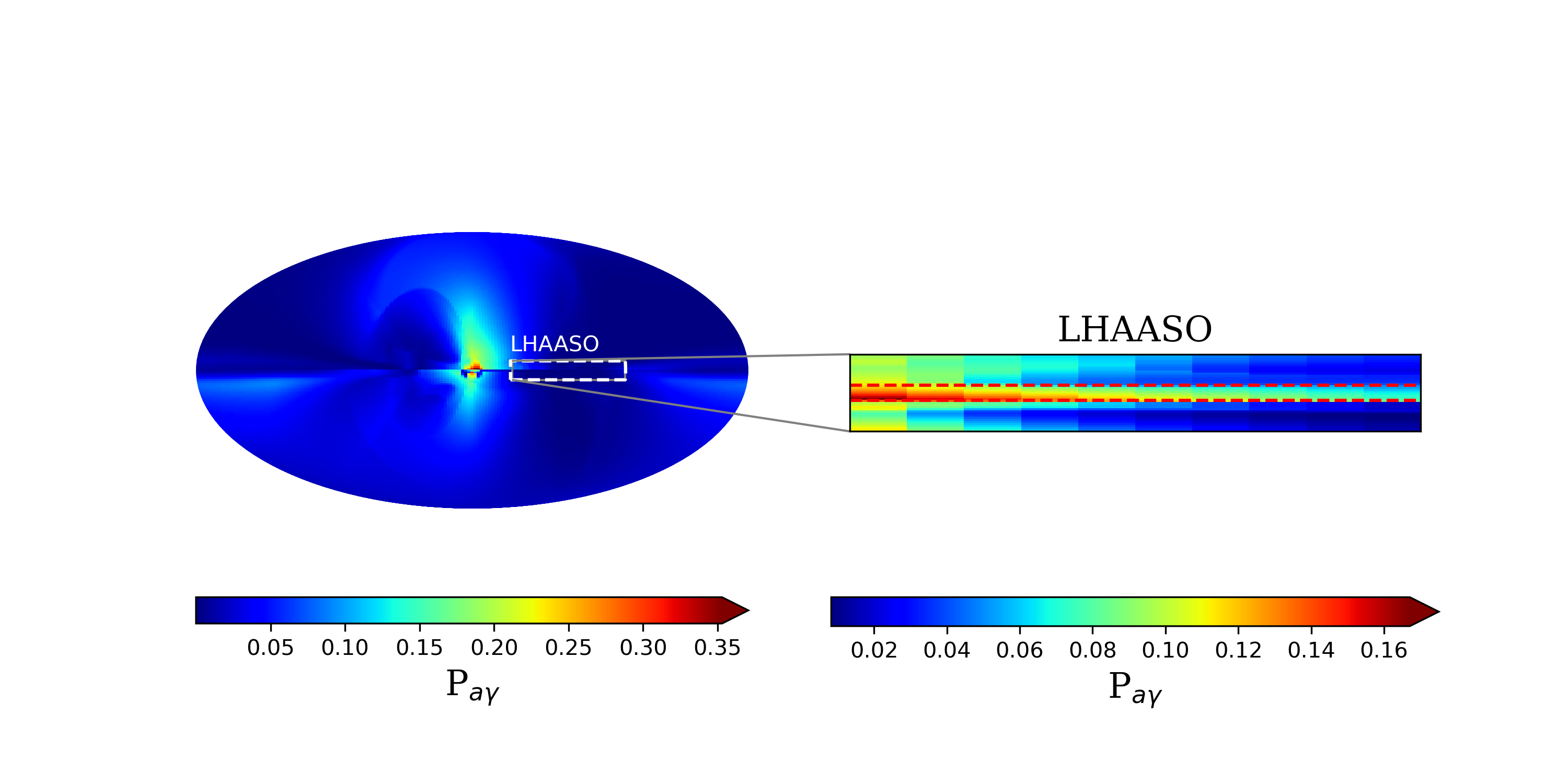}
    \caption{Skymap of the conversion probability in the Galaxy for $g_{a\gamma}=3\times 10^{-11}~\mathrm{GeV}^{-1}$, $m_{a}\ll 10^{-7}~\mathrm{eV}$ and $E=10$~TeV. 
    The regions probed by LHAASO is highlighted by the white dashed line and the region approximately defined by $b\in[-1^{\circ},1^{\circ}]$ is masked to avoid contamination by known sources in the Galactic plane.
    }
    \label{fig:skymap}
\end{figure*}

After ALPs and photons leave the host galaxy in which the source is placed, they propagate in the extragalactic space. At the energy we are considering ($E> 10$ TeV) the photons are completely absorbed due to the high intergalactic medium opacity. Instead, ALPs propagate unimpeded. Due to the uncertainty in the extragalactic magnetic field, we assume it sufficiently small in order to avoid  ALP-photon conversions in the extragalactic space (see, however, Ref. \cite{Montanino:2017ara} for possible effects).
When ALPs reach the edge of the Milky-Way, back-conversions into gamma-rays might occur in the Galactic $B$-field.
Then, the gamma-ray flux at Earth is given by 
\begin{equation}
    \frac{d\phi_{\gamma, \oplus}}{dE}=P_{a \gamma}^{\rm MW}     \,\frac{d\phi_a}{dE} \,\ ,
    \label{eq:flux_on_Earth}
\end{equation}
where $d\phi_{\gamma, \oplus}/dE$ is the photon flux reaching the Earth, $d\phi_a/dE$ is the flux obtained from Eq.~(\ref{eq:axion_flux}) and $P_{a \gamma}^{\rm MW}$ is the photon-ALP conversion probability in the Milky-Way. 

\begin{figure}
    \centering
    \includegraphics[scale=0.36]{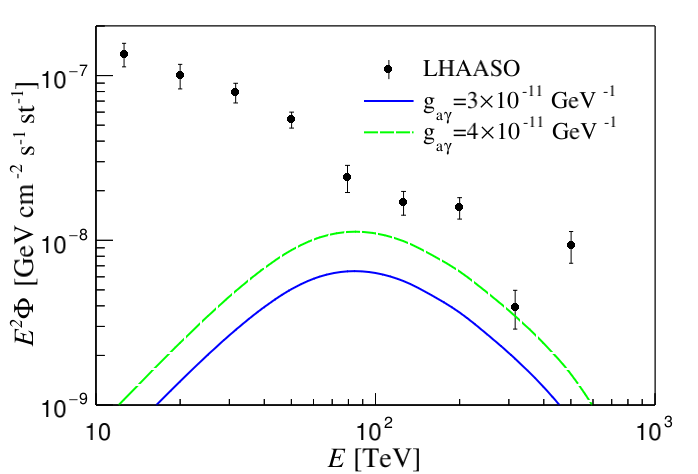}
    \caption{
    Expected photon flux from ALP conversions. 
    We assume the box model of the host galaxy $B$ field with $B_T=5 \times 10^{-6}$~G and the Fiducial case of Table~\ref{tab:RSFparameters} for the photon spectrum in source. We show cases with $g_{a\gamma}=3\times 10^{-11}~\mathrm{GeV^{-1}}$ (blue continuous line) and $g_{a\gamma}=4\times 10^{-11}~\mathrm{GeV^{-1}}$ (green dashed line).
    LHAASO data points are also shown.
    }
    \label{fig:LHAASO_data}
\end{figure}

We model the Galactic magnetic field as described by the Jansson-Farrar model~\cite{Jansson:2012pc} with the updated parameters given in Tab.~C.2 of Ref.~\cite{Planck:2016gdp} (``Jansson12c'' ordered fields) and an electron density in the Galaxy described by the model in Ref.~\cite{Cordes:2002wz}. The ALP propagation and mixing in the Galaxy is a purely 3-dimensional problem because of the highly non-trivial structure of the magnetic field (note that we neglect the small-scale turbulent field~\cite{Carenza:2021alz}). Therefore, both the photon polarization states play a role in the oscillation phenomenon. We have closely followed the technique described in Ref.~\cite{Horns:2012kw,Calore:2020tjw} to solve Eq.~(\ref{Eq:rho}) along a Galactic line of sight and obtain the back-conversion probability $P_{a\gamma}^{\rm MW}$ for the produced ALP. In Fig.~\ref{fig:skymap} we show the skymap of the conversion probability $P_{a \gamma}^{\rm MW}$ for $g_{a\gamma}=3\times 10^{-11}~\mathrm{GeV}^{-1}$, $m_{a}\ll 10^{-7}~\mathrm{eV}$ and $E=10$~TeV, as well as the region probed by LHAASO, corresponding to galactic longitude $25^{\circ}<l<100^{\circ}$ and latitude $1^{\circ}<|b|<5^{\circ}$.
The coordinates in the plot correspond to a Mollweide projection with positive longitudes on the right of the plot. It results that the averaged $P_{a\gamma}^{\rm MW} \sim 10^{-2}$ in the region of interest probed by LHAASO. Note that a strip of $1^{\circ}$ around the Galactic plane is not considered since the LHAASO measurement masked this region to avoid contamination by known sources, which are abundant in the Galactic plane.
Finally, in Fig.~\ref{fig:LHAASO_data} we show the produced photon flux in the Galaxy due to ALP conversions. For definiteness, we assume the box model of the host galaxy $B$ field with $B_T=5~\mu\mathrm{G}$,
and the Fiducial case of Table~\ref{tab:RSFparameters} for the photon spectrum in source.
We take $g_{a\gamma}=3\times 10^{-11}~\mathrm{GeV^{-1}}$ (blue continuous line) and $g_{a\gamma}=4\times 10^{-11}~\mathrm{GeV^{-1}}$ (green dashed line). LHAASO data points~\cite{Zhao:2021dqj} are also shown. In the next Section we will show how to obtain a bound in the plane $g_{a\gamma}$ \emph{vs} $m_a$ requiring that the produced photons flux does not exceed the experimental one.

\section{Analysis and results}
\label{Sec:Analysis and results}

To constraints the ALP parameter space, we perform a chi-squared analysis with
\begin{equation}
\chi^2 = \sum_{i=1}^N \left\{
    \begin{array}{l l}
        \left(\frac{\frac{d\phi^i_{\gamma, \oplus}}{dE}E_{i}^2 - \frac{d\phi^i_{\gamma, \rm exp}}{dE}E_{i}^2}{{\sigma(E_{i})}} \right)^2 & \quad \left(\frac{d\phi^i_{\gamma, \rm exp}}{dE} \leq \frac{d\phi^i_{\gamma, \oplus}}{dE} \right)  \\
        0 & \quad \left({\rm otherwise}\right)
    \end{array}\right.
    \label{eq:chi_quadro}
\end{equation}
where $N$ is the number of LHAASO data points, $E_i^2 d\phi^i_{\gamma, \rm exp}/dE$ are the experimental measurements and $\sigma(E_i)$ are the errors associated to those data.\\
The quantity defined in Eq.~(\ref{eq:chi_quadro}) follow an  half-$\chi^2$ distribution. Then we can exclude the values of $g_{a\gamma}$ for which $\chi^2>2.71$ to obtain bounds at $95\%$ confidence level (C.L.). We take as benchmark case the host galaxy magnetic field as a single box with $B_T= 5~\mathrm{\mu G}$, the Fiducial case of Tab.~\ref{tab:RSFparameters} and no photon background in the Milky-Way. In the Appendix we comment the dependence of our bound on the host galaxy magnetic field, on the assumption on the star formation rate and on the presence of a photon Galactic background. We summarize the impact of these factors on the final bound on $g_{a\gamma}$ in Table~\ref{tab:uncertanties}. We realize that the largest uncertainty on the bound is due to the unknown value of the magnetic field in the host galaxy, leading up to $\sim 50 \%$ of uncertainty.
\begin{table}
 \caption{Uncertainties on the ALP-photon coupling $g_{a\gamma}$ constraint, varying the condition of the benchmark case. The last row indicates the total uncertainty range when all sources are combined to form a most optimistic and most pessimistic scenario.}
    \centering
    \begin{tabular}{l c c}
    \hline
    Source of uncertainties &   Absolute [$10^{-11}~\mathrm{GeV^{-1}}$] &   Relative [\%] \\
    \hline 
         $Q_\nu$ and $Q_{\gamma}$ & $[4.25,5.00]$ & 17  \\
         $B$ field & $[4.90,7.40]$ & 51 \\
         photon bkg & $[4.35,4.85]$ & 11 \\
    \hline
    total & $[3.90,7.80]$ & 100 \\
    \hline
    \end{tabular}
    \label{tab:uncertanties}
\end{table}

Taking into account all the uncertainty we obtain a band in our exclusion plot shown in blue in Fig.~\ref{fig:bounds} at 95 \% C.L.
The strongest constraint gives $g_{a\gamma}< 3.9\times 10^{-11}~\mathrm{GeV^{-1}}$ for $m_a < 10^{-7}~\mathrm{eV}$
and corresponds to $B_T\sim 5~\mathrm{\mu G}$  (single box model), Upper case and photon background (light blue area).
Conversely the less restrictive bound gives  $g_{a\gamma}<7.8\times 10^{-11}~\mathrm{GeV^{-1}}$ for $m_a < 10^{-7}~\mathrm{eV}$ and corresponds to $B_T=2~\mathrm{\mu G}$ (cell model), Lower case without photon background (blue area). 
Finally, the black dashed curve corresponding to $g_{a\gamma}=4.8\times 10^{-11}~\mathrm{GeV^{-1}}$ is obtained for the benchmark case described above.
We notices that the bound is independent on the ALP mass for $m_a <\mathcal{O}(10^{-7}\rm{eV})$, while it deteriorates for higher values of the mass due to the mass suppression effect of the conversion probability. 

\section{Comparison of the bounds}
\label{sec:comparison}

\begin{figure*}[ht!]
\centering
    \includegraphics[scale=0.6]{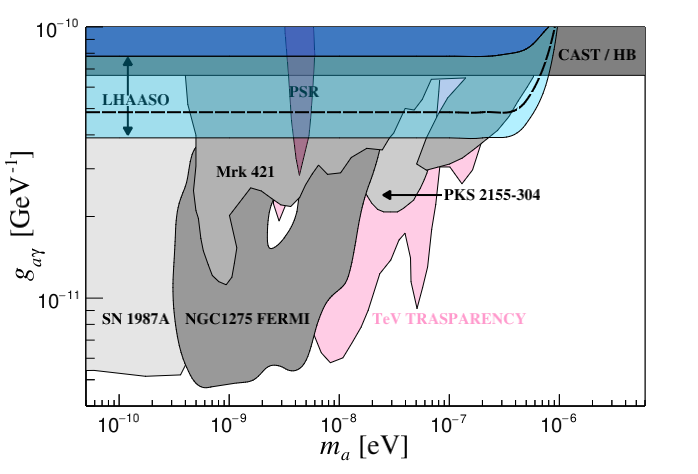}
    \caption{Exclusion plots in the parameter space $g_{a\gamma}$
    {\emph vs} $m_a$ from gamma-ray observations. The  LHAASO bounds are in blue, where the range dark-light blue depends on the assumptions of the magnetic field in the source and on the initial gamma-ray spectrum, and the dashed line represents our benchmark case.
    For comparison it is shown also the CAST bound on solar ALPs~\cite{CAST:2017uph}, comparable with the stellar bound on helium-buring stars~\cite{Ayala:2014pea}. Other astrophysical bounds are in grey: namely the SN 1987A~\cite{Payez:2014xsa}, the Fermi-LAT one on NGC 1275~\cite{Fermi-LAT:2016nkz}, the H.E.S.S. bound on PKS 2155-304~\cite{HESS:2013udx}, the Mrk 421~\cite{Li:2020pcn}. 
    It is shown also the region hinted by the spectral modulations of pulsars (PSR)~\cite{Pallathadka:2020vwu}. 
    In magenta it is shown   region where ALP conversions would affect the   transparency of TeV photons~\cite{Meyer:2013pny}.
    }
    \label{fig:bounds}
\end{figure*}

In this Section we compare our new bound with other ones in the same range of the ALP parameter space.
As custom, for comparison we take as reference the 
CAST bound on solar ALPs~\cite{CAST:2017uph}.
We realize that in the worst case our bound is slightly above  the CAST bound  (and to the stellar bound on helium-buring stars~\cite{Ayala:2014pea}), namely $g_{a\gamma} < 6.6 \times 10^{-11}~\mathrm{GeV^{-1}}$.
Conversely, in the most optimistic case we improve the CAST bound by a factor $\sim 2$.

In Fig.~~\ref{fig:bounds}, we also show other complementary bounds 
in the same region of the ALP parameter space from gamma-ray observations from astrophysical sources.
For ALPs with masses  $m_a \lesssim 10^{-9}$~eV, the strongest bound on $g_{a\gamma}$ is derived from the absence of gamma-rays from SN 1987A, giving $g_{a\gamma}< 5 \times 10^{-12}$~GeV$^{-1}$  for  $m_a \lesssim 10^{-10}$~eV~\cite{Payez:2014xsa}. A comparable bound on $g_{a\gamma}$ has been recently  extended in the mass range $0.5 \lesssim m_a \lesssim 5$~neV from the nonobservation in Fermi-LAT data of irregularities induced by photon-ALP conversions in the gamma-ray spectrum of NGC 1275, the central galaxy of the Perseus Cluster~\cite{TheFermi-LAT:2016zue}. Data from the H.E.S.S. observations of the distant BL Lac object PKS 2155-304 also limit $g_{a\gamma}< 2.1 \times 10^{-11}$~GeV$^{-1}$ for  $15 \lesssim m_a \lesssim 60$~neV~\cite{Abramowski:2013oea}. Finally, other limits have been obtained by the ARGO-YBJ and Fermi-LAT observations of Mrk 421, which find an upper limit on $g_{a\gamma}$ in the range $[2 \times 10^{-11}, 6 \times 10^{-11}]$~GeV$^{-1}$ for $5 \times 10^{-10} \lesssim m_a \lesssim 5 \times 10^{-7}$~eV~\cite{Li:2020pcn}.

However, these latter bounds strongly depend on severe uncertainties on the characterization of the magnetic field in the source and in the galaxy clusters. In this context, the choice of pure turbulent field in galaxy clusters has been recently criticized in Ref.~\cite{Libanov:2019fzq} 
(see also Ref.~\cite{Pallathadka:2020vwu}) showing that assuming a regular magnetic field the bound would be strongly relaxed, being confined above the CAST exclusion region.
As we have shown, since our bound is not based on the detection of irregular modulations of the gamma-ray spectrum, it is less sensitive to the uncertainty of the magnetic fields.

We also show the region (PSR) hinted by the spectral modulation observed in gamma-rays from Galactic pulsars and supernova remnants, that can be attributed to conversions into ALPs with a mass $m_a \sim 4 \times 10^{-9}$~eV and $g_{a\gamma}
\sim 2 \times 10^{-10}$~GeV$^{-1}$~\cite{Pallathadka:2020vwu}. Apparently, this region is in contrast with CAST bound and with SN 1987A. However, it has been proposed in Ref.~\cite{Pallathadka:2020vwu} that these bounds, relying on ALPs production in astrophysical plasmas might be evaded, assuming environmental effects suppressing the in-medium ALP-photon coupling. 
At this regard, our new bound  being based only on conversion effects would totally or partially constrain the 
PSR hinted area. 

Our bound is very similar to the recent one of  Ref.~\cite{Eckner:2022rwf} based on the diffuse gamma-ray signal measured by Tibet AS$\gamma$ and HAWC which constrain $g_{a\gamma} \lesssim 2.1 \times 10^{-11}$~GeV$^{-1}$ for $m_a < 10^{-7}$~eV. The main differences come from the the consideration of a photon background evaluated from the data of HAWC and Tibet AS$\gamma$~\cite{Luque:2022buq} which is not simultaneously compatible with the LHAASO data-set. 

Our bound still leaves a significant region of the parameter space that can be probed by the future Cherenkov Telescope Array~\cite{Meyer:2014gta,CTA:2020hii}, especially in connection with the region where ALP conversions would affect the TeV photon transparency (magenta region)~\cite{Meyer:2013pny}. However, at this regard, we mention that in Ref.~\cite{Dessert:2022yqq} recently it has been presented a  limit on $g_{a\gamma}$ from magnetic white dwarf polarization, that would be in tension with all the region hinted by the TeV photon anomalous transparency. 

\section{Conclusions}
\label{sec:conclusions}

The latest data of VHE gamma-rays measured by LHAASO have been used to constrain the ALP properties. Precisely, photons produced in extragalactic sources convert into ALPs in the magnetic field of the source forming a diffuse ALP flux that propagates unhindered (by contrast with photons that are absorbed) and re-convert into photon in the Galactic magnetic field. It is possible that this non-standard contribution to the VHE diffuse flux is sizable, leaving some imprints in the photon energy spectrum measured by LHAASO. Indeed, in the presence of ALPs one expects a larger photon flux compared to the standard case. We performed an analysis based on the latest LHAASO data in the energy range $10-500$~TeV to constrain the ALP-induced photon flux and then the ALP photon coupling.
We exclude axion-photon couplings $g_{a\gamma}>3.9-7.8\times 10^{-11}~\mathrm{GeV^{-1}}$ at the $95\%$ CL for $m_{a}\lesssim 4\times10^{-7}~\mathrm{eV}$.

At this regard, we have evaluated the changes in the assumption we have done (e.g. the module of the magnetic field  and the star formation rate model) finding  a factor $\sim 2$ of difference between the two extremal cases. Despite such a factor is not huge for an astrophysical bound, it changes the nature of our bound   from being competitive with the benchmark CAST bound to being completely inside the CAST excluded region.

Of course with a better characterization of the strength of the magnetic fields in the host galaxies one would strengthen the robustness of the bound.

Finally, we remark that  a significant part of ALP parameter space is left open to future gamma-ray experiments like CTA, as well as for forthcoming ALP searches, e.g. with  ALPS-II~\cite{Bahre:2013ywa} and IAXO~\cite{IAXO:2019mpb} experiments. Therefore, exciting times have to be expected in ALP searches due to the unprecedent sensitivity of future gamma-ray and axion experiments that will start soon.

\section*{Acknowledgements}
We thank Francesca Calore and Christopher Eckner for useful discussions.
This work of M.C., G.M.,  A.M., D.M.
was partially supported by the research grant number 2017W4HA7S ``NAT-NET: Neutrino and Astroparticle Theory Network'' under the program PRIN 2017 funded by the Italian Ministero dell'Università e della Ricerca (MUR). The authors also acknowledge the support by the research project TAsP (Theoretical Astroparticle Physics) funded by the Istituto Nazionale di Fisica Nucleare (INFN).\\
The work of P.C. is supported by the European Research Council under Grant No. 742104 and by the Swedish Research Council (VR) under grants  2018-03641 and 2019-02337.\\
The work of L.M. is supported by the Italian Istituto Nazionale di Fisica Nucleare (INFN) through the ``QGSKY'' project and by Ministero dell'Istruzione, Universit\`a e Ricerca (MIUR).\\
The work of D.F.G.F. is partially supported by the {\sc Villum Fonden} under project no.~29388.  This project has received funding from the European Union's Horizon 2020 research and innovation program under the Marie Sklodowska-Curie grant agreement No.~847523 ‘INTERACTIONS’.\\
The computational work has been executed on the IT resources of the ReCaS-Bari data center, which have been made available by two projects financed by the MIUR (Italian Ministry for Education, University and Re-search) in the "PON Ricerca e Competitività 2007-2013" Program: ReCaS (Azione I - Interventi di rafforzamento strutturale, PONa3\_00052, Avviso 254/Ric) and PRISMA (Asse II - Sostegno all'innovazione, PON04a2A)

\section*{Appendix A: Systematic uncertainty on the LHAASO upper limits
on $g_{a\gamma}$}
\label{sec:APP}

In this Appendix we comment about the 
systematic uncertainty on the LHAASO upper limits on $g_{a\gamma}$ due to the different assumptions of our work.
We take as benchmark case the host galaxy magnetic field 
as a single box with $\langle |\mathbf{B}_T|\rangle= 5~\mathrm{\mu G}$, and the Fiducial case for neutrino fluxes  in Table~\ref{tab:RSFparameters}.

\subsection{ Impact of star formation rate and original photon spectrum models}

\begin{figure}[t!]
    \centering
    \includegraphics[scale=0.36]{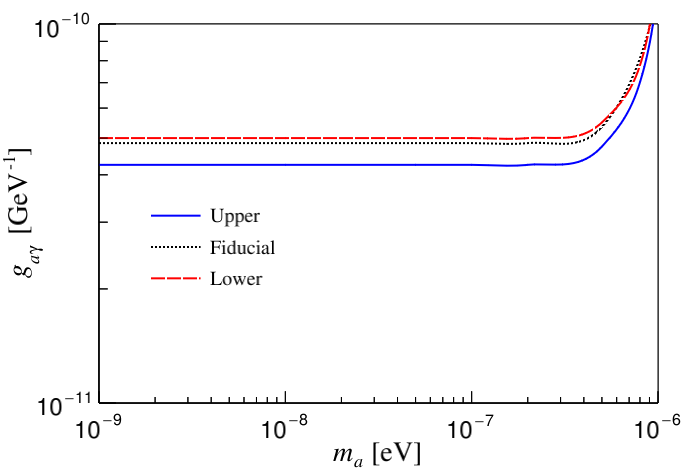}
    \caption{
     Variation in the bound on $g_{a\gamma}$ in dependence of the initial fluxes   described in Tab.~\ref{tab:RSFparameters}: Upper (blue continuous curve), Fiducial (black dotted curve)  and Lower (red dashed curve).
      }
    \label{fig:bounds_rho}
\end{figure}

We discuss the changes in the results obtained for the benchmark case, considering the Upper and Lower case in Tab.~\ref{tab:RSFparameters}. These cases differ for the $n_s$ and $Q_{\gamma}$ models as described in the table.
In Fig.~\ref{fig:bounds_rho} we show the difference in the constraints varying between the Lower, Fiducial and Upper case. As it is possible to see the bound is rather stable (there is only correction of $\sim 10\%$). This is because the ALP flux strongly depends on $g_{a\gamma}$ (in the perturbative regime it depends on $g_{a\gamma}^4$).

\subsection{Impact of host galaxy magnetic field}

\begin{figure}[t!]
    \centering
    \includegraphics[scale=0.36]{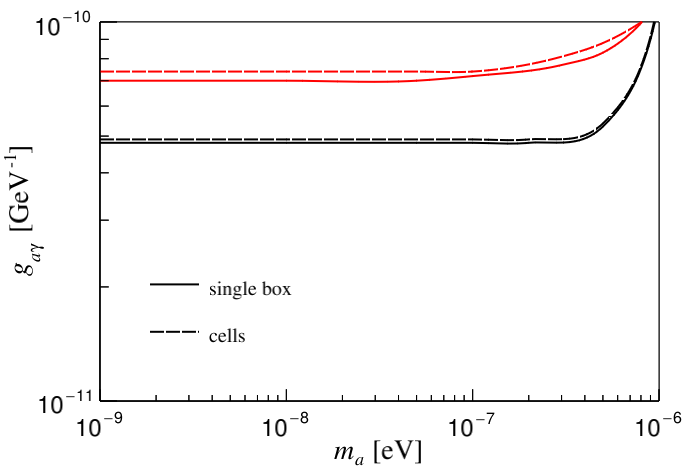}
    \caption{Variation in the bound due to the different choice of the magnetic field in the host galaxy. 
    The continuous curve refers to the box model, while the dashed curve corresponds to the cell model while the black curves refer the maximum bound obtained in each model and the red curves refer to the minimum bound. In each case the spread of the bound represents the variation of $B_T \in [2:8] \times 10^{-6}$~G.
    }
    \label{fig:bounds_cells}
\end{figure}

In Fig.~\ref{fig:bounds_cells} we show the impact on the 
exclusion plot for the benchmark case on the assumption of the magnetic field in the host galaxy. The continuous curve refers to the box model, while the dashed curve corresponds to the cell model. In each case the spread of the bound represents the variation of $B_T \in [2, 8] \times 10^{-6}$~G. The red curves shown the minimum constraints obtained for $B_T=2~\mathrm{\mu G}$, while the black curves are for the maximum constraints obtained for $B_T=5~\mathrm{\mu G}$ for the box model and $B_T=8~\mathrm{\mu G}$ for the cell model. 
It results that the impact on the variation of the host galaxy magnetic field model is a factor $\sim 1.5$ for both models.

\subsection{Impact of photon background}
\begin{figure}[t!]
    \centering
    \includegraphics[scale=0.45]{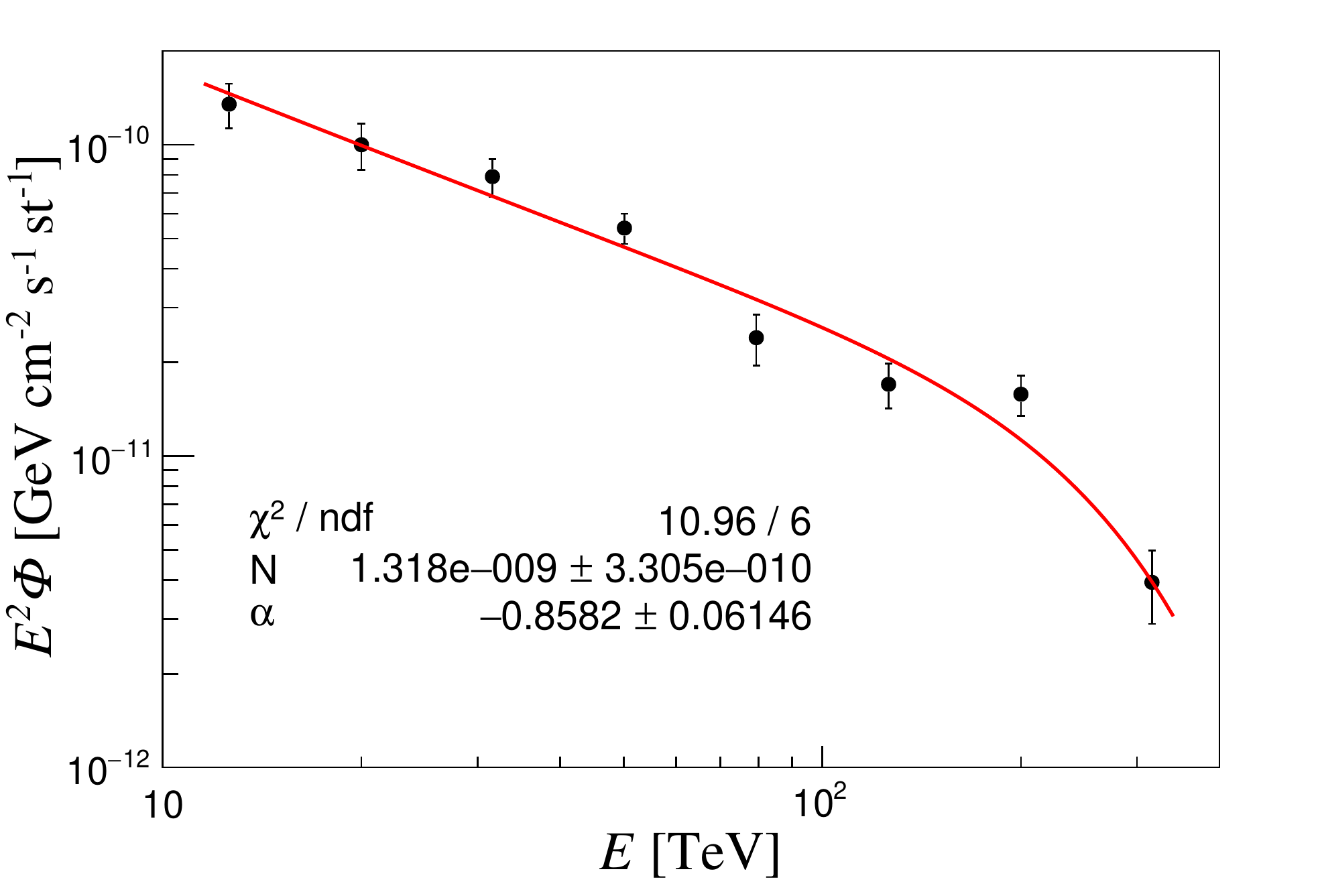}
    \caption{Fit of the LHAASO data points with the power-law background in Eq.~(\ref{eq:background}). The $\chi^2\simeq 11$, showing how it is only a fit acceptable into $2\sigma$ errors.}
    \label{fig:Fit0}
\end{figure}

\begin{figure}[t!]
    \centering
    \includegraphics[scale=0.36]{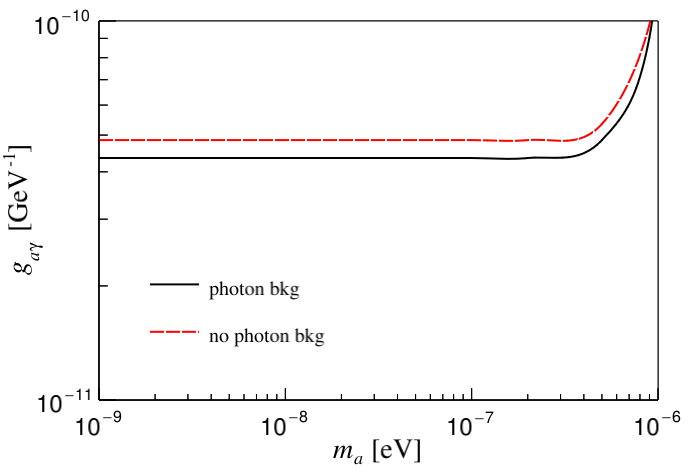}
    \caption{Impact on the constraint from the Fiducial case in the presence of a photon background. The case with no photon background in red dashed curve, while in the presence of a photon background we obtain the black curve.}
    \label{fig:bounds_nm}
\end{figure}

Here we discuss how the bound changes if we assume the presence of 
an additional gamma-ray flux background in the Milky-Way.
In this case the $\gamma$-ray flux on Earth can be written as:
\begin{equation}
  \frac{d\phi_{\gamma, \oplus}}{dE}=P_{a \gamma}^{\rm MW} \frac{d\phi_a}{dE} 
  + \frac{d\phi_{\gamma}^{\mathrm{bkg}}}{dE} \,\ ,
    \label{eq:flux_on_Earth_background}
\end{equation}
where $d\phi_{\gamma}^{\mathrm{bkg}}/dE$ is the background photon flux. We model this background flux as a power-law:
\begin{equation}
 \frac{d\phi_{\gamma}^{\mathrm{bkg}}}{dE}=NE^{\alpha}\times F(E) \,\ , 
 \label{eq:background}
\end{equation}
where $N$ and $\alpha$ are the two parameters of the power-law that have to be fitted and $F(E_{ \gamma})$ is a reduction factor caused by the CMB and SL+Infrared absorption. The chosen background seems to properly fit into a $2\sigma$ range all the points of the LHAASO data set except the last one. Therefore in our analysis, we focus only on the first eight experimental points. The reason is that, since we want to constrain an additional component to be added on top of this background, the results would not be reliable if the background itself were not realistic. We emphasize that eliminating one of the experimental points from our analysis is only conservative, since we are renouncing part of the information to constrain the model, and therefore the constraints we will draw are more robust.\\
In Fig.~\ref{fig:Fit0}, we show the result from the fit without considering the presence of ALPs, obtaining a $\chi^2_0=10.97$ over $6$ degrees of freedom.
When the ALP flux is added on top of the background component, we find that the fit always worsens. Therefore, we consider the solution without ALP as the best fit over the combined parameter space of the background and the ALP flux. Using as a test statistic $\chi^2-\chi^2_0$, and again accounting only for the upper fluctuations of the chi-squared, we exclude at $95\%$ confidence level values of $g_{a\gamma}$ such that $\chi^2-\chi^2_0>2.71$. Due to the high value of $\chi_0^2$, denoting no correct background assumption, and to the strong dependence of the axion flux on $g_{a\gamma}$, there is some $\mathcal{O}(10)~\%$ of difference with respect to the case without photon background as shown in  Fig.~\ref{fig:bounds_nm}.


\newpage
\bibliography{1main}

\providecommand{\noopsort}[1]{}\providecommand{\singleletter}[1]{#1}%
\begin{thebibliography}{113}%
\makeatletter
\providecommand \@ifxundefined [1]{%
 \@ifx{#1\undefined}
}%
\providecommand \@ifnum [1]{%
 \ifnum #1\expandafter \@firstoftwo
 \else \expandafter \@secondoftwo
 \fi
}%
\providecommand \@ifx [1]{%
 \ifx #1\expandafter \@firstoftwo
 \else \expandafter \@secondoftwo
 \fi
}%
\providecommand \natexlab [1]{#1}%
\providecommand \enquote  [1]{``#1''}%
\providecommand \bibnamefont  [1]{#1}%
\providecommand \bibfnamefont [1]{#1}%
\providecommand \citenamefont [1]{#1}%
\providecommand \href@noop [0]{\@secondoftwo}%
\providecommand \href [0]{\begingroup \@sanitize@url \@href}%
\providecommand \@href[1]{\@@startlink{#1}\@@href}%
\providecommand \@@href[1]{\endgroup#1\@@endlink}%
\providecommand \@sanitize@url [0]{\catcode `\\12\catcode `\$12\catcode
  `\&12\catcode `\#12\catcode `\^12\catcode `\_12\catcode `\%12\relax}%
\providecommand \@@startlink[1]{}%
\providecommand \@@endlink[0]{}%
\providecommand \url  [0]{\begingroup\@sanitize@url \@url }%
\providecommand \@url [1]{\endgroup\@href {#1}{\urlprefix }}%
\providecommand \urlprefix  [0]{URL }%
\providecommand \Eprint [0]{\href }%
\providecommand \doibase [0]{http://dx.doi.org/}%
\providecommand \selectlanguage [0]{\@gobble}%
\providecommand \bibinfo  [0]{\@secondoftwo}%
\providecommand \bibfield  [0]{\@secondoftwo}%
\providecommand \translation [1]{[#1]}%
\providecommand \BibitemOpen [0]{}%
\providecommand \bibitemStop [0]{}%
\providecommand \bibitemNoStop [0]{.\EOS\space}%
\providecommand \EOS [0]{\spacefactor3000\relax}%
\providecommand \BibitemShut  [1]{\csname bibitem#1\endcsname}%
\let\auto@bib@innerbib\@empty
\bibitem [{\citenamefont {Svrcek}\ and\ \citenamefont
  {Witten}(2006)}]{Svrcek:2006yi}%
  \BibitemOpen
  \bibfield  {author} {\bibinfo {author} {\bibfnamefont {Peter}\ \bibnamefont
  {Svrcek}}\ and\ \bibinfo {author} {\bibfnamefont {Edward}\ \bibnamefont
  {Witten}},\ }\bibfield  {title} {\enquote {\bibinfo {title} {{Axions In
  String Theory}},}\ }\href {\doibase 10.1088/1126-6708/2006/06/051} {\bibfield
   {journal} {\bibinfo  {journal} {JHEP}\ }\textbf {\bibinfo {volume} {06}},\
  \bibinfo {pages} {051} (\bibinfo {year} {2006})},\ \Eprint
  {http://arxiv.org/abs/hep-th/0605206} {arXiv:hep-th/0605206} \BibitemShut
  {NoStop}%
\bibitem [{\citenamefont {Arvanitaki}\ \emph {et~al.}(2010)\citenamefont
  {Arvanitaki}, \citenamefont {Dimopoulos}, \citenamefont {Dubovsky},
  \citenamefont {Kaloper},\ and\ \citenamefont
  {March-Russell}}]{Arvanitaki:2009fg}%
  \BibitemOpen
  \bibfield  {author} {\bibinfo {author} {\bibfnamefont {Asimina}\ \bibnamefont
  {Arvanitaki}}, \bibinfo {author} {\bibfnamefont {Savas}\ \bibnamefont
  {Dimopoulos}}, \bibinfo {author} {\bibfnamefont {Sergei}\ \bibnamefont
  {Dubovsky}}, \bibinfo {author} {\bibfnamefont {Nemanja}\ \bibnamefont
  {Kaloper}}, \ and\ \bibinfo {author} {\bibfnamefont {John}\ \bibnamefont
  {March-Russell}},\ }\bibfield  {title} {\enquote {\bibinfo {title} {{String
  Axiverse}},}\ }\href {\doibase 10.1103/PhysRevD.81.123530} {\bibfield
  {journal} {\bibinfo  {journal} {Phys. Rev. D}\ }\textbf {\bibinfo {volume}
  {81}},\ \bibinfo {pages} {123530} (\bibinfo {year} {2010})},\ \Eprint
  {http://arxiv.org/abs/0905.4720} {arXiv:0905.4720 [hep-th]} \BibitemShut
  {NoStop}%
\bibitem [{\citenamefont {Cicoli}\ \emph {et~al.}(2012)\citenamefont {Cicoli},
  \citenamefont {Goodsell},\ and\ \citenamefont {Ringwald}}]{Cicoli:2012sz}%
  \BibitemOpen
  \bibfield  {author} {\bibinfo {author} {\bibfnamefont {Michele}\ \bibnamefont
  {Cicoli}}, \bibinfo {author} {\bibfnamefont {Mark}\ \bibnamefont {Goodsell}},
  \ and\ \bibinfo {author} {\bibfnamefont {Andreas}\ \bibnamefont {Ringwald}},\
  }\bibfield  {title} {\enquote {\bibinfo {title} {{The type IIB string
  axiverse and its low-energy phenomenology}},}\ }\href {\doibase
  10.1007/JHEP10(2012)146} {\bibfield  {journal} {\bibinfo  {journal} {JHEP}\
  }\textbf {\bibinfo {volume} {10}},\ \bibinfo {pages} {146} (\bibinfo {year}
  {2012})},\ \Eprint {http://arxiv.org/abs/1206.0819} {arXiv:1206.0819
  [hep-th]} \BibitemShut {NoStop}%
\bibitem [{\citenamefont {Graham}\ \emph
  {et~al.}(2015{\natexlab{a}})\citenamefont {Graham}, \citenamefont {Kaplan},\
  and\ \citenamefont {Rajendran}}]{Graham:2015cka}%
  \BibitemOpen
  \bibfield  {author} {\bibinfo {author} {\bibfnamefont {Peter~W.}\
  \bibnamefont {Graham}}, \bibinfo {author} {\bibfnamefont {David~E.}\
  \bibnamefont {Kaplan}}, \ and\ \bibinfo {author} {\bibfnamefont {Surjeet}\
  \bibnamefont {Rajendran}},\ }\bibfield  {title} {\enquote {\bibinfo {title}
  {{Cosmological Relaxation of the Electroweak Scale}},}\ }\href {\doibase
  10.1103/PhysRevLett.115.221801} {\bibfield  {journal} {\bibinfo  {journal}
  {Phys. Rev. Lett.}\ }\textbf {\bibinfo {volume} {115}},\ \bibinfo {pages}
  {221801} (\bibinfo {year} {2015}{\natexlab{a}})},\ \Eprint
  {http://arxiv.org/abs/1504.07551} {arXiv:1504.07551 [hep-ph]} \BibitemShut
  {NoStop}%
\bibitem [{\citenamefont {Di~Luzio}\ \emph {et~al.}(2020)\citenamefont
  {Di~Luzio}, \citenamefont {Giannotti}, \citenamefont {Nardi},\ and\
  \citenamefont {Visinelli}}]{DiLuzio:2020wdo}%
  \BibitemOpen
  \bibfield  {author} {\bibinfo {author} {\bibfnamefont {Luca}\ \bibnamefont
  {Di~Luzio}}, \bibinfo {author} {\bibfnamefont {Maurizio}\ \bibnamefont
  {Giannotti}}, \bibinfo {author} {\bibfnamefont {Enrico}\ \bibnamefont
  {Nardi}}, \ and\ \bibinfo {author} {\bibfnamefont {Luca}\ \bibnamefont
  {Visinelli}},\ }\bibfield  {title} {\enquote {\bibinfo {title} {{The
  landscape of QCD axion models}},}\ }\href {\doibase
  10.1016/j.physrep.2020.06.002} {\bibfield  {journal} {\bibinfo  {journal}
  {Phys. Rept.}\ }\textbf {\bibinfo {volume} {870}},\ \bibinfo {pages} {1--117}
  (\bibinfo {year} {2020})},\ \Eprint {http://arxiv.org/abs/2003.01100}
  {arXiv:2003.01100 [hep-ph]} \BibitemShut {NoStop}%
\bibitem [{\citenamefont {Raffelt}\ and\ \citenamefont
  {Stodolsky}(1988)}]{Raffelt:1987im}%
  \BibitemOpen
  \bibfield  {author} {\bibinfo {author} {\bibfnamefont {Georg}\ \bibnamefont
  {Raffelt}}\ and\ \bibinfo {author} {\bibfnamefont {Leo}\ \bibnamefont
  {Stodolsky}},\ }\bibfield  {title} {\enquote {\bibinfo {title} {{Mixing of
  the Photon with Low Mass Particles}},}\ }\href {\doibase
  10.1103/PhysRevD.37.1237} {\bibfield  {journal} {\bibinfo  {journal} {Phys.
  Rev. D}\ }\textbf {\bibinfo {volume} {37}},\ \bibinfo {pages} {1237}
  (\bibinfo {year} {1988})}\BibitemShut {NoStop}%
\bibitem [{\citenamefont {Anselm}(1988)}]{Anselm:1987vj}%
  \BibitemOpen
  \bibfield  {author} {\bibinfo {author} {\bibfnamefont {Alexei~A.}\
  \bibnamefont {Anselm}},\ }\bibfield  {title} {\enquote {\bibinfo {title}
  {{Experimental Test for Arion \ensuremath{<}---\ensuremath{>} Photon
  Oscillations in a Homogeneous Constant Magnetic Field}},}\ }\href {\doibase
  10.1103/PhysRevD.37.2001} {\bibfield  {journal} {\bibinfo  {journal} {Phys.
  Rev. D}\ }\textbf {\bibinfo {volume} {37}},\ \bibinfo {pages} {2001}
  (\bibinfo {year} {1988})}\BibitemShut {NoStop}%
\bibitem [{\citenamefont {Graham}\ \emph
  {et~al.}(2015{\natexlab{b}})\citenamefont {Graham}, \citenamefont
  {Irastorza}, \citenamefont {Lamoreaux}, \citenamefont {Lindner},\ and\
  \citenamefont {van Bibber}}]{Graham:2015ouw}%
  \BibitemOpen
  \bibfield  {author} {\bibinfo {author} {\bibfnamefont {Peter~W.}\
  \bibnamefont {Graham}}, \bibinfo {author} {\bibfnamefont {Igor~G.}\
  \bibnamefont {Irastorza}}, \bibinfo {author} {\bibfnamefont {Steven~K.}\
  \bibnamefont {Lamoreaux}}, \bibinfo {author} {\bibfnamefont {Axel}\
  \bibnamefont {Lindner}}, \ and\ \bibinfo {author} {\bibfnamefont {Karl~A.}\
  \bibnamefont {van Bibber}},\ }\bibfield  {title} {\enquote {\bibinfo {title}
  {{Experimental Searches for the Axion and Axion-Like Particles}},}\ }\href
  {\doibase 10.1146/annurev-nucl-102014-022120} {\bibfield  {journal} {\bibinfo
   {journal} {Ann. Rev. Nucl. Part. Sci.}\ }\textbf {\bibinfo {volume} {65}},\
  \bibinfo {pages} {485--514} (\bibinfo {year} {2015}{\natexlab{b}})},\ \Eprint
  {http://arxiv.org/abs/1602.00039} {arXiv:1602.00039 [hep-ex]} \BibitemShut
  {NoStop}%
\bibitem [{\citenamefont {Irastorza}\ and\ \citenamefont
  {Redondo}(2018)}]{Irastorza:2018dyq}%
  \BibitemOpen
  \bibfield  {author} {\bibinfo {author} {\bibfnamefont {Igor~G.}\ \bibnamefont
  {Irastorza}}\ and\ \bibinfo {author} {\bibfnamefont {Javier}\ \bibnamefont
  {Redondo}},\ }\bibfield  {title} {\enquote {\bibinfo {title} {{New
  experimental approaches in the search for axion-like particles}},}\ }\href
  {\doibase 10.1016/j.ppnp.2018.05.003} {\bibfield  {journal} {\bibinfo
  {journal} {Prog. Part. Nucl. Phys.}\ }\textbf {\bibinfo {volume} {102}},\
  \bibinfo {pages} {89--159} (\bibinfo {year} {2018})},\ \Eprint
  {http://arxiv.org/abs/1801.08127} {arXiv:1801.08127 [hep-ph]} \BibitemShut
  {NoStop}%
\bibitem [{\citenamefont {Irastorza}(2021)}]{Irastorza:2021tdu}%
  \BibitemOpen
  \bibfield  {author} {\bibinfo {author} {\bibfnamefont {Igor~G.}\ \bibnamefont
  {Irastorza}},\ }\bibfield  {title} {\enquote {\bibinfo {title} {{An
  introduction to axions and their detection}},}\ }in\ \href {\doibase
  10.21468/SciPostPhysLectNotes.45} {\emph {\bibinfo {booktitle} {{Les Houches
  summer school on Dark Matter}}}}\ (\bibinfo {year} {2021})\ \Eprint
  {http://arxiv.org/abs/2109.07376} {arXiv:2109.07376 [hep-ph]} \BibitemShut
  {NoStop}%
\bibitem [{\citenamefont {Sikivie}(2021)}]{Sikivie:2020zpn}%
  \BibitemOpen
  \bibfield  {author} {\bibinfo {author} {\bibfnamefont {Pierre}\ \bibnamefont
  {Sikivie}},\ }\bibfield  {title} {\enquote {\bibinfo {title} {{Invisible
  Axion Search Methods}},}\ }\href {\doibase 10.1103/RevModPhys.93.015004}
  {\bibfield  {journal} {\bibinfo  {journal} {Rev. Mod. Phys.}\ }\textbf
  {\bibinfo {volume} {93}},\ \bibinfo {pages} {015004} (\bibinfo {year}
  {2021})},\ \Eprint {http://arxiv.org/abs/2003.02206} {arXiv:2003.02206
  [hep-ph]} \BibitemShut {NoStop}%
\bibitem [{\citenamefont {Mirizzi}\ \emph {et~al.}(2007)\citenamefont
  {Mirizzi}, \citenamefont {Raffelt},\ and\ \citenamefont
  {Serpico}}]{Mirizzi:2007hr}%
  \BibitemOpen
  \bibfield  {author} {\bibinfo {author} {\bibfnamefont {Alessandro}\
  \bibnamefont {Mirizzi}}, \bibinfo {author} {\bibfnamefont {Georg~G.}\
  \bibnamefont {Raffelt}}, \ and\ \bibinfo {author} {\bibfnamefont
  {Pasquale~D.}\ \bibnamefont {Serpico}},\ }\bibfield  {title} {\enquote
  {\bibinfo {title} {{Signatures of Axion-Like Particles in the Spectra of TeV
  Gamma-Ray Sources}},}\ }\href {\doibase 10.1103/PhysRevD.76.023001}
  {\bibfield  {journal} {\bibinfo  {journal} {Phys. Rev. D}\ }\textbf {\bibinfo
  {volume} {76}},\ \bibinfo {pages} {023001} (\bibinfo {year} {2007})},\
  \Eprint {http://arxiv.org/abs/0704.3044} {arXiv:0704.3044 [astro-ph]}
  \BibitemShut {NoStop}%
\bibitem [{\citenamefont {De~Angelis}\ \emph {et~al.}(2008)\citenamefont
  {De~Angelis}, \citenamefont {Mansutti},\ and\ \citenamefont
  {Roncadelli}}]{DeAngelis:2007wiw}%
  \BibitemOpen
  \bibfield  {author} {\bibinfo {author} {\bibfnamefont {Alessandro}\
  \bibnamefont {De~Angelis}}, \bibinfo {author} {\bibfnamefont {Oriana}\
  \bibnamefont {Mansutti}}, \ and\ \bibinfo {author} {\bibfnamefont {Marco}\
  \bibnamefont {Roncadelli}},\ }\bibfield  {title} {\enquote {\bibinfo {title}
  {{Axion-Like Particles, Cosmic Magnetic Fields and Gamma-Ray
  Astrophysics}},}\ }\href {\doibase 10.1016/j.physletb.2007.12.012} {\bibfield
   {journal} {\bibinfo  {journal} {Phys. Lett. B}\ }\textbf {\bibinfo {volume}
  {659}},\ \bibinfo {pages} {847--855} (\bibinfo {year} {2008})},\ \Eprint
  {http://arxiv.org/abs/0707.2695} {arXiv:0707.2695 [astro-ph]} \BibitemShut
  {NoStop}%
\bibitem [{\citenamefont {De~Angelis}\ \emph {et~al.}(2007)\citenamefont
  {De~Angelis}, \citenamefont {Roncadelli},\ and\ \citenamefont
  {Mansutti}}]{DeAngelis:2007dqd}%
  \BibitemOpen
  \bibfield  {author} {\bibinfo {author} {\bibfnamefont {Alessandro}\
  \bibnamefont {De~Angelis}}, \bibinfo {author} {\bibfnamefont {Marco}\
  \bibnamefont {Roncadelli}}, \ and\ \bibinfo {author} {\bibfnamefont {Oriana}\
  \bibnamefont {Mansutti}},\ }\bibfield  {title} {\enquote {\bibinfo {title}
  {{Evidence for a new light spin-zero boson from cosmological gamma-ray
  propagation?}}}\ }\href {\doibase 10.1103/PhysRevD.76.121301} {\bibfield
  {journal} {\bibinfo  {journal} {Phys. Rev. D}\ }\textbf {\bibinfo {volume}
  {76}},\ \bibinfo {pages} {121301} (\bibinfo {year} {2007})},\ \Eprint
  {http://arxiv.org/abs/0707.4312} {arXiv:0707.4312 [astro-ph]} \BibitemShut
  {NoStop}%
\bibitem [{\citenamefont {Hooper}\ and\ \citenamefont
  {Serpico}(2007)}]{Hooper:2007bq}%
  \BibitemOpen
  \bibfield  {author} {\bibinfo {author} {\bibfnamefont {Dan}\ \bibnamefont
  {Hooper}}\ and\ \bibinfo {author} {\bibfnamefont {Pasquale~D.}\ \bibnamefont
  {Serpico}},\ }\bibfield  {title} {\enquote {\bibinfo {title} {{Detecting
  Axion-Like Particles With Gamma Ray Telescopes}},}\ }\href {\doibase
  10.1103/PhysRevLett.99.231102} {\bibfield  {journal} {\bibinfo  {journal}
  {Phys. Rev. Lett.}\ }\textbf {\bibinfo {volume} {99}},\ \bibinfo {pages}
  {231102} (\bibinfo {year} {2007})},\ \Eprint {http://arxiv.org/abs/0706.3203}
  {arXiv:0706.3203 [hep-ph]} \BibitemShut {NoStop}%
\bibitem [{\citenamefont {Simet}\ \emph {et~al.}(2008)\citenamefont {Simet},
  \citenamefont {Hooper},\ and\ \citenamefont {Serpico}}]{Simet:2007sa}%
  \BibitemOpen
  \bibfield  {author} {\bibinfo {author} {\bibfnamefont {Melanie}\ \bibnamefont
  {Simet}}, \bibinfo {author} {\bibfnamefont {Dan}\ \bibnamefont {Hooper}}, \
  and\ \bibinfo {author} {\bibfnamefont {Pasquale~D.}\ \bibnamefont
  {Serpico}},\ }\bibfield  {title} {\enquote {\bibinfo {title} {{The Milky Way
  as a Kiloparsec-Scale Axionscope}},}\ }\href {\doibase
  10.1103/PhysRevD.77.063001} {\bibfield  {journal} {\bibinfo  {journal} {Phys.
  Rev. D}\ }\textbf {\bibinfo {volume} {77}},\ \bibinfo {pages} {063001}
  (\bibinfo {year} {2008})},\ \Eprint {http://arxiv.org/abs/0712.2825}
  {arXiv:0712.2825 [astro-ph]} \BibitemShut {NoStop}%
\bibitem [{\citenamefont {De~Angelis}\ \emph {et~al.}(2011)\citenamefont
  {De~Angelis}, \citenamefont {Galanti},\ and\ \citenamefont
  {Roncadelli}}]{DeAngelis:2011id}%
  \BibitemOpen
  \bibfield  {author} {\bibinfo {author} {\bibfnamefont {Alessandro}\
  \bibnamefont {De~Angelis}}, \bibinfo {author} {\bibfnamefont {Giorgio}\
  \bibnamefont {Galanti}}, \ and\ \bibinfo {author} {\bibfnamefont {Marco}\
  \bibnamefont {Roncadelli}},\ }\bibfield  {title} {\enquote {\bibinfo {title}
  {{Relevance of axion-like particles for very-high-energy astrophysics}},}\
  }\href {\doibase 10.1103/PhysRevD.84.105030} {\bibfield  {journal} {\bibinfo
  {journal} {Phys. Rev. D}\ }\textbf {\bibinfo {volume} {84}},\ \bibinfo
  {pages} {105030} (\bibinfo {year} {2011})},\ \bibinfo {note} {[Erratum:
  Phys.Rev.D 87, 109903 (2013)]},\ \Eprint {http://arxiv.org/abs/1106.1132}
  {arXiv:1106.1132 [astro-ph.HE]} \BibitemShut {NoStop}%
\bibitem [{\citenamefont {Mirizzi}\ and\ \citenamefont
  {Montanino}(2009)}]{Mirizzi:2009aj}%
  \BibitemOpen
  \bibfield  {author} {\bibinfo {author} {\bibfnamefont {Alessandro}\
  \bibnamefont {Mirizzi}}\ and\ \bibinfo {author} {\bibfnamefont {Daniele}\
  \bibnamefont {Montanino}},\ }\bibfield  {title} {\enquote {\bibinfo {title}
  {{Stochastic conversions of TeV photons into axion-like particles in
  extragalactic magnetic fields}},}\ }\href {\doibase
  10.1088/1475-7516/2009/12/004} {\bibfield  {journal} {\bibinfo  {journal}
  {JCAP}\ }\textbf {\bibinfo {volume} {12}},\ \bibinfo {pages} {004} (\bibinfo
  {year} {2009})},\ \Eprint {http://arxiv.org/abs/0911.0015} {arXiv:0911.0015
  [astro-ph.HE]} \BibitemShut {NoStop}%
\bibitem [{\citenamefont {Horns}\ \emph {et~al.}(2012)\citenamefont {Horns},
  \citenamefont {Maccione}, \citenamefont {Meyer}, \citenamefont {Mirizzi},
  \citenamefont {Montanino},\ and\ \citenamefont {Roncadelli}}]{Horns:2012kw}%
  \BibitemOpen
  \bibfield  {author} {\bibinfo {author} {\bibfnamefont {Dieter}\ \bibnamefont
  {Horns}}, \bibinfo {author} {\bibfnamefont {Luca}\ \bibnamefont {Maccione}},
  \bibinfo {author} {\bibfnamefont {Manuel}\ \bibnamefont {Meyer}}, \bibinfo
  {author} {\bibfnamefont {Alessandro}\ \bibnamefont {Mirizzi}}, \bibinfo
  {author} {\bibfnamefont {Daniele}\ \bibnamefont {Montanino}}, \ and\ \bibinfo
  {author} {\bibfnamefont {Marco}\ \bibnamefont {Roncadelli}},\ }\bibfield
  {title} {\enquote {\bibinfo {title} {{Hardening of TeV gamma spectrum of AGNs
  in galaxy clusters by conversions of photons into axion-like particles}},}\
  }\href {\doibase 10.1103/PhysRevD.86.075024} {\bibfield  {journal} {\bibinfo
  {journal} {Phys. Rev. D}\ }\textbf {\bibinfo {volume} {86}},\ \bibinfo
  {pages} {075024} (\bibinfo {year} {2012})},\ \Eprint
  {http://arxiv.org/abs/1207.0776} {arXiv:1207.0776 [astro-ph.HE]} \BibitemShut
  {NoStop}%
\bibitem [{\citenamefont {Meyer}\ \emph {et~al.}(2013)\citenamefont {Meyer},
  \citenamefont {Horns},\ and\ \citenamefont {Raue}}]{Meyer:2013pny}%
  \BibitemOpen
  \bibfield  {author} {\bibinfo {author} {\bibfnamefont {Manuel}\ \bibnamefont
  {Meyer}}, \bibinfo {author} {\bibfnamefont {Dieter}\ \bibnamefont {Horns}}, \
  and\ \bibinfo {author} {\bibfnamefont {Martin}\ \bibnamefont {Raue}},\
  }\bibfield  {title} {\enquote {\bibinfo {title} {{First lower limits on the
  photon-axion-like particle coupling from very high energy gamma-ray
  observations}},}\ }\href {\doibase 10.1103/PhysRevD.87.035027} {\bibfield
  {journal} {\bibinfo  {journal} {Phys. Rev. D}\ }\textbf {\bibinfo {volume}
  {87}},\ \bibinfo {pages} {035027} (\bibinfo {year} {2013})},\ \Eprint
  {http://arxiv.org/abs/1302.1208} {arXiv:1302.1208 [astro-ph.HE]} \BibitemShut
  {NoStop}%
\bibitem [{\citenamefont {Abramowski}\ \emph
  {et~al.}(2013{\natexlab{a}})\citenamefont {Abramowski} \emph
  {et~al.}}]{HESS:2013udx}%
  \BibitemOpen
  \bibfield  {author} {\bibinfo {author} {\bibfnamefont {A.}~\bibnamefont
  {Abramowski}} \emph {et~al.} (\bibinfo {collaboration} {H.E.S.S.}),\
  }\bibfield  {title} {\enquote {\bibinfo {title} {{Constraints on axionlike
  particles with H.E.S.S. from the irregularity of the PKS 2155-304 energy
  spectrum}},}\ }\href {\doibase 10.1103/PhysRevD.88.102003} {\bibfield
  {journal} {\bibinfo  {journal} {Phys. Rev. D}\ }\textbf {\bibinfo {volume}
  {88}},\ \bibinfo {pages} {102003} (\bibinfo {year} {2013}{\natexlab{a}})},\
  \Eprint {http://arxiv.org/abs/1311.3148} {arXiv:1311.3148 [astro-ph.HE]}
  \BibitemShut {NoStop}%
\bibitem [{\citenamefont {Meyer}\ \emph {et~al.}(2014)\citenamefont {Meyer},
  \citenamefont {Montanino},\ and\ \citenamefont {Conrad}}]{Meyer:2014epa}%
  \BibitemOpen
  \bibfield  {author} {\bibinfo {author} {\bibfnamefont {Manuel}\ \bibnamefont
  {Meyer}}, \bibinfo {author} {\bibfnamefont {Daniele}\ \bibnamefont
  {Montanino}}, \ and\ \bibinfo {author} {\bibfnamefont {Jan}\ \bibnamefont
  {Conrad}},\ }\bibfield  {title} {\enquote {\bibinfo {title} {{On detecting
  oscillations of gamma rays into axion-like particles in turbulent and
  coherent magnetic fields}},}\ }\href {\doibase 10.1088/1475-7516/2014/09/003}
  {\bibfield  {journal} {\bibinfo  {journal} {JCAP}\ }\textbf {\bibinfo
  {volume} {09}},\ \bibinfo {pages} {003} (\bibinfo {year} {2014})},\ \Eprint
  {http://arxiv.org/abs/1406.5972} {arXiv:1406.5972 [astro-ph.HE]} \BibitemShut
  {NoStop}%
\bibitem [{\citenamefont {Meyer}\ and\ \citenamefont
  {Conrad}(2014)}]{Meyer:2014gta}%
  \BibitemOpen
  \bibfield  {author} {\bibinfo {author} {\bibfnamefont {Manuel}\ \bibnamefont
  {Meyer}}\ and\ \bibinfo {author} {\bibfnamefont {J.}~\bibnamefont {Conrad}},\
  }\bibfield  {title} {\enquote {\bibinfo {title} {{Sensitivity of the
  Cherenkov Telescope Array to the detection of axion-like particles at high
  gamma-ray opacities}},}\ }\href {\doibase 10.1088/1475-7516/2014/12/016}
  {\bibfield  {journal} {\bibinfo  {journal} {JCAP}\ }\textbf {\bibinfo
  {volume} {12}},\ \bibinfo {pages} {016} (\bibinfo {year} {2014})},\ \Eprint
  {http://arxiv.org/abs/1410.1556} {arXiv:1410.1556 [astro-ph.HE]} \BibitemShut
  {NoStop}%
\bibitem [{\citenamefont {Ajello}\ \emph
  {et~al.}(2016{\natexlab{a}})\citenamefont {Ajello} \emph
  {et~al.}}]{Fermi-LAT:2016nkz}%
  \BibitemOpen
  \bibfield  {author} {\bibinfo {author} {\bibfnamefont {M.}~\bibnamefont
  {Ajello}} \emph {et~al.} (\bibinfo {collaboration} {Fermi-LAT}),\ }\bibfield
  {title} {\enquote {\bibinfo {title} {{Search for Spectral Irregularities due
  to Photon\textendash{}Axionlike-Particle Oscillations with the Fermi Large
  Area Telescope}},}\ }\href {\doibase 10.1103/PhysRevLett.116.161101}
  {\bibfield  {journal} {\bibinfo  {journal} {Phys. Rev. Lett.}\ }\textbf
  {\bibinfo {volume} {116}},\ \bibinfo {pages} {161101} (\bibinfo {year}
  {2016}{\natexlab{a}})},\ \Eprint {http://arxiv.org/abs/1603.06978}
  {arXiv:1603.06978 [astro-ph.HE]} \BibitemShut {NoStop}%
\bibitem [{\citenamefont {Montanino}\ \emph {et~al.}(2017)\citenamefont
  {Montanino}, \citenamefont {Vazza}, \citenamefont {Mirizzi},\ and\
  \citenamefont {Viel}}]{Montanino:2017ara}%
  \BibitemOpen
  \bibfield  {author} {\bibinfo {author} {\bibfnamefont {Daniele}\ \bibnamefont
  {Montanino}}, \bibinfo {author} {\bibfnamefont {Franco}\ \bibnamefont
  {Vazza}}, \bibinfo {author} {\bibfnamefont {Alessandro}\ \bibnamefont
  {Mirizzi}}, \ and\ \bibinfo {author} {\bibfnamefont {Matteo}\ \bibnamefont
  {Viel}},\ }\bibfield  {title} {\enquote {\bibinfo {title} {{Enhancing the
  Spectral Hardening of Cosmic TeV Photons by Mixing with Axionlike Particles
  in the Magnetized Cosmic Web}},}\ }\href {\doibase
  10.1103/PhysRevLett.119.101101} {\bibfield  {journal} {\bibinfo  {journal}
  {Phys. Rev. Lett.}\ }\textbf {\bibinfo {volume} {119}},\ \bibinfo {pages}
  {101101} (\bibinfo {year} {2017})},\ \Eprint
  {http://arxiv.org/abs/1703.07314} {arXiv:1703.07314 [astro-ph.HE]}
  \BibitemShut {NoStop}%
\bibitem [{\citenamefont {Zhang}\ \emph {et~al.}(2018)\citenamefont {Zhang},
  \citenamefont {Liang}, \citenamefont {Li}, \citenamefont {Liao},
  \citenamefont {Feng}, \citenamefont {Yuan}, \citenamefont {Fan},\ and\
  \citenamefont {Ren}}]{Zhang:2018wpc}%
  \BibitemOpen
  \bibfield  {author} {\bibinfo {author} {\bibfnamefont {Cun}\ \bibnamefont
  {Zhang}}, \bibinfo {author} {\bibfnamefont {Yun-Feng}\ \bibnamefont {Liang}},
  \bibinfo {author} {\bibfnamefont {Shang}\ \bibnamefont {Li}}, \bibinfo
  {author} {\bibfnamefont {Neng-Hui}\ \bibnamefont {Liao}}, \bibinfo {author}
  {\bibfnamefont {Lei}\ \bibnamefont {Feng}}, \bibinfo {author} {\bibfnamefont
  {Qiang}\ \bibnamefont {Yuan}}, \bibinfo {author} {\bibfnamefont {Yi-Zhong}\
  \bibnamefont {Fan}}, \ and\ \bibinfo {author} {\bibfnamefont {Zhong-Zhou}\
  \bibnamefont {Ren}},\ }\bibfield  {title} {\enquote {\bibinfo {title} {{New
  bounds on axionlike particles from the Fermi Large Area Telescope observation
  of PKS 2155-304}},}\ }\href {\doibase 10.1103/PhysRevD.97.063009} {\bibfield
  {journal} {\bibinfo  {journal} {Phys. Rev. D}\ }\textbf {\bibinfo {volume}
  {97}},\ \bibinfo {pages} {063009} (\bibinfo {year} {2018})},\ \Eprint
  {http://arxiv.org/abs/1802.08420} {arXiv:1802.08420 [hep-ph]} \BibitemShut
  {NoStop}%
\bibitem [{\citenamefont {Xia}\ \emph {et~al.}(2018)\citenamefont {Xia},
  \citenamefont {Zhang}, \citenamefont {Liang}, \citenamefont {Feng},
  \citenamefont {Yuan}, \citenamefont {Fan},\ and\ \citenamefont
  {Wu}}]{Xia:2018xbt}%
  \BibitemOpen
  \bibfield  {author} {\bibinfo {author} {\bibfnamefont {Zi-Qing}\ \bibnamefont
  {Xia}}, \bibinfo {author} {\bibfnamefont {Cun}\ \bibnamefont {Zhang}},
  \bibinfo {author} {\bibfnamefont {Yun-Feng}\ \bibnamefont {Liang}}, \bibinfo
  {author} {\bibfnamefont {Lei}\ \bibnamefont {Feng}}, \bibinfo {author}
  {\bibfnamefont {Qiang}\ \bibnamefont {Yuan}}, \bibinfo {author}
  {\bibfnamefont {Yi-Zhong}\ \bibnamefont {Fan}}, \ and\ \bibinfo {author}
  {\bibfnamefont {Jian}\ \bibnamefont {Wu}},\ }\bibfield  {title} {\enquote
  {\bibinfo {title} {{Searching for spectral oscillations due to
  photon-axionlike particle conversion using the Fermi-LAT observations of
  bright supernova remnants}},}\ }\href {\doibase 10.1103/PhysRevD.97.063003}
  {\bibfield  {journal} {\bibinfo  {journal} {Phys. Rev. D}\ }\textbf {\bibinfo
  {volume} {97}},\ \bibinfo {pages} {063003} (\bibinfo {year} {2018})},\
  \Eprint {http://arxiv.org/abs/1801.01646} {arXiv:1801.01646 [astro-ph.HE]}
  \BibitemShut {NoStop}%
\bibitem [{\citenamefont {Majumdar}\ \emph {et~al.}(2018)\citenamefont
  {Majumdar}, \citenamefont {Calore},\ and\ \citenamefont
  {Horns}}]{Majumdar:2018sbv}%
  \BibitemOpen
  \bibfield  {author} {\bibinfo {author} {\bibfnamefont {Jhilik}\ \bibnamefont
  {Majumdar}}, \bibinfo {author} {\bibfnamefont {Francesca}\ \bibnamefont
  {Calore}}, \ and\ \bibinfo {author} {\bibfnamefont {Dieter}\ \bibnamefont
  {Horns}},\ }\bibfield  {title} {\enquote {\bibinfo {title} {{Search for
  gamma-ray spectral modulations in Galactic pulsars}},}\ }\href {\doibase
  10.1088/1475-7516/2018/04/048} {\bibfield  {journal} {\bibinfo  {journal}
  {JCAP}\ }\textbf {\bibinfo {volume} {04}},\ \bibinfo {pages} {048} (\bibinfo
  {year} {2018})},\ \Eprint {http://arxiv.org/abs/1801.08813} {arXiv:1801.08813
  [hep-ph]} \BibitemShut {NoStop}%
\bibitem [{\citenamefont {Galanti}\ \emph {et~al.}(2019)\citenamefont
  {Galanti}, \citenamefont {Tavecchio}, \citenamefont {Roncadelli},\ and\
  \citenamefont {Evoli}}]{Galanti:2018upl}%
  \BibitemOpen
  \bibfield  {author} {\bibinfo {author} {\bibfnamefont {Giorgio}\ \bibnamefont
  {Galanti}}, \bibinfo {author} {\bibfnamefont {Fabrizio}\ \bibnamefont
  {Tavecchio}}, \bibinfo {author} {\bibfnamefont {Marco}\ \bibnamefont
  {Roncadelli}}, \ and\ \bibinfo {author} {\bibfnamefont {Carmelo}\
  \bibnamefont {Evoli}},\ }\bibfield  {title} {\enquote {\bibinfo {title}
  {{Blazar VHE spectral alterations induced by photon\textendash{}ALP
  oscillations}},}\ }\href {\doibase 10.1093/mnras/stz1144} {\bibfield
  {journal} {\bibinfo  {journal} {Mon. Not. Roy. Astron. Soc.}\ }\textbf
  {\bibinfo {volume} {487}},\ \bibinfo {pages} {123--132} (\bibinfo {year}
  {2019})},\ \Eprint {http://arxiv.org/abs/1811.03548} {arXiv:1811.03548
  [astro-ph.HE]} \BibitemShut {NoStop}%
\bibitem [{\citenamefont {Liang}\ \emph {et~al.}(2019)\citenamefont {Liang},
  \citenamefont {Zhang}, \citenamefont {Xia}, \citenamefont {Feng},
  \citenamefont {Yuan},\ and\ \citenamefont {Fan}}]{Liang:2018mqm}%
  \BibitemOpen
  \bibfield  {author} {\bibinfo {author} {\bibfnamefont {Yun-Feng}\
  \bibnamefont {Liang}}, \bibinfo {author} {\bibfnamefont {Cun}\ \bibnamefont
  {Zhang}}, \bibinfo {author} {\bibfnamefont {Zi-Qing}\ \bibnamefont {Xia}},
  \bibinfo {author} {\bibfnamefont {Lei}\ \bibnamefont {Feng}}, \bibinfo
  {author} {\bibfnamefont {Qiang}\ \bibnamefont {Yuan}}, \ and\ \bibinfo
  {author} {\bibfnamefont {Yi-Zhong}\ \bibnamefont {Fan}},\ }\bibfield  {title}
  {\enquote {\bibinfo {title} {{Constraints on axion-like particle properties
  with TeV gamma-ray observations of Galactic sources}},}\ }\href {\doibase
  10.1088/1475-7516/2019/06/042} {\bibfield  {journal} {\bibinfo  {journal}
  {JCAP}\ }\textbf {\bibinfo {volume} {06}},\ \bibinfo {pages} {042} (\bibinfo
  {year} {2019})},\ \Eprint {http://arxiv.org/abs/1804.07186} {arXiv:1804.07186
  [hep-ph]} \BibitemShut {NoStop}%
\bibitem [{\citenamefont {Bu}\ and\ \citenamefont {Li}(2019)}]{Bu:2019qqg}%
  \BibitemOpen
  \bibfield  {author} {\bibinfo {author} {\bibfnamefont {Jia}\ \bibnamefont
  {Bu}}\ and\ \bibinfo {author} {\bibfnamefont {Ya-Ping}\ \bibnamefont {Li}},\
  }\bibfield  {title} {\enquote {\bibinfo {title} {{Constraints on axionlike
  particles with different magnetic field models from the PKS 2155-304 energy
  spectrum}},}\ }\href {\doibase 10.1088/1674-4527/19/10/154} {\  (\bibinfo
  {year} {2019}),\ 10.1088/1674-4527/19/10/154},\ \Eprint
  {http://arxiv.org/abs/1906.00357} {arXiv:1906.00357 [astro-ph.HE]}
  \BibitemShut {NoStop}%
\bibitem [{\citenamefont {Cheng}\ \emph {et~al.}(2020)\citenamefont {Cheng},
  \citenamefont {He}, \citenamefont {Liang}, \citenamefont {Lu},\ and\
  \citenamefont {Liang}}]{Cheng:2020bhr}%
  \BibitemOpen
  \bibfield  {author} {\bibinfo {author} {\bibfnamefont {Ji-Gui}\ \bibnamefont
  {Cheng}}, \bibinfo {author} {\bibfnamefont {Ya-Jun}\ \bibnamefont {He}},
  \bibinfo {author} {\bibfnamefont {Yun-Feng}\ \bibnamefont {Liang}}, \bibinfo
  {author} {\bibfnamefont {Rui-Jing}\ \bibnamefont {Lu}}, \ and\ \bibinfo
  {author} {\bibfnamefont {En-Wei}\ \bibnamefont {Liang}},\ }\bibfield  {title}
  {\enquote {\bibinfo {title} {{Revisiting the analysis of axion-like particles
  with the Fermi-LAT gamma-ray observation of NGC1275}},}\ }\href@noop {} {\
  (\bibinfo {year} {2020})},\ \Eprint {http://arxiv.org/abs/2010.12396}
  {arXiv:2010.12396 [astro-ph.HE]} \BibitemShut {NoStop}%
\bibitem [{\citenamefont {Carenza}\ \emph {et~al.}(2021)\citenamefont
  {Carenza}, \citenamefont {Evoli}, \citenamefont {Giannotti}, \citenamefont
  {Mirizzi},\ and\ \citenamefont {Montanino}}]{Carenza:2021alz}%
  \BibitemOpen
  \bibfield  {author} {\bibinfo {author} {\bibfnamefont {Pierluca}\
  \bibnamefont {Carenza}}, \bibinfo {author} {\bibfnamefont {Carmelo}\
  \bibnamefont {Evoli}}, \bibinfo {author} {\bibfnamefont {Maurizio}\
  \bibnamefont {Giannotti}}, \bibinfo {author} {\bibfnamefont {Alessandro}\
  \bibnamefont {Mirizzi}}, \ and\ \bibinfo {author} {\bibfnamefont {Daniele}\
  \bibnamefont {Montanino}},\ }\bibfield  {title} {\enquote {\bibinfo {title}
  {{Turbulent axion-photon conversions in the Milky~Way}},}\ }\href {\doibase
  10.1103/PhysRevD.104.023003} {\bibfield  {journal} {\bibinfo  {journal}
  {Phys. Rev. D}\ }\textbf {\bibinfo {volume} {104}},\ \bibinfo {pages}
  {023003} (\bibinfo {year} {2021})},\ \Eprint
  {http://arxiv.org/abs/2104.13935} {arXiv:2104.13935 [hep-ph]} \BibitemShut
  {NoStop}%
\bibitem [{\citenamefont {Marsh}\ \emph {et~al.}(2021)\citenamefont {Marsh},
  \citenamefont {Matthews}, \citenamefont {Reynolds},\ and\ \citenamefont
  {Carenza}}]{Marsh:2021ajy}%
  \BibitemOpen
  \bibfield  {author} {\bibinfo {author} {\bibfnamefont {M.~C.~David}\
  \bibnamefont {Marsh}}, \bibinfo {author} {\bibfnamefont {James~H.}\
  \bibnamefont {Matthews}}, \bibinfo {author} {\bibfnamefont {Christopher}\
  \bibnamefont {Reynolds}}, \ and\ \bibinfo {author} {\bibfnamefont {Pierluca}\
  \bibnamefont {Carenza}},\ }\bibfield  {title} {\enquote {\bibinfo {title}
  {{The Fourier formalism for relativistic axion-photon conversion, with
  astrophysical applications}},}\ }\href@noop {} {\  (\bibinfo {year}
  {2021})},\ \Eprint {http://arxiv.org/abs/2107.08040} {arXiv:2107.08040
  [hep-ph]} \BibitemShut {NoStop}%
\bibitem [{\citenamefont {Reyn\'es}\ \emph {et~al.}(2021)\citenamefont
  {Reyn\'es}, \citenamefont {Matthews}, \citenamefont {Reynolds}, \citenamefont
  {Russell}, \citenamefont {Smith},\ and\ \citenamefont
  {Marsh}}]{Reynes:2021bpe}%
  \BibitemOpen
  \bibfield  {author} {\bibinfo {author} {\bibfnamefont {J\'ulia~Sisk}\
  \bibnamefont {Reyn\'es}}, \bibinfo {author} {\bibfnamefont {James~H.}\
  \bibnamefont {Matthews}}, \bibinfo {author} {\bibfnamefont {Christopher~S.}\
  \bibnamefont {Reynolds}}, \bibinfo {author} {\bibfnamefont {Helen~R.}\
  \bibnamefont {Russell}}, \bibinfo {author} {\bibfnamefont {Robyn~N.}\
  \bibnamefont {Smith}}, \ and\ \bibinfo {author} {\bibfnamefont {M.~C.~David}\
  \bibnamefont {Marsh}},\ }\bibfield  {title} {\enquote {\bibinfo {title} {{New
  constraints on light axion-like particles using Chandra transmission grating
  spectroscopy of the powerful cluster-hosted quasar H1821+643}},}\ }\href
  {\doibase 10.1093/mnras/stab3464} {\bibfield  {journal} {\bibinfo  {journal}
  {Mon. Not. Roy. Astron. Soc.}\ }\textbf {\bibinfo {volume} {510}},\ \bibinfo
  {pages} {1264--1277} (\bibinfo {year} {2021})},\ \Eprint
  {http://arxiv.org/abs/2109.03261} {arXiv:2109.03261 [astro-ph.HE]}
  \BibitemShut {NoStop}%
\bibitem [{\citenamefont {Matthews}\ \emph {et~al.}(2022)\citenamefont
  {Matthews}, \citenamefont {Reynolds}, \citenamefont {Marsh}, \citenamefont
  {Sisk-Reyn\'es},\ and\ \citenamefont {Rodman}}]{Matthews:2022gqi}%
  \BibitemOpen
  \bibfield  {author} {\bibinfo {author} {\bibfnamefont {James~H.}\
  \bibnamefont {Matthews}}, \bibinfo {author} {\bibfnamefont {Christopher~S.}\
  \bibnamefont {Reynolds}}, \bibinfo {author} {\bibfnamefont {M.~C.~David}\
  \bibnamefont {Marsh}}, \bibinfo {author} {\bibfnamefont {J\'ulia}\
  \bibnamefont {Sisk-Reyn\'es}}, \ and\ \bibinfo {author} {\bibfnamefont
  {Payton~E.}\ \bibnamefont {Rodman}},\ }\bibfield  {title} {\enquote {\bibinfo
  {title} {{How Do Magnetic Field Models Affect Astrophysical Limits on Light
  Axion-like Particles? An X-Ray Case Study with NGC 1275}},}\ }\href {\doibase
  10.3847/1538-4357/ac5625} {\bibfield  {journal} {\bibinfo  {journal}
  {Astrophys. J.}\ }\textbf {\bibinfo {volume} {930}},\ \bibinfo {pages} {90}
  (\bibinfo {year} {2022})},\ \Eprint {http://arxiv.org/abs/2202.08875}
  {arXiv:2202.08875 [astro-ph.HE]} \BibitemShut {NoStop}%
\bibitem [{\citenamefont {Jacobsen}\ \emph {et~al.}(2022)\citenamefont
  {Jacobsen}, \citenamefont {Linden},\ and\ \citenamefont
  {Freese}}]{Jacobsen:2022swa}%
  \BibitemOpen
  \bibfield  {author} {\bibinfo {author} {\bibfnamefont {Sunniva}\ \bibnamefont
  {Jacobsen}}, \bibinfo {author} {\bibfnamefont {Tim}\ \bibnamefont {Linden}},
  \ and\ \bibinfo {author} {\bibfnamefont {Katherine}\ \bibnamefont {Freese}},\
  }\bibfield  {title} {\enquote {\bibinfo {title} {{Constraining Axion-Like
  Particles with HAWC Observations of TeV Blazars}},}\ }\href@noop {} {\
  (\bibinfo {year} {2022})},\ \Eprint {http://arxiv.org/abs/2203.04332}
  {arXiv:2203.04332 [hep-ph]} \BibitemShut {NoStop}%
\bibitem [{\citenamefont {Galanti}\ and\ \citenamefont
  {Roncadelli}(2022)}]{Galanti:2022ijh}%
  \BibitemOpen
  \bibfield  {author} {\bibinfo {author} {\bibfnamefont {Giorgio}\ \bibnamefont
  {Galanti}}\ and\ \bibinfo {author} {\bibfnamefont {Marco}\ \bibnamefont
  {Roncadelli}},\ }\bibfield  {title} {\enquote {\bibinfo {title} {{Axion-like
  Particles Implications for High-Energy Astrophysics}},}\ }\href {\doibase
  10.3390/universe8050253} {\bibfield  {journal} {\bibinfo  {journal}
  {Universe}\ }\textbf {\bibinfo {volume} {8}},\ \bibinfo {pages} {253}
  (\bibinfo {year} {2022})},\ \Eprint {http://arxiv.org/abs/2205.00940}
  {arXiv:2205.00940 [hep-ph]} \BibitemShut {NoStop}%
\bibitem [{\citenamefont {Aartsen}\ \emph
  {et~al.}(2013{\natexlab{a}})\citenamefont {Aartsen} \emph
  {et~al.}}]{IceCube:2013cdw}%
  \BibitemOpen
  \bibfield  {author} {\bibinfo {author} {\bibfnamefont {M.~G.}\ \bibnamefont
  {Aartsen}} \emph {et~al.} (\bibinfo {collaboration} {IceCube}),\ }\bibfield
  {title} {\enquote {\bibinfo {title} {{First observation of PeV-energy
  neutrinos with IceCube}},}\ }\href {\doibase 10.1103/PhysRevLett.111.021103}
  {\bibfield  {journal} {\bibinfo  {journal} {Phys. Rev. Lett.}\ }\textbf
  {\bibinfo {volume} {111}},\ \bibinfo {pages} {021103} (\bibinfo {year}
  {2013}{\natexlab{a}})},\ \Eprint {http://arxiv.org/abs/1304.5356}
  {arXiv:1304.5356 [astro-ph.HE]} \BibitemShut {NoStop}%
\bibitem [{\citenamefont {Aartsen}\ \emph
  {et~al.}(2013{\natexlab{b}})\citenamefont {Aartsen} \emph
  {et~al.}}]{IceCube:2013low}%
  \BibitemOpen
  \bibfield  {author} {\bibinfo {author} {\bibfnamefont {M.~G.}\ \bibnamefont
  {Aartsen}} \emph {et~al.} (\bibinfo {collaboration} {IceCube}),\ }\bibfield
  {title} {\enquote {\bibinfo {title} {{Evidence for High-Energy
  Extraterrestrial Neutrinos at the IceCube Detector}},}\ }\href {\doibase
  10.1126/science.1242856} {\bibfield  {journal} {\bibinfo  {journal}
  {Science}\ }\textbf {\bibinfo {volume} {342}},\ \bibinfo {pages} {1242856}
  (\bibinfo {year} {2013}{\natexlab{b}})},\ \Eprint
  {http://arxiv.org/abs/1311.5238} {arXiv:1311.5238 [astro-ph.HE]} \BibitemShut
  {NoStop}%
\bibitem [{\citenamefont {Aartsen}\ \emph {et~al.}(2014)\citenamefont {Aartsen}
  \emph {et~al.}}]{IceCube:2014stg}%
  \BibitemOpen
  \bibfield  {author} {\bibinfo {author} {\bibfnamefont {M.~G.}\ \bibnamefont
  {Aartsen}} \emph {et~al.} (\bibinfo {collaboration} {IceCube}),\ }\bibfield
  {title} {\enquote {\bibinfo {title} {{Observation of High-Energy
  Astrophysical Neutrinos in Three Years of IceCube Data}},}\ }\href {\doibase
  10.1103/PhysRevLett.113.101101} {\bibfield  {journal} {\bibinfo  {journal}
  {Phys. Rev. Lett.}\ }\textbf {\bibinfo {volume} {113}},\ \bibinfo {pages}
  {101101} (\bibinfo {year} {2014})},\ \Eprint {http://arxiv.org/abs/1405.5303}
  {arXiv:1405.5303 [astro-ph.HE]} \BibitemShut {NoStop}%
\bibitem [{\citenamefont {Aartsen}\ \emph {et~al.}(2015)\citenamefont {Aartsen}
  \emph {et~al.}}]{IceCube:2015qii}%
  \BibitemOpen
  \bibfield  {author} {\bibinfo {author} {\bibfnamefont {M.~G.}\ \bibnamefont
  {Aartsen}} \emph {et~al.} (\bibinfo {collaboration} {IceCube}),\ }\bibfield
  {title} {\enquote {\bibinfo {title} {{Evidence for Astrophysical Muon
  Neutrinos from the Northern Sky with IceCube}},}\ }\href {\doibase
  10.1103/PhysRevLett.115.081102} {\bibfield  {journal} {\bibinfo  {journal}
  {Phys. Rev. Lett.}\ }\textbf {\bibinfo {volume} {115}},\ \bibinfo {pages}
  {081102} (\bibinfo {year} {2015})},\ \Eprint
  {http://arxiv.org/abs/1507.04005} {arXiv:1507.04005 [astro-ph.HE]}
  \BibitemShut {NoStop}%
\bibitem [{\citenamefont {Aartsen}\ \emph {et~al.}(2016)\citenamefont {Aartsen}
  \emph {et~al.}}]{IceCube:2016umi}%
  \BibitemOpen
  \bibfield  {author} {\bibinfo {author} {\bibfnamefont {M.~G.}\ \bibnamefont
  {Aartsen}} \emph {et~al.} (\bibinfo {collaboration} {IceCube}),\ }\bibfield
  {title} {\enquote {\bibinfo {title} {{Observation and Characterization of a
  Cosmic Muon Neutrino Flux from the Northern Hemisphere using six years of
  IceCube data}},}\ }\href {\doibase 10.3847/0004-637X/833/1/3} {\bibfield
  {journal} {\bibinfo  {journal} {Astrophys. J.}\ }\textbf {\bibinfo {volume}
  {833}},\ \bibinfo {pages} {3} (\bibinfo {year} {2016})},\ \Eprint
  {http://arxiv.org/abs/1607.08006} {arXiv:1607.08006 [astro-ph.HE]}
  \BibitemShut {NoStop}%
\bibitem [{\citenamefont {Ahlers}\ and\ \citenamefont
  {Halzen}(2018)}]{Ahlers:2018fkn}%
  \BibitemOpen
  \bibfield  {author} {\bibinfo {author} {\bibfnamefont {Markus}\ \bibnamefont
  {Ahlers}}\ and\ \bibinfo {author} {\bibfnamefont {Francis}\ \bibnamefont
  {Halzen}},\ }\bibfield  {title} {\enquote {\bibinfo {title} {{Opening a New
  Window onto the Universe with IceCube}},}\ }\href {\doibase
  10.1016/j.ppnp.2018.05.001} {\bibfield  {journal} {\bibinfo  {journal} {Prog.
  Part. Nucl. Phys.}\ }\textbf {\bibinfo {volume} {102}},\ \bibinfo {pages}
  {73--88} (\bibinfo {year} {2018})},\ \Eprint
  {http://arxiv.org/abs/1805.11112} {arXiv:1805.11112 [astro-ph.HE]}
  \BibitemShut {NoStop}%
\bibitem [{\citenamefont {Abbasi}\ \emph {et~al.}(2021)\citenamefont {Abbasi}
  \emph {et~al.}}]{IceCube:2020wum}%
  \BibitemOpen
  \bibfield  {author} {\bibinfo {author} {\bibfnamefont {R.}~\bibnamefont
  {Abbasi}} \emph {et~al.} (\bibinfo {collaboration} {IceCube}),\ }\bibfield
  {title} {\enquote {\bibinfo {title} {{The IceCube high-energy starting event
  sample: Description and flux characterization with 7.5 years of data}},}\
  }\href {\doibase 10.1103/PhysRevD.104.022002} {\bibfield  {journal} {\bibinfo
   {journal} {Phys. Rev. D}\ }\textbf {\bibinfo {volume} {104}},\ \bibinfo
  {pages} {022002} (\bibinfo {year} {2021})},\ \Eprint
  {http://arxiv.org/abs/2011.03545} {arXiv:2011.03545 [astro-ph.HE]}
  \BibitemShut {NoStop}%
\bibitem [{\citenamefont {Vogel}\ \emph {et~al.}(2019)\citenamefont {Vogel},
  \citenamefont {Laha},\ and\ \citenamefont {Meyer}}]{Vogel:2017fmc}%
  \BibitemOpen
  \bibfield  {author} {\bibinfo {author} {\bibfnamefont {Hendrik}\ \bibnamefont
  {Vogel}}, \bibinfo {author} {\bibfnamefont {Ranjan}\ \bibnamefont {Laha}}, \
  and\ \bibinfo {author} {\bibfnamefont {Manuel}\ \bibnamefont {Meyer}},\
  }\bibfield  {title} {\enquote {\bibinfo {title} {{Diffuse axion-like particle
  searches}},}\ }\href {\doibase 10.22323/1.337.0091} {\bibfield  {journal}
  {\bibinfo  {journal} {PoS}\ }\textbf {\bibinfo {volume} {NOW2018}},\ \bibinfo
  {pages} {091} (\bibinfo {year} {2019})},\ \Eprint
  {http://arxiv.org/abs/1712.01839} {arXiv:1712.01839 [hep-ph]} \BibitemShut
  {NoStop}%
\bibitem [{\citenamefont {Eckner}\ and\ \citenamefont
  {Calore}(2022)}]{Eckner:2022rwf}%
  \BibitemOpen
  \bibfield  {author} {\bibinfo {author} {\bibfnamefont {Christopher}\
  \bibnamefont {Eckner}}\ and\ \bibinfo {author} {\bibfnamefont {Francesca}\
  \bibnamefont {Calore}},\ }\bibfield  {title} {\enquote {\bibinfo {title}
  {{First constraints on axion-like particles from Galactic sub-PeV gamma
  rays}},}\ }\href@noop {} {\  (\bibinfo {year} {2022})},\ \Eprint
  {http://arxiv.org/abs/2204.12487} {arXiv:2204.12487 [astro-ph.HE]}
  \BibitemShut {NoStop}%
\bibitem [{\citenamefont {Addazi}\ \emph {et~al.}(2022)\citenamefont {Addazi}
  \emph {et~al.}}]{LHAASO:2019qtb}%
  \BibitemOpen
  \bibfield  {author} {\bibinfo {author} {\bibfnamefont {Andrea}\ \bibnamefont
  {Addazi}} \emph {et~al.} (\bibinfo {collaboration} {LHAASO}),\ }\bibfield
  {title} {\enquote {\bibinfo {title} {{The Large High Altitude Air Shower
  Observatory (LHAASO) Science Book (2021 Edition)}},}\ }\href@noop {}
  {\bibfield  {journal} {\bibinfo  {journal} {Chin. Phys. C}\ }\textbf
  {\bibinfo {volume} {46}},\ \bibinfo {pages} {035001--035007} (\bibinfo {year}
  {2022})},\ \Eprint {http://arxiv.org/abs/1905.02773} {arXiv:1905.02773
  [astro-ph.HE]} \BibitemShut {NoStop}%
\bibitem [{\citenamefont {Zhao}\ \emph {et~al.}(2021)\citenamefont {Zhao},
  \citenamefont {Zhang}, \citenamefont {Zhang},\ and\ \citenamefont
  {Yuan}}]{Zhao:2021dqj}%
  \BibitemOpen
  \bibfield  {author} {\bibinfo {author} {\bibfnamefont {Shiping}\ \bibnamefont
  {Zhao}}, \bibinfo {author} {\bibfnamefont {R.}~\bibnamefont {Zhang}},
  \bibinfo {author} {\bibfnamefont {Y.}~\bibnamefont {Zhang}}, \ and\ \bibinfo
  {author} {\bibfnamefont {Q.}~\bibnamefont {Yuan}} (\bibinfo {collaboration}
  {LHAASO}),\ }\bibfield  {title} {\enquote {\bibinfo {title} {{Measurement of
  the diffuse gamma-ray emission from Galactic plane with LHAASO-KM2A}},}\
  }\href {\doibase 10.22323/1.395.0859} {\bibfield  {journal} {\bibinfo
  {journal} {PoS}\ }\textbf {\bibinfo {volume} {ICRC2021}},\ \bibinfo {pages}
  {859} (\bibinfo {year} {2021})}\BibitemShut {NoStop}%
\bibitem [{\citenamefont {Murase}\ \emph {et~al.}(2016)\citenamefont {Murase},
  \citenamefont {Guetta},\ and\ \citenamefont {Ahlers}}]{Murase:2015xka}%
  \BibitemOpen
  \bibfield  {author} {\bibinfo {author} {\bibfnamefont {Kohta}\ \bibnamefont
  {Murase}}, \bibinfo {author} {\bibfnamefont {Dafne}\ \bibnamefont {Guetta}},
  \ and\ \bibinfo {author} {\bibfnamefont {Markus}\ \bibnamefont {Ahlers}},\
  }\bibfield  {title} {\enquote {\bibinfo {title} {{Hidden Cosmic-Ray
  Accelerators as an Origin of TeV-PeV Cosmic Neutrinos}},}\ }\href {\doibase
  10.1103/PhysRevLett.116.071101} {\bibfield  {journal} {\bibinfo  {journal}
  {Phys. Rev. Lett.}\ }\textbf {\bibinfo {volume} {116}},\ \bibinfo {pages}
  {071101} (\bibinfo {year} {2016})},\ \Eprint
  {http://arxiv.org/abs/1509.00805} {arXiv:1509.00805 [astro-ph.HE]}
  \BibitemShut {NoStop}%
\bibitem [{\citenamefont {Capanema}\ \emph {et~al.}(2020)\citenamefont
  {Capanema}, \citenamefont {Esmaili},\ and\ \citenamefont
  {Murase}}]{Capanema:2020rjj}%
  \BibitemOpen
  \bibfield  {author} {\bibinfo {author} {\bibfnamefont {Antonio}\ \bibnamefont
  {Capanema}}, \bibinfo {author} {\bibfnamefont {Arman}\ \bibnamefont
  {Esmaili}}, \ and\ \bibinfo {author} {\bibfnamefont {Kohta}\ \bibnamefont
  {Murase}},\ }\bibfield  {title} {\enquote {\bibinfo {title} {{New constraints
  on the origin of medium-energy neutrinos observed by IceCube}},}\ }\href
  {\doibase 10.1103/PhysRevD.101.103012} {\bibfield  {journal} {\bibinfo
  {journal} {Phys. Rev. D}\ }\textbf {\bibinfo {volume} {101}},\ \bibinfo
  {pages} {103012} (\bibinfo {year} {2020})},\ \Eprint
  {http://arxiv.org/abs/2002.07192} {arXiv:2002.07192 [hep-ph]} \BibitemShut
  {NoStop}%
\bibitem [{\citenamefont {Capanema}\ \emph {et~al.}(2021)\citenamefont
  {Capanema}, \citenamefont {Esmaili},\ and\ \citenamefont
  {Serpico}}]{Capanema:2020oet}%
  \BibitemOpen
  \bibfield  {author} {\bibinfo {author} {\bibfnamefont {Antonio}\ \bibnamefont
  {Capanema}}, \bibinfo {author} {\bibfnamefont {Arman}\ \bibnamefont
  {Esmaili}}, \ and\ \bibinfo {author} {\bibfnamefont {Pasquale~Dario}\
  \bibnamefont {Serpico}},\ }\bibfield  {title} {\enquote {\bibinfo {title}
  {{Where do IceCube neutrinos come from? Hints from the diffuse gamma-ray
  flux}},}\ }\href {\doibase 10.1088/1475-7516/2021/02/037} {\bibfield
  {journal} {\bibinfo  {journal} {JCAP}\ }\textbf {\bibinfo {volume} {02}},\
  \bibinfo {pages} {037} (\bibinfo {year} {2021})},\ \Eprint
  {http://arxiv.org/abs/2007.07911} {arXiv:2007.07911 [hep-ph]} \BibitemShut
  {NoStop}%
\bibitem [{\citenamefont {Loeb}\ and\ \citenamefont
  {Waxman}(2006)}]{Loeb:2006tw}%
  \BibitemOpen
  \bibfield  {author} {\bibinfo {author} {\bibfnamefont {Abraham}\ \bibnamefont
  {Loeb}}\ and\ \bibinfo {author} {\bibfnamefont {Eli}\ \bibnamefont
  {Waxman}},\ }\bibfield  {title} {\enquote {\bibinfo {title} {{The Cumulative
  background of high energy neutrinos from starburst galaxies}},}\ }\href
  {\doibase 10.1088/1475-7516/2006/05/003} {\bibfield  {journal} {\bibinfo
  {journal} {JCAP}\ }\textbf {\bibinfo {volume} {05}},\ \bibinfo {pages} {003}
  (\bibinfo {year} {2006})},\ \Eprint {http://arxiv.org/abs/astro-ph/0601695}
  {arXiv:astro-ph/0601695} \BibitemShut {NoStop}%
\bibitem [{\citenamefont {Thompson}\ \emph {et~al.}(2006)\citenamefont
  {Thompson}, \citenamefont {Quataert}, \citenamefont {Waxman},\ and\
  \citenamefont {Loeb}}]{Thompson:2006np}%
  \BibitemOpen
  \bibfield  {author} {\bibinfo {author} {\bibfnamefont {Todd~A.}\ \bibnamefont
  {Thompson}}, \bibinfo {author} {\bibfnamefont {Eliot}\ \bibnamefont
  {Quataert}}, \bibinfo {author} {\bibfnamefont {Eli}\ \bibnamefont {Waxman}},
  \ and\ \bibinfo {author} {\bibfnamefont {Abraham}\ \bibnamefont {Loeb}},\
  }\bibfield  {title} {\enquote {\bibinfo {title} {{Assessing The Starburst
  Contribution to the Gamma-Ray and Neutrino Backgrounds}},}\ }\href@noop {} {\
   (\bibinfo {year} {2006})},\ \Eprint {http://arxiv.org/abs/astro-ph/0608699}
  {arXiv:astro-ph/0608699} \BibitemShut {NoStop}%
\bibitem [{\citenamefont {Tamborra}\ \emph {et~al.}(2014)\citenamefont
  {Tamborra}, \citenamefont {Ando},\ and\ \citenamefont
  {Murase}}]{Tamborra:2014xia}%
  \BibitemOpen
  \bibfield  {author} {\bibinfo {author} {\bibfnamefont {Irene}\ \bibnamefont
  {Tamborra}}, \bibinfo {author} {\bibfnamefont {Shin'ichiro}\ \bibnamefont
  {Ando}}, \ and\ \bibinfo {author} {\bibfnamefont {Kohta}\ \bibnamefont
  {Murase}},\ }\bibfield  {title} {\enquote {\bibinfo {title} {{Star-forming
  galaxies as the origin of diffuse high-energy backgrounds: Gamma-ray and
  neutrino connections, and implications for starburst history}},}\ }\href
  {\doibase 10.1088/1475-7516/2014/09/043} {\bibfield  {journal} {\bibinfo
  {journal} {JCAP}\ }\textbf {\bibinfo {volume} {09}},\ \bibinfo {pages} {043}
  (\bibinfo {year} {2014})},\ \Eprint {http://arxiv.org/abs/1404.1189}
  {arXiv:1404.1189 [astro-ph.HE]} \BibitemShut {NoStop}%
\bibitem [{\citenamefont {Chang}\ and\ \citenamefont
  {Wang}(2014)}]{Chang:2014hua}%
  \BibitemOpen
  \bibfield  {author} {\bibinfo {author} {\bibfnamefont {Xiao-Chuan}\
  \bibnamefont {Chang}}\ and\ \bibinfo {author} {\bibfnamefont {Xiang-Yu}\
  \bibnamefont {Wang}},\ }\bibfield  {title} {\enquote {\bibinfo {title} {{The
  diffuse gamma-ray flux associated with sub-PeV/PeV neutrinos from starburst
  galaxies}},}\ }\href {\doibase 10.1088/0004-637X/793/2/131} {\bibfield
  {journal} {\bibinfo  {journal} {Astrophys. J.}\ }\textbf {\bibinfo {volume}
  {793}},\ \bibinfo {pages} {131} (\bibinfo {year} {2014})},\ \Eprint
  {http://arxiv.org/abs/1406.1099} {arXiv:1406.1099 [astro-ph.HE]} \BibitemShut
  {NoStop}%
\bibitem [{\citenamefont {Senno}\ \emph {et~al.}(2015)\citenamefont {Senno},
  \citenamefont {M\'esz\'aros}, \citenamefont {Murase}, \citenamefont
  {Baerwald},\ and\ \citenamefont {Rees}}]{Senno:2015tra}%
  \BibitemOpen
  \bibfield  {author} {\bibinfo {author} {\bibfnamefont {Nicholas}\
  \bibnamefont {Senno}}, \bibinfo {author} {\bibfnamefont {Peter}\ \bibnamefont
  {M\'esz\'aros}}, \bibinfo {author} {\bibfnamefont {Kohta}\ \bibnamefont
  {Murase}}, \bibinfo {author} {\bibfnamefont {Philipp}\ \bibnamefont
  {Baerwald}}, \ and\ \bibinfo {author} {\bibfnamefont {Martin~J.}\
  \bibnamefont {Rees}},\ }\bibfield  {title} {\enquote {\bibinfo {title}
  {{Extragalactic star-forming galaxies with hypernovae and supernovae as
  high-energy neutrino and gamma-ray sources: the case of the 10 TeV neutrino
  data}},}\ }\href {\doibase 10.1088/0004-637X/806/1/24} {\bibfield  {journal}
  {\bibinfo  {journal} {Astrophys. J.}\ }\textbf {\bibinfo {volume} {806}},\
  \bibinfo {pages} {24} (\bibinfo {year} {2015})},\ \Eprint
  {http://arxiv.org/abs/1501.04934} {arXiv:1501.04934 [astro-ph.HE]}
  \BibitemShut {NoStop}%
\bibitem [{\citenamefont {Chakraborty}\ and\ \citenamefont
  {Izaguirre}(2015)}]{Chakraborty:2015sta}%
  \BibitemOpen
  \bibfield  {author} {\bibinfo {author} {\bibfnamefont {Sovan}\ \bibnamefont
  {Chakraborty}}\ and\ \bibinfo {author} {\bibfnamefont {Ignacio}\ \bibnamefont
  {Izaguirre}},\ }\bibfield  {title} {\enquote {\bibinfo {title} {{Diffuse
  neutrinos from extragalactic supernova remnants: Dominating the 100 TeV
  IceCube flux}},}\ }\href {\doibase 10.1016/j.physletb.2015.04.032} {\bibfield
   {journal} {\bibinfo  {journal} {Phys. Lett. B}\ }\textbf {\bibinfo {volume}
  {745}},\ \bibinfo {pages} {35--39} (\bibinfo {year} {2015})},\ \Eprint
  {http://arxiv.org/abs/1501.02615} {arXiv:1501.02615 [hep-ph]} \BibitemShut
  {NoStop}%
\bibitem [{\citenamefont {Peretti}\ \emph {et~al.}(2019)\citenamefont
  {Peretti}, \citenamefont {Blasi}, \citenamefont {Aharonian},\ and\
  \citenamefont {Morlino}}]{Peretti:2018tmo}%
  \BibitemOpen
  \bibfield  {author} {\bibinfo {author} {\bibfnamefont {Enrico}\ \bibnamefont
  {Peretti}}, \bibinfo {author} {\bibfnamefont {Pasquale}\ \bibnamefont
  {Blasi}}, \bibinfo {author} {\bibfnamefont {Felix}\ \bibnamefont
  {Aharonian}}, \ and\ \bibinfo {author} {\bibfnamefont {Giovanni}\
  \bibnamefont {Morlino}},\ }\bibfield  {title} {\enquote {\bibinfo {title}
  {{Cosmic ray transport and radiative processes in nuclei of starburst
  galaxies}},}\ }\href {\doibase 10.1093/mnras/stz1161} {\bibfield  {journal}
  {\bibinfo  {journal} {Mon. Not. Roy. Astron. Soc.}\ }\textbf {\bibinfo
  {volume} {487}},\ \bibinfo {pages} {168--180} (\bibinfo {year} {2019})},\
  \Eprint {http://arxiv.org/abs/1812.01996} {arXiv:1812.01996 [astro-ph.HE]}
  \BibitemShut {NoStop}%
\bibitem [{\citenamefont {Peretti}\ \emph {et~al.}(2020)\citenamefont
  {Peretti}, \citenamefont {Blasi}, \citenamefont {Aharonian}, \citenamefont
  {Morlino},\ and\ \citenamefont {Cristofari}}]{Peretti:2019vsj}%
  \BibitemOpen
  \bibfield  {author} {\bibinfo {author} {\bibfnamefont {Enrico}\ \bibnamefont
  {Peretti}}, \bibinfo {author} {\bibfnamefont {Pasquale}\ \bibnamefont
  {Blasi}}, \bibinfo {author} {\bibfnamefont {Felix}\ \bibnamefont
  {Aharonian}}, \bibinfo {author} {\bibfnamefont {Giovanni}\ \bibnamefont
  {Morlino}}, \ and\ \bibinfo {author} {\bibfnamefont {Pierre}\ \bibnamefont
  {Cristofari}},\ }\bibfield  {title} {\enquote {\bibinfo {title}
  {{Contribution of starburst nuclei to the diffuse gamma-ray and neutrino
  flux}},}\ }\href {\doibase 10.1093/mnras/staa698} {\bibfield  {journal}
  {\bibinfo  {journal} {Mon. Not. Roy. Astron. Soc.}\ }\textbf {\bibinfo
  {volume} {493}},\ \bibinfo {pages} {5880--5891} (\bibinfo {year} {2020})},\
  \Eprint {http://arxiv.org/abs/1911.06163} {arXiv:1911.06163 [astro-ph.HE]}
  \BibitemShut {NoStop}%
\bibitem [{\citenamefont {Ambrosone}\ \emph
  {et~al.}(2021{\natexlab{a}})\citenamefont {Ambrosone}, \citenamefont
  {Chianese}, \citenamefont {Fiorillo}, \citenamefont {Marinelli},
  \citenamefont {Miele},\ and\ \citenamefont {Pisanti}}]{Ambrosone:2020evo}%
  \BibitemOpen
  \bibfield  {author} {\bibinfo {author} {\bibfnamefont {Antonio}\ \bibnamefont
  {Ambrosone}}, \bibinfo {author} {\bibfnamefont {Marco}\ \bibnamefont
  {Chianese}}, \bibinfo {author} {\bibfnamefont {Damiano F.~G.}\ \bibnamefont
  {Fiorillo}}, \bibinfo {author} {\bibfnamefont {Antonio}\ \bibnamefont
  {Marinelli}}, \bibinfo {author} {\bibfnamefont {Gennaro}\ \bibnamefont
  {Miele}}, \ and\ \bibinfo {author} {\bibfnamefont {Ofelia}\ \bibnamefont
  {Pisanti}},\ }\bibfield  {title} {\enquote {\bibinfo {title} {{Starburst
  galaxies strike back: a multi-messenger analysis with Fermi-LAT and IceCube
  data}},}\ }\href {\doibase 10.1093/mnras/stab659} {\bibfield  {journal}
  {\bibinfo  {journal} {Mon. Not. Roy. Astron. Soc.}\ }\textbf {\bibinfo
  {volume} {503}},\ \bibinfo {pages} {4032--4049} (\bibinfo {year}
  {2021}{\natexlab{a}})},\ \Eprint {http://arxiv.org/abs/2011.02483}
  {arXiv:2011.02483 [astro-ph.HE]} \BibitemShut {NoStop}%
\bibitem [{\citenamefont {Ambrosone}\ \emph
  {et~al.}(2021{\natexlab{b}})\citenamefont {Ambrosone}, \citenamefont
  {Chianese}, \citenamefont {Fiorillo}, \citenamefont {Marinelli},\ and\
  \citenamefont {Miele}}]{Ambrosone:2021aaw}%
  \BibitemOpen
  \bibfield  {author} {\bibinfo {author} {\bibfnamefont {Antonio}\ \bibnamefont
  {Ambrosone}}, \bibinfo {author} {\bibfnamefont {Marco}\ \bibnamefont
  {Chianese}}, \bibinfo {author} {\bibfnamefont {Damiano F.~G.}\ \bibnamefont
  {Fiorillo}}, \bibinfo {author} {\bibfnamefont {Antonio}\ \bibnamefont
  {Marinelli}}, \ and\ \bibinfo {author} {\bibfnamefont {Gennaro}\ \bibnamefont
  {Miele}},\ }\bibfield  {title} {\enquote {\bibinfo {title} {{Could Nearby
  Star-forming Galaxies Light Up the Pointlike Neutrino Sky?}}}\ }\href
  {\doibase 10.3847/2041-8213/ac25ff} {\bibfield  {journal} {\bibinfo
  {journal} {Astrophys. J. Lett.}\ }\textbf {\bibinfo {volume} {919}},\
  \bibinfo {pages} {L32} (\bibinfo {year} {2021}{\natexlab{b}})},\ \Eprint
  {http://arxiv.org/abs/2106.13248} {arXiv:2106.13248 [astro-ph.HE]}
  \BibitemShut {NoStop}%
\bibitem [{\citenamefont {Stecker}\ and\ \citenamefont
  {Salamon}(1996)}]{Stecker:1995th}%
  \BibitemOpen
  \bibfield  {author} {\bibinfo {author} {\bibfnamefont {F.~W.}\ \bibnamefont
  {Stecker}}\ and\ \bibinfo {author} {\bibfnamefont {M.~H.}\ \bibnamefont
  {Salamon}},\ }\bibfield  {title} {\enquote {\bibinfo {title} {{High-energy
  neutrinos from quasars}},}\ }\href {\doibase 10.1007/BF00195044} {\bibfield
  {journal} {\bibinfo  {journal} {Space Sci. Rev.}\ }\textbf {\bibinfo {volume}
  {75}},\ \bibinfo {pages} {341--355} (\bibinfo {year} {1996})},\ \Eprint
  {http://arxiv.org/abs/astro-ph/9501064} {arXiv:astro-ph/9501064} \BibitemShut
  {NoStop}%
\bibitem [{\citenamefont {Atoyan}\ and\ \citenamefont
  {Dermer}(2001)}]{Atoyan:2001ey}%
  \BibitemOpen
  \bibfield  {author} {\bibinfo {author} {\bibfnamefont {Armen}\ \bibnamefont
  {Atoyan}}\ and\ \bibinfo {author} {\bibfnamefont {Charles~D.}\ \bibnamefont
  {Dermer}},\ }\bibfield  {title} {\enquote {\bibinfo {title} {{High-energy
  neutrinos from photomeson processes in blazars}},}\ }\href {\doibase
  10.1103/PhysRevLett.87.221102} {\bibfield  {journal} {\bibinfo  {journal}
  {Phys. Rev. Lett.}\ }\textbf {\bibinfo {volume} {87}},\ \bibinfo {pages}
  {221102} (\bibinfo {year} {2001})},\ \Eprint
  {http://arxiv.org/abs/astro-ph/0108053} {arXiv:astro-ph/0108053} \BibitemShut
  {NoStop}%
\bibitem [{\citenamefont {Murase}\ \emph {et~al.}(2014)\citenamefont {Murase},
  \citenamefont {Inoue},\ and\ \citenamefont {Dermer}}]{Murase:2014foa}%
  \BibitemOpen
  \bibfield  {author} {\bibinfo {author} {\bibfnamefont {Kohta}\ \bibnamefont
  {Murase}}, \bibinfo {author} {\bibfnamefont {Yoshiyuki}\ \bibnamefont
  {Inoue}}, \ and\ \bibinfo {author} {\bibfnamefont {Charles~D.}\ \bibnamefont
  {Dermer}},\ }\bibfield  {title} {\enquote {\bibinfo {title} {{Diffuse
  Neutrino Intensity from the Inner Jets of Active Galactic Nuclei: Impacts of
  External Photon Fields and the Blazar Sequence}},}\ }\href {\doibase
  10.1103/PhysRevD.90.023007} {\bibfield  {journal} {\bibinfo  {journal} {Phys.
  Rev. D}\ }\textbf {\bibinfo {volume} {90}},\ \bibinfo {pages} {023007}
  (\bibinfo {year} {2014})},\ \Eprint {http://arxiv.org/abs/1403.4089}
  {arXiv:1403.4089 [astro-ph.HE]} \BibitemShut {NoStop}%
\bibitem [{\citenamefont {Padovani}\ \emph {et~al.}(2016)\citenamefont
  {Padovani}, \citenamefont {Resconi}, \citenamefont {Giommi}, \citenamefont
  {Arsioli},\ and\ \citenamefont {Chang}}]{Padovani:2016wwn}%
  \BibitemOpen
  \bibfield  {author} {\bibinfo {author} {\bibfnamefont {P.}~\bibnamefont
  {Padovani}}, \bibinfo {author} {\bibfnamefont {E.}~\bibnamefont {Resconi}},
  \bibinfo {author} {\bibfnamefont {P.}~\bibnamefont {Giommi}}, \bibinfo
  {author} {\bibfnamefont {B.}~\bibnamefont {Arsioli}}, \ and\ \bibinfo
  {author} {\bibfnamefont {Y.~L.}\ \bibnamefont {Chang}},\ }\bibfield  {title}
  {\enquote {\bibinfo {title} {{Extreme blazars as counterparts of IceCube
  astrophysical neutrinos}},}\ }\href {\doibase 10.1093/mnras/stw228}
  {\bibfield  {journal} {\bibinfo  {journal} {Mon. Not. Roy. Astron. Soc.}\
  }\textbf {\bibinfo {volume} {457}},\ \bibinfo {pages} {3582--3592} (\bibinfo
  {year} {2016})},\ \Eprint {http://arxiv.org/abs/1601.06550} {arXiv:1601.06550
  [astro-ph.HE]} \BibitemShut {NoStop}%
\bibitem [{\citenamefont {Alvarez-Muniz}\ and\ \citenamefont
  {Meszaros}(2004)}]{Alvarez-Muniz:2004xlu}%
  \BibitemOpen
  \bibfield  {author} {\bibinfo {author} {\bibfnamefont {Jaime}\ \bibnamefont
  {Alvarez-Muniz}}\ and\ \bibinfo {author} {\bibfnamefont {Peter}\ \bibnamefont
  {Meszaros}},\ }\bibfield  {title} {\enquote {\bibinfo {title} {{High energy
  neutrinos from radio-quiet AGNs}},}\ }\href {\doibase
  10.1103/PhysRevD.70.123001} {\bibfield  {journal} {\bibinfo  {journal} {Phys.
  Rev. D}\ }\textbf {\bibinfo {volume} {70}},\ \bibinfo {pages} {123001}
  (\bibinfo {year} {2004})},\ \Eprint {http://arxiv.org/abs/astro-ph/0409034}
  {arXiv:astro-ph/0409034} \BibitemShut {NoStop}%
\bibitem [{\citenamefont {Pe'er}\ \emph {et~al.}(2009)\citenamefont {Pe'er},
  \citenamefont {Murase},\ and\ \citenamefont {Meszaros}}]{Peer:2009vnw}%
  \BibitemOpen
  \bibfield  {author} {\bibinfo {author} {\bibfnamefont {Asaf}\ \bibnamefont
  {Pe'er}}, \bibinfo {author} {\bibfnamefont {Kohta}\ \bibnamefont {Murase}}, \
  and\ \bibinfo {author} {\bibfnamefont {Peter}\ \bibnamefont {Meszaros}},\
  }\bibfield  {title} {\enquote {\bibinfo {title} {{Radio Quiet AGNs as
  Possible Sources of Ultra-high Energy Cosmic Rays}},}\ }\href {\doibase
  10.1103/PhysRevD.80.123018} {\bibfield  {journal} {\bibinfo  {journal} {Phys.
  Rev. D}\ }\textbf {\bibinfo {volume} {80}},\ \bibinfo {pages} {123018}
  (\bibinfo {year} {2009})},\ \Eprint {http://arxiv.org/abs/0911.1776}
  {arXiv:0911.1776 [astro-ph.HE]} \BibitemShut {NoStop}%
\bibitem [{\citenamefont {Palladino}\ \emph {et~al.}(2019)\citenamefont
  {Palladino}, \citenamefont {Rodrigues}, \citenamefont {Gao},\ and\
  \citenamefont {Winter}}]{Palladino:2018lov}%
  \BibitemOpen
  \bibfield  {author} {\bibinfo {author} {\bibfnamefont {Andrea}\ \bibnamefont
  {Palladino}}, \bibinfo {author} {\bibfnamefont {Xavier}\ \bibnamefont
  {Rodrigues}}, \bibinfo {author} {\bibfnamefont {Shan}\ \bibnamefont {Gao}}, \
  and\ \bibinfo {author} {\bibfnamefont {Walter}\ \bibnamefont {Winter}},\
  }\bibfield  {title} {\enquote {\bibinfo {title} {{Interpretation of the
  diffuse astrophysical neutrino flux in terms of the blazar sequence}},}\
  }\href {\doibase 10.3847/1538-4357/aaf507} {\bibfield  {journal} {\bibinfo
  {journal} {Astrophys. J.}\ }\textbf {\bibinfo {volume} {871}},\ \bibinfo
  {pages} {41} (\bibinfo {year} {2019})},\ \Eprint
  {http://arxiv.org/abs/1806.04769} {arXiv:1806.04769 [astro-ph.HE]}
  \BibitemShut {NoStop}%
\bibitem [{\citenamefont {Paczynski}\ and\ \citenamefont
  {Xu}(1994)}]{Paczynski:1994uv}%
  \BibitemOpen
  \bibfield  {author} {\bibinfo {author} {\bibfnamefont {B.}~\bibnamefont
  {Paczynski}}\ and\ \bibinfo {author} {\bibfnamefont {G.~H.}\ \bibnamefont
  {Xu}},\ }\bibfield  {title} {\enquote {\bibinfo {title} {{Neutrino bursts
  from gamma-ray bursts}},}\ }\href {\doibase 10.1086/174178} {\bibfield
  {journal} {\bibinfo  {journal} {Astrophys. J.}\ }\textbf {\bibinfo {volume}
  {427}},\ \bibinfo {pages} {708--713} (\bibinfo {year} {1994})}\BibitemShut
  {NoStop}%
\bibitem [{\citenamefont {Waxman}\ and\ \citenamefont
  {Bahcall}(1997)}]{Waxman:1997ti}%
  \BibitemOpen
  \bibfield  {author} {\bibinfo {author} {\bibfnamefont {Eli}\ \bibnamefont
  {Waxman}}\ and\ \bibinfo {author} {\bibfnamefont {John~N.}\ \bibnamefont
  {Bahcall}},\ }\bibfield  {title} {\enquote {\bibinfo {title} {{High-energy
  neutrinos from cosmological gamma-ray burst fireballs}},}\ }\href {\doibase
  10.1103/PhysRevLett.78.2292} {\bibfield  {journal} {\bibinfo  {journal}
  {Phys. Rev. Lett.}\ }\textbf {\bibinfo {volume} {78}},\ \bibinfo {pages}
  {2292--2295} (\bibinfo {year} {1997})},\ \Eprint
  {http://arxiv.org/abs/astro-ph/9701231} {arXiv:astro-ph/9701231} \BibitemShut
  {NoStop}%
\bibitem [{\citenamefont {Murase}\ and\ \citenamefont
  {Nagataki}(2006)}]{Murase:2005hy}%
  \BibitemOpen
  \bibfield  {author} {\bibinfo {author} {\bibfnamefont {Kohta}\ \bibnamefont
  {Murase}}\ and\ \bibinfo {author} {\bibfnamefont {Shigehiro}\ \bibnamefont
  {Nagataki}},\ }\bibfield  {title} {\enquote {\bibinfo {title} {{High energy
  neutrino emission and neutrino background from gamma-ray bursts in the
  internal shock model}},}\ }\href {\doibase 10.1103/PhysRevD.73.063002}
  {\bibfield  {journal} {\bibinfo  {journal} {Phys. Rev. D}\ }\textbf {\bibinfo
  {volume} {73}},\ \bibinfo {pages} {063002} (\bibinfo {year} {2006})},\
  \Eprint {http://arxiv.org/abs/astro-ph/0512275} {arXiv:astro-ph/0512275}
  \BibitemShut {NoStop}%
\bibitem [{\citenamefont {Baerwald}\ \emph {et~al.}(2011)\citenamefont
  {Baerwald}, \citenamefont {Hummer},\ and\ \citenamefont
  {Winter}}]{Baerwald:2010fk}%
  \BibitemOpen
  \bibfield  {author} {\bibinfo {author} {\bibfnamefont {Philipp}\ \bibnamefont
  {Baerwald}}, \bibinfo {author} {\bibfnamefont {Svenja}\ \bibnamefont
  {Hummer}}, \ and\ \bibinfo {author} {\bibfnamefont {Walter}\ \bibnamefont
  {Winter}},\ }\bibfield  {title} {\enquote {\bibinfo {title} {{Magnetic Field
  and Flavor Effects on the Gamma-Ray Burst Neutrino Flux}},}\ }\href {\doibase
  10.1103/PhysRevD.83.067303} {\bibfield  {journal} {\bibinfo  {journal} {Phys.
  Rev. D}\ }\textbf {\bibinfo {volume} {83}},\ \bibinfo {pages} {067303}
  (\bibinfo {year} {2011})},\ \Eprint {http://arxiv.org/abs/1009.4010}
  {arXiv:1009.4010 [astro-ph.HE]} \BibitemShut {NoStop}%
\bibitem [{\citenamefont {Bustamante}\ \emph {et~al.}(2017)\citenamefont
  {Bustamante}, \citenamefont {Murase}, \citenamefont {Winter},\ and\
  \citenamefont {Heinze}}]{Bustamante:2016wpu}%
  \BibitemOpen
  \bibfield  {author} {\bibinfo {author} {\bibfnamefont {Mauricio}\
  \bibnamefont {Bustamante}}, \bibinfo {author} {\bibfnamefont {Kohta}\
  \bibnamefont {Murase}}, \bibinfo {author} {\bibfnamefont {Walter}\
  \bibnamefont {Winter}}, \ and\ \bibinfo {author} {\bibfnamefont {Jonas}\
  \bibnamefont {Heinze}},\ }\bibfield  {title} {\enquote {\bibinfo {title}
  {{Multi-messenger light curves from gamma-ray bursts in the internal shock
  model}},}\ }\href {\doibase 10.3847/1538-4357/837/1/33} {\bibfield  {journal}
  {\bibinfo  {journal} {Astrophys. J.}\ }\textbf {\bibinfo {volume} {837}},\
  \bibinfo {pages} {33} (\bibinfo {year} {2017})},\ \Eprint
  {http://arxiv.org/abs/1606.02325} {arXiv:1606.02325 [astro-ph.HE]}
  \BibitemShut {NoStop}%
\bibitem [{\citenamefont {Abbasi}\ \emph {et~al.}(2022)\citenamefont {Abbasi}
  \emph {et~al.}}]{IceCube:2021uhz}%
  \BibitemOpen
  \bibfield  {author} {\bibinfo {author} {\bibfnamefont {R.}~\bibnamefont
  {Abbasi}} \emph {et~al.} (\bibinfo {collaboration} {IceCube}),\ }\bibfield
  {title} {\enquote {\bibinfo {title} {{Improved Characterization of the
  Astrophysical Muon\textendash{}neutrino Flux with 9.5 Years of IceCube
  Data}},}\ }\href {\doibase 10.3847/1538-4357/ac4d29} {\bibfield  {journal}
  {\bibinfo  {journal} {Astrophys. J.}\ }\textbf {\bibinfo {volume} {928}},\
  \bibinfo {pages} {50} (\bibinfo {year} {2022})},\ \Eprint
  {http://arxiv.org/abs/2111.10299} {arXiv:2111.10299 [astro-ph.HE]}
  \BibitemShut {NoStop}%
\bibitem [{\citenamefont {Aartsen}\ \emph {et~al.}(2020)\citenamefont {Aartsen}
  \emph {et~al.}}]{Aartsen:2020aqd}%
  \BibitemOpen
  \bibfield  {author} {\bibinfo {author} {\bibfnamefont {M.~G.}\ \bibnamefont
  {Aartsen}} \emph {et~al.} (\bibinfo {collaboration} {IceCube}),\ }\bibfield
  {title} {\enquote {\bibinfo {title} {{Characteristics of the diffuse
  astrophysical electron and tau neutrino flux with six years of IceCube high
  energy cascade data}},}\ }\href {\doibase 10.1103/PhysRevLett.125.121104}
  {\bibfield  {journal} {\bibinfo  {journal} {Phys. Rev. Lett.}\ }\textbf
  {\bibinfo {volume} {125}},\ \bibinfo {pages} {121104} (\bibinfo {year}
  {2020})},\ \Eprint {http://arxiv.org/abs/2001.09520} {arXiv:2001.09520
  [astro-ph.HE]} \BibitemShut {NoStop}%
\bibitem [{\citenamefont {Abbasi}\ \emph {et~al.}(2020)\citenamefont {Abbasi}
  \emph {et~al.}}]{Abbasi:2020jmh}%
  \BibitemOpen
  \bibfield  {author} {\bibinfo {author} {\bibfnamefont {R.}~\bibnamefont
  {Abbasi}} \emph {et~al.} (\bibinfo {collaboration} {IceCube}),\ }\bibfield
  {title} {\enquote {\bibinfo {title} {{The IceCube high-energy starting event
  sample: Description and flux characterization with 7.5 years of data}},}\
  }\href@noop {} {\  (\bibinfo {year} {2020})},\ \Eprint
  {http://arxiv.org/abs/2011.03545} {arXiv:2011.03545 [astro-ph.HE]}
  \BibitemShut {NoStop}%
\bibitem [{\citenamefont {Yuksel}\ \emph {et~al.}(2008)\citenamefont {Yuksel},
  \citenamefont {Kistler}, \citenamefont {Beacom},\ and\ \citenamefont
  {Hopkins}}]{Yuksel:2008cu}%
  \BibitemOpen
  \bibfield  {author} {\bibinfo {author} {\bibfnamefont {Hasan}\ \bibnamefont
  {Yuksel}}, \bibinfo {author} {\bibfnamefont {Matthew~D.}\ \bibnamefont
  {Kistler}}, \bibinfo {author} {\bibfnamefont {John~F.}\ \bibnamefont
  {Beacom}}, \ and\ \bibinfo {author} {\bibfnamefont {Andrew~M.}\ \bibnamefont
  {Hopkins}},\ }\bibfield  {title} {\enquote {\bibinfo {title} {{Revealing the
  High-Redshift Star Formation Rate with Gamma-Ray Bursts}},}\ }\href {\doibase
  10.1086/591449} {\bibfield  {journal} {\bibinfo  {journal} {Astrophys. J.
  Lett.}\ }\textbf {\bibinfo {volume} {683}},\ \bibinfo {pages} {L5--L8}
  (\bibinfo {year} {2008})},\ \Eprint {http://arxiv.org/abs/0804.4008}
  {arXiv:0804.4008 [astro-ph]} \BibitemShut {NoStop}%
\bibitem [{\citenamefont {Horiuchi}\ \emph {et~al.}(2009)\citenamefont
  {Horiuchi}, \citenamefont {Beacom},\ and\ \citenamefont
  {Dwek}}]{Horiuchi:2008jz}%
  \BibitemOpen
  \bibfield  {author} {\bibinfo {author} {\bibfnamefont {Shunsaku}\
  \bibnamefont {Horiuchi}}, \bibinfo {author} {\bibfnamefont {John~F.}\
  \bibnamefont {Beacom}}, \ and\ \bibinfo {author} {\bibfnamefont {Eli}\
  \bibnamefont {Dwek}},\ }\bibfield  {title} {\enquote {\bibinfo {title} {{The
  Diffuse Supernova Neutrino Background is detectable in Super-Kamiokande}},}\
  }\href {\doibase 10.1103/PhysRevD.79.083013} {\bibfield  {journal} {\bibinfo
  {journal} {Phys. Rev. D}\ }\textbf {\bibinfo {volume} {79}},\ \bibinfo
  {pages} {083013} (\bibinfo {year} {2009})},\ \Eprint
  {http://arxiv.org/abs/0812.3157} {arXiv:0812.3157 [astro-ph]} \BibitemShut
  {NoStop}%
\bibitem [{\citenamefont {Halzen}\ \emph {et~al.}(2019)\citenamefont {Halzen},
  \citenamefont {Kheirandish}, \citenamefont {Weisgarber},\ and\ \citenamefont
  {Wakely}}]{Halzen:2018iak}%
  \BibitemOpen
  \bibfield  {author} {\bibinfo {author} {\bibfnamefont {Francis}\ \bibnamefont
  {Halzen}}, \bibinfo {author} {\bibfnamefont {Ali}\ \bibnamefont
  {Kheirandish}}, \bibinfo {author} {\bibfnamefont {Thomas}\ \bibnamefont
  {Weisgarber}}, \ and\ \bibinfo {author} {\bibfnamefont {Scott~P.}\
  \bibnamefont {Wakely}},\ }\bibfield  {title} {\enquote {\bibinfo {title} {{On
  the Neutrino Flares from the Direction of TXS 0506+056}},}\ }\href {\doibase
  10.3847/2041-8213/ab0d27} {\bibfield  {journal} {\bibinfo  {journal}
  {Astrophys. J. Lett.}\ }\textbf {\bibinfo {volume} {874}},\ \bibinfo {pages}
  {L9} (\bibinfo {year} {2019})},\ \Eprint {http://arxiv.org/abs/1811.07439}
  {arXiv:1811.07439 [astro-ph.HE]} \BibitemShut {NoStop}%
\bibitem [{\citenamefont {Kartavtsev}\ \emph {et~al.}(2017)\citenamefont
  {Kartavtsev}, \citenamefont {Raffelt},\ and\ \citenamefont
  {Vogel}}]{Kartavtsev:2016doq}%
  \BibitemOpen
  \bibfield  {author} {\bibinfo {author} {\bibfnamefont {A.}~\bibnamefont
  {Kartavtsev}}, \bibinfo {author} {\bibfnamefont {G.}~\bibnamefont {Raffelt}},
  \ and\ \bibinfo {author} {\bibfnamefont {H.}~\bibnamefont {Vogel}},\
  }\bibfield  {title} {\enquote {\bibinfo {title} {{Extragalactic photon-ALP
  conversion at CTA energies}},}\ }\href {\doibase
  10.1088/1475-7516/2017/01/024} {\bibfield  {journal} {\bibinfo  {journal}
  {JCAP}\ }\textbf {\bibinfo {volume} {01}},\ \bibinfo {pages} {024} (\bibinfo
  {year} {2017})},\ \Eprint {http://arxiv.org/abs/1611.04526} {arXiv:1611.04526
  [astro-ph.HE]} \BibitemShut {NoStop}%
\bibitem [{\citenamefont {Mirizzi}\ \emph {et~al.}(2005)\citenamefont
  {Mirizzi}, \citenamefont {Raffelt},\ and\ \citenamefont
  {Serpico}}]{Mirizzi:2005ng}%
  \BibitemOpen
  \bibfield  {author} {\bibinfo {author} {\bibfnamefont {Alessandro}\
  \bibnamefont {Mirizzi}}, \bibinfo {author} {\bibfnamefont {Georg~G.}\
  \bibnamefont {Raffelt}}, \ and\ \bibinfo {author} {\bibfnamefont
  {Pasquale~D.}\ \bibnamefont {Serpico}},\ }\bibfield  {title} {\enquote
  {\bibinfo {title} {{Photon-axion conversion as a mechanism for supernova
  dimming: Limits from CMB spectral distortion}},}\ }\href {\doibase
  10.1103/PhysRevD.72.023501} {\bibfield  {journal} {\bibinfo  {journal} {Phys.
  Rev. D}\ }\textbf {\bibinfo {volume} {72}},\ \bibinfo {pages} {023501}
  (\bibinfo {year} {2005})},\ \Eprint {http://arxiv.org/abs/astro-ph/0506078}
  {arXiv:astro-ph/0506078} \BibitemShut {NoStop}%
\bibitem [{\citenamefont {Mirizzi}\ \emph {et~al.}(2008)\citenamefont
  {Mirizzi}, \citenamefont {Raffelt},\ and\ \citenamefont
  {Serpico}}]{Mirizzi:2006zy}%
  \BibitemOpen
  \bibfield  {author} {\bibinfo {author} {\bibfnamefont {Alessandro}\
  \bibnamefont {Mirizzi}}, \bibinfo {author} {\bibfnamefont {Georg~G.}\
  \bibnamefont {Raffelt}}, \ and\ \bibinfo {author} {\bibfnamefont
  {Pasquale~D.}\ \bibnamefont {Serpico}},\ }\bibfield  {title} {\enquote
  {\bibinfo {title} {{Photon-axion conversion in intergalactic magnetic fields
  and cosmological consequences}},}\ }\href {\doibase
  10.1007/978-3-540-73518-2_7} {\bibfield  {journal} {\bibinfo  {journal}
  {Lect. Notes Phys.}\ }\textbf {\bibinfo {volume} {741}},\ \bibinfo {pages}
  {115--134} (\bibinfo {year} {2008})},\ \Eprint
  {http://arxiv.org/abs/astro-ph/0607415} {arXiv:astro-ph/0607415} \BibitemShut
  {NoStop}%
\bibitem [{\citenamefont {Dobrynina}\ \emph {et~al.}(2015)\citenamefont
  {Dobrynina}, \citenamefont {Kartavtsev},\ and\ \citenamefont
  {Raffelt}}]{Dobrynina:2014qba}%
  \BibitemOpen
  \bibfield  {author} {\bibinfo {author} {\bibfnamefont {Alexandra}\
  \bibnamefont {Dobrynina}}, \bibinfo {author} {\bibfnamefont {Alexander}\
  \bibnamefont {Kartavtsev}}, \ and\ \bibinfo {author} {\bibfnamefont {Georg}\
  \bibnamefont {Raffelt}},\ }\bibfield  {title} {\enquote {\bibinfo {title}
  {{Photon-photon dispersion of TeV gamma rays and its role for photon-ALP
  conversion}},}\ }\href {\doibase 10.1103/PhysRevD.91.083003} {\bibfield
  {journal} {\bibinfo  {journal} {Phys. Rev. D}\ }\textbf {\bibinfo {volume}
  {91}},\ \bibinfo {pages} {083003} (\bibinfo {year} {2015})},\ \bibinfo {note}
  {[Erratum: Phys.Rev.D 95, 109905 (2017)]},\ \Eprint
  {http://arxiv.org/abs/1412.4777} {arXiv:1412.4777 [astro-ph.HE]} \BibitemShut
  {NoStop}%
\bibitem [{\citenamefont {Gould}\ and\ \citenamefont
  {Schreder}(1967)}]{Gould:1967zzb}%
  \BibitemOpen
  \bibfield  {author} {\bibinfo {author} {\bibfnamefont {Robert~J.}\
  \bibnamefont {Gould}}\ and\ \bibinfo {author} {\bibfnamefont {Gerard~P.}\
  \bibnamefont {Schreder}},\ }\bibfield  {title} {\enquote {\bibinfo {title}
  {{Pair Production in Photon-Photon Collisions}},}\ }\href {\doibase
  10.1103/PhysRev.155.1404} {\bibfield  {journal} {\bibinfo  {journal} {Phys.
  Rev.}\ }\textbf {\bibinfo {volume} {155}},\ \bibinfo {pages} {1404--1407}
  (\bibinfo {year} {1967})}\BibitemShut {NoStop}%
\bibitem [{\citenamefont {Heitler}(1936)}]{Heitler:1936jqw}%
  \BibitemOpen
  \bibfield  {author} {\bibinfo {author} {\bibfnamefont {W.}~\bibnamefont
  {Heitler}},\ }\href@noop {} {\emph {\bibinfo {title} {{The quantum theory of
  radiation}}}},\ \bibinfo {series} {International Series of Monographs on
  Physics}, Vol.~\bibinfo {volume} {5}\ (\bibinfo  {publisher} {Oxford
  University Press},\ \bibinfo {address} {Oxford},\ \bibinfo {year}
  {1936})\BibitemShut {NoStop}%
\bibitem [{\citenamefont {Porter}\ \emph {et~al.}(2017)\citenamefont {Porter},
  \citenamefont {Johannesson},\ and\ \citenamefont
  {Moskalenko}}]{Porter:2017vaa}%
  \BibitemOpen
  \bibfield  {author} {\bibinfo {author} {\bibfnamefont {Troy~A.}\ \bibnamefont
  {Porter}}, \bibinfo {author} {\bibfnamefont {Gudlaugur}\ \bibnamefont
  {Johannesson}}, \ and\ \bibinfo {author} {\bibfnamefont {Igor~V.}\
  \bibnamefont {Moskalenko}},\ }\bibfield  {title} {\enquote {\bibinfo {title}
  {{High-Energy Gamma Rays from the Milky Way: Three-Dimensional Spatial Models
  for the Cosmic-Ray and Radiation Field Densities in the Interstellar
  Medium}},}\ }\href {\doibase 10.3847/1538-4357/aa844d} {\bibfield  {journal}
  {\bibinfo  {journal} {Astrophys. J.}\ }\textbf {\bibinfo {volume} {846}},\
  \bibinfo {pages} {67} (\bibinfo {year} {2017})},\ \Eprint
  {http://arxiv.org/abs/1708.00816} {arXiv:1708.00816 [astro-ph.HE]}
  \BibitemShut {NoStop}%
\bibitem [{\citenamefont {Moss}\ and\ \citenamefont
  {Shukurov}(1996)}]{moss1996turbulence}%
  \BibitemOpen
  \bibfield  {author} {\bibinfo {author} {\bibfnamefont {David}\ \bibnamefont
  {Moss}}\ and\ \bibinfo {author} {\bibfnamefont {Anvar}\ \bibnamefont
  {Shukurov}},\ }\bibfield  {title} {\enquote {\bibinfo {title} {Turbulence and
  magnetic fields in elliptical galaxies},}\ }\href@noop {} {\bibfield
  {journal} {\bibinfo  {journal} {Monthly Notices of the Royal Astronomical
  Society}\ }\textbf {\bibinfo {volume} {279}},\ \bibinfo {pages} {229--239}
  (\bibinfo {year} {1996})}\BibitemShut {NoStop}%
\bibitem [{\citenamefont {Tavecchio}\ \emph {et~al.}(2012)\citenamefont
  {Tavecchio}, \citenamefont {Roncadelli}, \citenamefont {Galanti},\ and\
  \citenamefont {Bonnoli}}]{Tavecchio:2012um}%
  \BibitemOpen
  \bibfield  {author} {\bibinfo {author} {\bibfnamefont {Fabrizio}\
  \bibnamefont {Tavecchio}}, \bibinfo {author} {\bibfnamefont {Marco}\
  \bibnamefont {Roncadelli}}, \bibinfo {author} {\bibfnamefont {Giorgio}\
  \bibnamefont {Galanti}}, \ and\ \bibinfo {author} {\bibfnamefont {Giacomo}\
  \bibnamefont {Bonnoli}},\ }\bibfield  {title} {\enquote {\bibinfo {title}
  {{Evidence for an axion-like particle from PKS 1222+216?}}}\ }\href {\doibase
  10.1103/PhysRevD.86.085036} {\bibfield  {journal} {\bibinfo  {journal} {Phys.
  Rev. D}\ }\textbf {\bibinfo {volume} {86}},\ \bibinfo {pages} {085036}
  (\bibinfo {year} {2012})},\ \Eprint {http://arxiv.org/abs/1202.6529}
  {arXiv:1202.6529 [astro-ph.HE]} \BibitemShut {NoStop}%
\bibitem [{\citenamefont {Fletcher}(2011)}]{Fletcher:2011fn}%
  \BibitemOpen
  \bibfield  {author} {\bibinfo {author} {\bibfnamefont {Andrew}\ \bibnamefont
  {Fletcher}},\ }\bibfield  {title} {\enquote {\bibinfo {title} {{Magnetic
  fields in nearby galaxies}},}\ }\href@noop {} {\bibfield  {journal} {\bibinfo
   {journal} {ASP Conf. Ser.}\ }\textbf {\bibinfo {volume} {438}},\ \bibinfo
  {pages} {197--210} (\bibinfo {year} {2011})},\ \Eprint
  {http://arxiv.org/abs/1104.2427} {arXiv:1104.2427 [astro-ph.CO]} \BibitemShut
  {NoStop}%
\bibitem [{\citenamefont {Beck}\ and\ \citenamefont
  {Wielebinski}(2013)}]{Beck:2013bxa}%
  \BibitemOpen
  \bibfield  {author} {\bibinfo {author} {\bibfnamefont {Rainer}\ \bibnamefont
  {Beck}}\ and\ \bibinfo {author} {\bibfnamefont {Richard}\ \bibnamefont
  {Wielebinski}},\ }\bibfield  {title} {\enquote {\bibinfo {title} {{Magnetic
  Fields in the Milky Way and in Galaxies}},}\ }\href {\doibase
  10.1007/978-94-007-5612-0\_13} {\  (\bibinfo {year} {2013}),\
  10.1007/978-94-007-5612-0\_13},\ \Eprint {http://arxiv.org/abs/1302.5663}
  {arXiv:1302.5663 [astro-ph.GA]} \BibitemShut {NoStop}%
\bibitem [{\citenamefont {Pakmor}\ \emph {et~al.}(2017)\citenamefont {Pakmor},
  \citenamefont {G{\'{o} }mez}, \citenamefont {Grand}, \citenamefont
  {Marinacci}, \citenamefont {Simpson}, \citenamefont {Springel}, \citenamefont
  {Campbell}, \citenamefont {Frenk}, \citenamefont {Guillet}, \citenamefont
  {Pfrommer},\ and\ \citenamefont {White}}]{Pakmor_2017}%
  \BibitemOpen
  \bibfield  {author} {\bibinfo {author} {\bibfnamefont {Rüdiger}\
  \bibnamefont {Pakmor}}, \bibinfo {author} {\bibfnamefont {Facundo~A.}\
  \bibnamefont {G{\'{o} }mez}}, \bibinfo {author} {\bibfnamefont {Robert
  J.~J.}\ \bibnamefont {Grand}}, \bibinfo {author} {\bibfnamefont {Federico}\
  \bibnamefont {Marinacci}}, \bibinfo {author} {\bibfnamefont {Christine~M.}\
  \bibnamefont {Simpson}}, \bibinfo {author} {\bibfnamefont {Volker}\
  \bibnamefont {Springel}}, \bibinfo {author} {\bibfnamefont {David J.~R.}\
  \bibnamefont {Campbell}}, \bibinfo {author} {\bibfnamefont {Carlos~S.}\
  \bibnamefont {Frenk}}, \bibinfo {author} {\bibfnamefont {Thomas}\
  \bibnamefont {Guillet}}, \bibinfo {author} {\bibfnamefont {Christoph}\
  \bibnamefont {Pfrommer}}, \ and\ \bibinfo {author} {\bibfnamefont {Simon
  D.~M.}\ \bibnamefont {White}},\ }\bibfield  {title} {\enquote {\bibinfo
  {title} {Magnetic field formation in the milky way like disc galaxies of the
  auriga project},}\ }\href {\doibase 10.1093/mnras/stx1074} {\bibfield
  {journal} {\bibinfo  {journal} {Monthly Notices of the Royal Astronomical
  Society}\ }\textbf {\bibinfo {volume} {469}},\ \bibinfo {pages} {3185--3199}
  (\bibinfo {year} {2017})}\BibitemShut {NoStop}%
\bibitem [{\citenamefont {Bassan}\ \emph {et~al.}(2010)\citenamefont {Bassan},
  \citenamefont {Mirizzi},\ and\ \citenamefont {Roncadelli}}]{Bassan:2010ya}%
  \BibitemOpen
  \bibfield  {author} {\bibinfo {author} {\bibfnamefont {Nicola}\ \bibnamefont
  {Bassan}}, \bibinfo {author} {\bibfnamefont {Alessandro}\ \bibnamefont
  {Mirizzi}}, \ and\ \bibinfo {author} {\bibfnamefont {Marco}\ \bibnamefont
  {Roncadelli}},\ }\bibfield  {title} {\enquote {\bibinfo {title} {{Axion-like
  particle effects on the polarization of cosmic high-energy gamma sources}},}\
  }\href {\doibase 10.1088/1475-7516/2010/05/010} {\bibfield  {journal}
  {\bibinfo  {journal} {JCAP}\ }\textbf {\bibinfo {volume} {05}},\ \bibinfo
  {pages} {010} (\bibinfo {year} {2010})},\ \Eprint
  {http://arxiv.org/abs/1001.5267} {arXiv:1001.5267 [astro-ph.HE]} \BibitemShut
  {NoStop}%
\bibitem [{\citenamefont {Krause}(2019)}]{galaxies7020054}%
  \BibitemOpen
  \bibfield  {author} {\bibinfo {author} {\bibfnamefont {Marita}\ \bibnamefont
  {Krause}},\ }\bibfield  {title} {\enquote {\bibinfo {title} {Magnetic fields
  and halos in spiral galaxies},}\ }\href {\doibase 10.3390/galaxies7020054}
  {\bibfield  {journal} {\bibinfo  {journal} {Galaxies}\ }\textbf {\bibinfo
  {volume} {7}} (\bibinfo {year} {2019}),\ 10.3390/galaxies7020054}\BibitemShut
  {NoStop}%
\bibitem [{\citenamefont {Rodrigues}\ \emph {et~al.}(2018)\citenamefont
  {Rodrigues}, \citenamefont {Chamandy}, \citenamefont {Shukurov},
  \citenamefont {Baugh},\ and\ \citenamefont {Taylor}}]{10.1093/mnras/sty3270}%
  \BibitemOpen
  \bibfield  {author} {\bibinfo {author} {\bibfnamefont {L~F~S}\ \bibnamefont
  {Rodrigues}}, \bibinfo {author} {\bibfnamefont {L}~\bibnamefont {Chamandy}},
  \bibinfo {author} {\bibfnamefont {A}~\bibnamefont {Shukurov}}, \bibinfo
  {author} {\bibfnamefont {C~M}\ \bibnamefont {Baugh}}, \ and\ \bibinfo
  {author} {\bibfnamefont {A~R}\ \bibnamefont {Taylor}},\ }\bibfield  {title}
  {\enquote {\bibinfo {title} {{Evolution of galactic magnetic fields}},}\
  }\href {\doibase 10.1093/mnras/sty3270} {\bibfield  {journal} {\bibinfo
  {journal} {Monthly Notices of the Royal Astronomical Society}\ }\textbf
  {\bibinfo {volume} {483}},\ \bibinfo {pages} {2424--2440} (\bibinfo {year}
  {2018})},\ \Eprint
  {http://arxiv.org/abs/https://academic.oup.com/mnras/article-pdf/483/2/2424/27184716/sty3270.pdf}
  {https://academic.oup.com/mnras/article-pdf/483/2/2424/27184716/sty3270.pdf}
  \BibitemShut {NoStop}%
\bibitem [{\citenamefont {Adams}\ and\ \citenamefont
  {Laughlin}(1997)}]{Adams:1996xe}%
  \BibitemOpen
  \bibfield  {author} {\bibinfo {author} {\bibfnamefont {Fred~C.}\ \bibnamefont
  {Adams}}\ and\ \bibinfo {author} {\bibfnamefont {Gregory}\ \bibnamefont
  {Laughlin}},\ }\bibfield  {title} {\enquote {\bibinfo {title} {{A Dying
  universe: The Long term fate and evolution of astrophysical objects}},}\
  }\href {\doibase 10.1103/RevModPhys.69.337} {\bibfield  {journal} {\bibinfo
  {journal} {Rev. Mod. Phys.}\ }\textbf {\bibinfo {volume} {69}},\ \bibinfo
  {pages} {337--372} (\bibinfo {year} {1997})},\ \Eprint
  {http://arxiv.org/abs/astro-ph/9701131} {arXiv:astro-ph/9701131} \BibitemShut
  {NoStop}%
\bibitem [{\citenamefont {Jansson}\ and\ \citenamefont
  {Farrar}(2012)}]{Jansson:2012pc}%
  \BibitemOpen
  \bibfield  {author} {\bibinfo {author} {\bibfnamefont {Ronnie}\ \bibnamefont
  {Jansson}}\ and\ \bibinfo {author} {\bibfnamefont {Glennys~R.}\ \bibnamefont
  {Farrar}},\ }\bibfield  {title} {\enquote {\bibinfo {title} {{A New Model of
  the Galactic Magnetic Field}},}\ }\href {\doibase 10.1088/0004-637X/757/1/14}
  {\bibfield  {journal} {\bibinfo  {journal} {Astrophys. J.}\ }\textbf
  {\bibinfo {volume} {757}},\ \bibinfo {pages} {14} (\bibinfo {year} {2012})},\
  \Eprint {http://arxiv.org/abs/1204.3662} {arXiv:1204.3662 [astro-ph.GA]}
  \BibitemShut {NoStop}%
\bibitem [{\citenamefont {Adam}\ \emph {et~al.}(2016)\citenamefont {Adam} \emph
  {et~al.}}]{Planck:2016gdp}%
  \BibitemOpen
  \bibfield  {author} {\bibinfo {author} {\bibfnamefont {R.}~\bibnamefont
  {Adam}} \emph {et~al.} (\bibinfo {collaboration} {Planck}),\ }\bibfield
  {title} {\enquote {\bibinfo {title} {{Planck intermediate results.}: {XLII.
  Large-scale Galactic magnetic fields}},}\ }\href {\doibase
  10.1051/0004-6361/201528033} {\bibfield  {journal} {\bibinfo  {journal}
  {Astron. Astrophys.}\ }\textbf {\bibinfo {volume} {596}},\ \bibinfo {pages}
  {A103} (\bibinfo {year} {2016})},\ \Eprint {http://arxiv.org/abs/1601.00546}
  {arXiv:1601.00546 [astro-ph.GA]} \BibitemShut {NoStop}%
\bibitem [{\citenamefont {Cordes}\ and\ \citenamefont
  {Lazio}(2002)}]{Cordes:2002wz}%
  \BibitemOpen
  \bibfield  {author} {\bibinfo {author} {\bibfnamefont {James~M.}\
  \bibnamefont {Cordes}}\ and\ \bibinfo {author} {\bibfnamefont {T.~J.~W.}\
  \bibnamefont {Lazio}},\ }\bibfield  {title} {\enquote {\bibinfo {title}
  {{NE2001. 1. A New model for the galactic distribution of free electrons and
  its fluctuations}},}\ }\href@noop {} {\  (\bibinfo {year} {2002})},\ \Eprint
  {http://arxiv.org/abs/astro-ph/0207156} {arXiv:astro-ph/0207156} \BibitemShut
  {NoStop}%
\bibitem [{\citenamefont {Calore}\ \emph {et~al.}(2020)\citenamefont {Calore},
  \citenamefont {Carenza}, \citenamefont {Giannotti}, \citenamefont {Jaeckel},\
  and\ \citenamefont {Mirizzi}}]{Calore:2020tjw}%
  \BibitemOpen
  \bibfield  {author} {\bibinfo {author} {\bibfnamefont {Francesca}\
  \bibnamefont {Calore}}, \bibinfo {author} {\bibfnamefont {Pierluca}\
  \bibnamefont {Carenza}}, \bibinfo {author} {\bibfnamefont {Maurizio}\
  \bibnamefont {Giannotti}}, \bibinfo {author} {\bibfnamefont {Joerg}\
  \bibnamefont {Jaeckel}}, \ and\ \bibinfo {author} {\bibfnamefont
  {Alessandro}\ \bibnamefont {Mirizzi}},\ }\bibfield  {title} {\enquote
  {\bibinfo {title} {{Bounds on axionlike particles from the diffuse supernova
  flux}},}\ }\href {\doibase 10.1103/PhysRevD.102.123005} {\bibfield  {journal}
  {\bibinfo  {journal} {Phys. Rev. D}\ }\textbf {\bibinfo {volume} {102}},\
  \bibinfo {pages} {123005} (\bibinfo {year} {2020})},\ \Eprint
  {http://arxiv.org/abs/2008.11741} {arXiv:2008.11741 [hep-ph]} \BibitemShut
  {NoStop}%
\bibitem [{\citenamefont {Anastassopoulos}\ \emph {et~al.}(2017)\citenamefont
  {Anastassopoulos} \emph {et~al.}}]{CAST:2017uph}%
  \BibitemOpen
  \bibfield  {author} {\bibinfo {author} {\bibfnamefont {V.}~\bibnamefont
  {Anastassopoulos}} \emph {et~al.} (\bibinfo {collaboration} {CAST}),\
  }\bibfield  {title} {\enquote {\bibinfo {title} {{New CAST Limit on the
  Axion-Photon Interaction}},}\ }\href {\doibase 10.1038/nphys4109} {\bibfield
  {journal} {\bibinfo  {journal} {Nature Phys.}\ }\textbf {\bibinfo {volume}
  {13}},\ \bibinfo {pages} {584--590} (\bibinfo {year} {2017})},\ \Eprint
  {http://arxiv.org/abs/1705.02290} {arXiv:1705.02290 [hep-ex]} \BibitemShut
  {NoStop}%
\bibitem [{\citenamefont {Ayala}\ \emph {et~al.}(2014)\citenamefont {Ayala},
  \citenamefont {Dom\'\i{}nguez}, \citenamefont {Giannotti}, \citenamefont
  {Mirizzi},\ and\ \citenamefont {Straniero}}]{Ayala:2014pea}%
  \BibitemOpen
  \bibfield  {author} {\bibinfo {author} {\bibfnamefont {Adrian}\ \bibnamefont
  {Ayala}}, \bibinfo {author} {\bibfnamefont {Inma}\ \bibnamefont
  {Dom\'\i{}nguez}}, \bibinfo {author} {\bibfnamefont {Maurizio}\ \bibnamefont
  {Giannotti}}, \bibinfo {author} {\bibfnamefont {Alessandro}\ \bibnamefont
  {Mirizzi}}, \ and\ \bibinfo {author} {\bibfnamefont {Oscar}\ \bibnamefont
  {Straniero}},\ }\bibfield  {title} {\enquote {\bibinfo {title} {{Revisiting
  the bound on axion-photon coupling from Globular Clusters}},}\ }\href
  {\doibase 10.1103/PhysRevLett.113.191302} {\bibfield  {journal} {\bibinfo
  {journal} {Phys. Rev. Lett.}\ }\textbf {\bibinfo {volume} {113}},\ \bibinfo
  {pages} {191302} (\bibinfo {year} {2014})},\ \Eprint
  {http://arxiv.org/abs/1406.6053} {arXiv:1406.6053 [astro-ph.SR]} \BibitemShut
  {NoStop}%
\bibitem [{\citenamefont {Payez}\ \emph {et~al.}(2015)\citenamefont {Payez},
  \citenamefont {Evoli}, \citenamefont {Fischer}, \citenamefont {Giannotti},
  \citenamefont {Mirizzi},\ and\ \citenamefont {Ringwald}}]{Payez:2014xsa}%
  \BibitemOpen
  \bibfield  {author} {\bibinfo {author} {\bibfnamefont {Alexandre}\
  \bibnamefont {Payez}}, \bibinfo {author} {\bibfnamefont {Carmelo}\
  \bibnamefont {Evoli}}, \bibinfo {author} {\bibfnamefont {Tobias}\
  \bibnamefont {Fischer}}, \bibinfo {author} {\bibfnamefont {Maurizio}\
  \bibnamefont {Giannotti}}, \bibinfo {author} {\bibfnamefont {Alessandro}\
  \bibnamefont {Mirizzi}}, \ and\ \bibinfo {author} {\bibfnamefont {Andreas}\
  \bibnamefont {Ringwald}},\ }\bibfield  {title} {\enquote {\bibinfo {title}
  {{Revisiting the SN1987A gamma-ray limit on ultralight axion-like
  particles}},}\ }\href {\doibase 10.1088/1475-7516/2015/02/006} {\bibfield
  {journal} {\bibinfo  {journal} {JCAP}\ }\textbf {\bibinfo {volume} {02}},\
  \bibinfo {pages} {006} (\bibinfo {year} {2015})},\ \Eprint
  {http://arxiv.org/abs/1410.3747} {arXiv:1410.3747 [astro-ph.HE]} \BibitemShut
  {NoStop}%
\bibitem [{\citenamefont {Li}\ \emph {et~al.}(2021)\citenamefont {Li},
  \citenamefont {Guo}, \citenamefont {Bi}, \citenamefont {Lin},\ and\
  \citenamefont {Yin}}]{Li:2020pcn}%
  \BibitemOpen
  \bibfield  {author} {\bibinfo {author} {\bibfnamefont {Hai-Jun}\ \bibnamefont
  {Li}}, \bibinfo {author} {\bibfnamefont {Jun-Guang}\ \bibnamefont {Guo}},
  \bibinfo {author} {\bibfnamefont {Xiao-Jun}\ \bibnamefont {Bi}}, \bibinfo
  {author} {\bibfnamefont {Su-Jie}\ \bibnamefont {Lin}}, \ and\ \bibinfo
  {author} {\bibfnamefont {Peng-Fei}\ \bibnamefont {Yin}},\ }\bibfield  {title}
  {\enquote {\bibinfo {title} {{Limits on axion-like particles from Mrk 421
  with 4.5-year period observations by ARGO-YBJ and Fermi-LAT}},}\ }\href
  {\doibase 10.1103/PhysRevD.103.083003} {\bibfield  {journal} {\bibinfo
  {journal} {Phys. Rev. D}\ }\textbf {\bibinfo {volume} {103}},\ \bibinfo
  {pages} {083003} (\bibinfo {year} {2021})},\ \Eprint
  {http://arxiv.org/abs/2008.09464} {arXiv:2008.09464 [astro-ph.HE]}
  \BibitemShut {NoStop}%
\bibitem [{\citenamefont {Pallathadka}\ \emph {et~al.}(2020)\citenamefont
  {Pallathadka}, \citenamefont {Calore}, \citenamefont {Carenza}, \citenamefont
  {Giannotti}, \citenamefont {Horns}, \citenamefont {Majumdar}, \citenamefont
  {Mirizzi}, \citenamefont {Ringwald}, \citenamefont {Sokolov},\ and\
  \citenamefont {Stief}}]{Pallathadka:2020vwu}%
  \BibitemOpen
  \bibfield  {author} {\bibinfo {author} {\bibfnamefont {Gautham~Adamane}\
  \bibnamefont {Pallathadka}}, \bibinfo {author} {\bibfnamefont {Francesca}\
  \bibnamefont {Calore}}, \bibinfo {author} {\bibfnamefont {Pierluca}\
  \bibnamefont {Carenza}}, \bibinfo {author} {\bibfnamefont {Maurizio}\
  \bibnamefont {Giannotti}}, \bibinfo {author} {\bibfnamefont {Dieter}\
  \bibnamefont {Horns}}, \bibinfo {author} {\bibfnamefont {Jhilik}\
  \bibnamefont {Majumdar}}, \bibinfo {author} {\bibfnamefont {Alessandro}\
  \bibnamefont {Mirizzi}}, \bibinfo {author} {\bibfnamefont {Andreas}\
  \bibnamefont {Ringwald}}, \bibinfo {author} {\bibfnamefont {Anton}\
  \bibnamefont {Sokolov}}, \ and\ \bibinfo {author} {\bibfnamefont {Franziska}\
  \bibnamefont {Stief}},\ }\bibfield  {title} {\enquote {\bibinfo {title}
  {{Reconciling hints on axion-like-particles from high-energy gamma rays with
  stellar bounds}},}\ }\href@noop {} {\  (\bibinfo {year} {2020})},\ \Eprint
  {http://arxiv.org/abs/2008.08100} {arXiv:2008.08100 [hep-ph]} \BibitemShut
  {NoStop}%
\bibitem [{\citenamefont {Ajello}\ \emph
  {et~al.}(2016{\natexlab{b}})\citenamefont {Ajello} \emph
  {et~al.}}]{TheFermi-LAT:2016zue}%
  \BibitemOpen
  \bibfield  {author} {\bibinfo {author} {\bibfnamefont {M.}~\bibnamefont
  {Ajello}} \emph {et~al.} (\bibinfo {collaboration} {Fermi-LAT}),\ }\bibfield
  {title} {\enquote {\bibinfo {title} {{Search for Spectral Irregularities due
  to Photon\textendash{}Axionlike-Particle Oscillations with the Fermi Large
  Area Telescope}},}\ }\href {\doibase 10.1103/PhysRevLett.116.161101}
  {\bibfield  {journal} {\bibinfo  {journal} {Phys. Rev. Lett.}\ }\textbf
  {\bibinfo {volume} {116}},\ \bibinfo {pages} {161101} (\bibinfo {year}
  {2016}{\natexlab{b}})},\ \Eprint {http://arxiv.org/abs/1603.06978}
  {arXiv:1603.06978 [astro-ph.HE]} \BibitemShut {NoStop}%
\bibitem [{\citenamefont {Abramowski}\ \emph
  {et~al.}(2013{\natexlab{b}})\citenamefont {Abramowski} \emph
  {et~al.}}]{Abramowski:2013oea}%
  \BibitemOpen
  \bibfield  {author} {\bibinfo {author} {\bibfnamefont {A.}~\bibnamefont
  {Abramowski}} \emph {et~al.} (\bibinfo {collaboration} {H.E.S.S.}),\
  }\bibfield  {title} {\enquote {\bibinfo {title} {{Constraints on axionlike
  particles with H.E.S.S. from the irregularity of the PKS 2155-304 energy
  spectrum}},}\ }\href {\doibase 10.1103/PhysRevD.88.102003} {\bibfield
  {journal} {\bibinfo  {journal} {Phys. Rev. D}\ }\textbf {\bibinfo {volume}
  {88}},\ \bibinfo {pages} {102003} (\bibinfo {year} {2013}{\natexlab{b}})},\
  \Eprint {http://arxiv.org/abs/1311.3148} {arXiv:1311.3148 [astro-ph.HE]}
  \BibitemShut {NoStop}%
\bibitem [{\citenamefont {Libanov}\ and\ \citenamefont
  {Troitsky}(2020)}]{Libanov:2019fzq}%
  \BibitemOpen
  \bibfield  {author} {\bibinfo {author} {\bibfnamefont {Maxim}\ \bibnamefont
  {Libanov}}\ and\ \bibinfo {author} {\bibfnamefont {Sergey}\ \bibnamefont
  {Troitsky}},\ }\bibfield  {title} {\enquote {\bibinfo {title} {{On the impact
  of magnetic-field models in galaxy clusters on constraints on axion-like
  particles from the lack of irregularities in high-energy spectra of
  astrophysical sources}},}\ }\href {\doibase 10.1016/j.physletb.2020.135252}
  {\bibfield  {journal} {\bibinfo  {journal} {Phys. Lett. B}\ }\textbf
  {\bibinfo {volume} {802}},\ \bibinfo {pages} {135252} (\bibinfo {year}
  {2020})},\ \Eprint {http://arxiv.org/abs/1908.03084} {arXiv:1908.03084
  [astro-ph.HE]} \BibitemShut {NoStop}%
\bibitem [{\citenamefont {Luque}\ \emph {et~al.}(2022)\citenamefont {Luque},
  \citenamefont {Gaggero}, \citenamefont {Grasso}, \citenamefont {Fornieri},
  \citenamefont {Egberts}, \citenamefont {Steppa},\ and\ \citenamefont
  {Evoli}}]{Luque:2022buq}%
  \BibitemOpen
  \bibfield  {author} {\bibinfo {author} {\bibfnamefont {Pedro De la~Torre}\
  \bibnamefont {Luque}}, \bibinfo {author} {\bibfnamefont {Daniele}\
  \bibnamefont {Gaggero}}, \bibinfo {author} {\bibfnamefont {Dario}\
  \bibnamefont {Grasso}}, \bibinfo {author} {\bibfnamefont {Ottavio}\
  \bibnamefont {Fornieri}}, \bibinfo {author} {\bibfnamefont {Kathrin}\
  \bibnamefont {Egberts}}, \bibinfo {author} {\bibfnamefont {Constantin}\
  \bibnamefont {Steppa}}, \ and\ \bibinfo {author} {\bibfnamefont {Carmelo}\
  \bibnamefont {Evoli}},\ }\bibfield  {title} {\enquote {\bibinfo {title}
  {{Galactic diffuse gamma rays meet the PeV frontier}},}\ }\href@noop {} {\
  (\bibinfo {year} {2022})},\ \Eprint {http://arxiv.org/abs/2203.15759}
  {arXiv:2203.15759 [astro-ph.HE]} \BibitemShut {NoStop}%
\bibitem [{\citenamefont {Abdalla}\ \emph {et~al.}(2021)\citenamefont {Abdalla}
  \emph {et~al.}}]{CTA:2020hii}%
  \BibitemOpen
  \bibfield  {author} {\bibinfo {author} {\bibfnamefont {H.}~\bibnamefont
  {Abdalla}} \emph {et~al.} (\bibinfo {collaboration} {CTA}),\ }\bibfield
  {title} {\enquote {\bibinfo {title} {{Sensitivity of the Cherenkov Telescope
  Array for probing cosmology and fundamental physics with gamma-ray
  propagation}},}\ }\href {\doibase 10.1088/1475-7516/2021/02/048} {\bibfield
  {journal} {\bibinfo  {journal} {JCAP}\ }\textbf {\bibinfo {volume} {02}},\
  \bibinfo {pages} {048} (\bibinfo {year} {2021})},\ \Eprint
  {http://arxiv.org/abs/2010.01349} {arXiv:2010.01349 [astro-ph.HE]}
  \BibitemShut {NoStop}%
\bibitem [{\citenamefont {Dessert}\ \emph {et~al.}(2022)\citenamefont
  {Dessert}, \citenamefont {Dunsky},\ and\ \citenamefont
  {Safdi}}]{Dessert:2022yqq}%
  \BibitemOpen
  \bibfield  {author} {\bibinfo {author} {\bibfnamefont {Christopher}\
  \bibnamefont {Dessert}}, \bibinfo {author} {\bibfnamefont {David}\
  \bibnamefont {Dunsky}}, \ and\ \bibinfo {author} {\bibfnamefont
  {Benjamin~R.}\ \bibnamefont {Safdi}},\ }\bibfield  {title} {\enquote
  {\bibinfo {title} {{Upper limit on the axion-photon coupling from magnetic
  white dwarf polarization}},}\ }\href@noop {} {\  (\bibinfo {year} {2022})},\
  \Eprint {http://arxiv.org/abs/2203.04319} {arXiv:2203.04319 [hep-ph]}
  \BibitemShut {NoStop}%
\bibitem [{\citenamefont {B{\"a}hre}\ \emph {et~al.}(2013)\citenamefont
  {B{\"a}hre}, \citenamefont {D{\"o}brich}, \citenamefont
  {Dreyling-Eschweiler}, \citenamefont {Ghazaryan}, \citenamefont {Hodajerdi}
  \emph {et~al.}}]{Bahre:2013ywa}%
  \BibitemOpen
  \bibfield  {author} {\bibinfo {author} {\bibfnamefont {Robin}\ \bibnamefont
  {B{\"a}hre}}, \bibinfo {author} {\bibfnamefont {B.}~\bibnamefont
  {D{\"o}brich}}, \bibinfo {author} {\bibfnamefont {Jan}\ \bibnamefont
  {Dreyling-Eschweiler}}, \bibinfo {author} {\bibfnamefont {Samvel}\
  \bibnamefont {Ghazaryan}}, \bibinfo {author} {\bibfnamefont {Reza}\
  \bibnamefont {Hodajerdi}},  \emph {et~al.},\ }\bibfield  {title} {\enquote
  {\bibinfo {title} {{Any light particle search II --Technical Design
  Report}},}\ }\href {\doibase 10.1088/1748-0221/8/09/T09001} {\bibfield
  {journal} {\bibinfo  {journal} {JINST}\ }\textbf {\bibinfo {volume} {8}},\
  \bibinfo {pages} {T09001} (\bibinfo {year} {2013})},\ \Eprint
  {http://arxiv.org/abs/1302.5647} {arXiv:1302.5647 [physics.ins-det]}
  \BibitemShut {NoStop}%
\bibitem [{\citenamefont {Armengaud}\ \emph {et~al.}(2019)\citenamefont
  {Armengaud} \emph {et~al.}}]{IAXO:2019mpb}%
  \BibitemOpen
  \bibfield  {author} {\bibinfo {author} {\bibfnamefont {E.}~\bibnamefont
  {Armengaud}} \emph {et~al.} (\bibinfo {collaboration} {IAXO}),\ }\bibfield
  {title} {\enquote {\bibinfo {title} {{Physics potential of the International
  Axion Observatory (IAXO)}},}\ }\href {\doibase 10.1088/1475-7516/2019/06/047}
  {\bibfield  {journal} {\bibinfo  {journal} {JCAP}\ }\textbf {\bibinfo
  {volume} {06}},\ \bibinfo {pages} {047} (\bibinfo {year} {2019})},\ \Eprint
  {http://arxiv.org/abs/1904.09155} {arXiv:1904.09155 [hep-ph]} \BibitemShut
  {NoStop}%
\end{thebibliography}%

\end{document}